\documentclass{aa} 
\usepackage[varg]{txfonts}

\usepackage{graphicx}%,subfigure}
\usepackage{epsfig}
\usepackage{color}
\usepackage{amssymb}
\usepackage{txfonts}
\usepackage{lscape}
\usepackage{float}
\usepackage{amsmath}
\usepackage{color}
\usepackage{longtable}
\usepackage{journal_names}
\usepackage{rotating}
\usepackage{pdflscape}
\usepackage{dcolumn}
\usepackage{caption}
\usepackage{subcaption}
\usepackage[nameinlink]{cleveref}

\usepackage{color,soul}
\usepackage{amssymb}
\usepackage{lscape}
\usepackage{float}
\usepackage{enumitem}
\usepackage{mathtools}
\DeclareMathOperator{\MyProd}{\scalebox{1.4}{$\mathrm{I\kern-0.2ex I}$}}
\usepackage{subcaption}
\usepackage{caption}
\usepackage{xcolor}
\usepackage[export]{adjustbox}
\DeclareCaptionLabelFormat{continued}{Figure \ref{postage}}
\newcommand*{\thead}[1]{\multicolumn{1}{c}{ #1}}

\def\la{\mathrel{\hbox{\rlap{\hbox{\lower4pt\hbox{$\sim$}}}\hbox{$<$}}}}
\def\ga{\mathrel{\hbox{\rlap{\hbox{\lower4pt\hbox{$\sim$}}}\hbox{$>$}}}}

\def\arcmin{\hbox{$^\prime$}}
\def\arcsec{\hbox{$^{\prime\prime}$}}

\newcommand{\dg}{^{\circ}}

\newcommand{\MSUN}{${\rm M}_\odot$}

\newcommand{\kms}{{\,km\,s$^{-1}$}}

\newcommand{\HI}{\mbox{\normalsize H\thinspace\footnotesize I}}

\newcommand*{\rom}[1]{\expandafter\@slowromancap\romannumeral #1@}

\begin{document}

\title{A near-infrared study of the obscured 3C129 galaxy cluster}

\author{M. Ramatsoku\inst{1,2,3}\fnmsep\thanks{m.ramatsoku@ru.ac.za} \and  M.A.W Verheijen\inst{1} \and  R.C. Kraan-Korteweg\inst{2} \and T.H. Jarrett\inst{2} \and K. Said\inst{5}  \and A.C. Schr\"{o}der\inst{4} }

\institute{Kapteyn Astronomical Institute, University of Groningen, Landleven 12, 9747 AV, Groningen, The Netherlands 
         \and
             Department of Astronomy, University of Cape Town, Private Bag X3, Rondebosch 7701, South Africa
          \and 
            Department of Physics and Electronics, Rhodes University, PO Box 94, Makhanda, 6140, South Africa 
          \and
             South African Astronomical Observatory (SAAO), PO Box 9, 7935 Observatory, Cape Town, South Africa
           \and 
             School of Mathematics and Physics, University of Queensland, Brisbane, QLD 4072, Australia
             }

 \date{Received 05 May 2020; accepted 31 August 2020}

% \abstract{}{}{}{}{} 
% 5 {} token are mandatory
 
  \abstract{We present a catalogue of 261 new infrared selected members of the 3C\,129 galaxy cluster. The cluster, located at $z \approx$ 0.02, forms part of the Perseus-Pisces filament and is obscured at optical wavelengths due to its location in the zone of avoidance. We identified these galaxies using the $J-$ and $K-$band imaging data provided by the UKIDSS Galactic Plane Survey within an area with a radius of $1.1\dg$ centred on the X-ray emission of the cluster at $\ell, b \approx 160.52\dg, 0.27\dg$. A total of 26 of the identified galaxy members have known redshifts 24 of which are from our 2016 Westerbork \HI\ survey and two are from optical spectroscopy. An analysis of the galaxy density at the core of the 3C\,129 cluster shows it to be less dense than the Coma and Norma clusters, but comparable to the galaxy density in the core of the Perseus cluster. From an assessment of the spatial and velocity distributions of the 3C\,129 cluster galaxies that have redshifts, we derived a velocity of $cz =  5227 \pm 171$ \kms\ and $\sigma = 1097 \pm 252$ \kms\ for the main cluster, with a substructure in the cluster outskirts at $cz = 6923 \pm 71$ \kms\ with $\sigma = 422 \pm 100$ \kms. The presence of this substructure is consistent with previous claims based on the X-ray analysis that the cluster is not yet virialised and may have undergone a recent merger.}

\keywords{atlases: catalogs:  infrared - galaxies: galaxy clusters - individual}

\maketitle
%
%________________________________________________________________

\section{Introduction}\label{intro}
At the nodes of large filamentary structures in the Universe, clusters of galaxies are continually growing and evolving as they accrete matter from the surrounding large-scale structures (\citealp{Ouchi2005}, \citealp{Springel2005}, \citealp{Muldrew2015}). Galaxy clusters are the largest gravitationally bound systems (\citealp{Borgani2001}, \citealp{Allen2004}, \citealp{Vikhlinin2009}, \citealp{Mantz2010}) and critical cosmological probes. Much of our present understanding on galaxy formation and evolution has been based on observing galaxies in cluster environments (\citealp{Bell2004}, \citealp{Miller2009}). Nearby galaxy clusters have particularly shaped our understanding because we are able to study them with level of detail that would not be possible for clusters at high redshifts (eg., \citealp{Drinkwater2001}, \citealp{Chung2009}, \citealp{Hammer2010}, \citealp{Karachentsev2014}, \citealp{Weinzirl2014}).

Our attention was drawn to the nearby 3C\,129 cluster when it was uncovered as an overdensity in our Nan\c{c}ay \HI\ survey of bright 2MASS Extended Sources Catalogue (2MASX; \citealp{Jarrett2000}) galaxies in the zone of avoidance (\citealp{vandriel09}, \citealp{Ramatsoku2014}, \citealp{KraanKorteweg2018}). It is part of the expansive Perseus-Pisces supercluster (PPS) at a redshift of $z \approx 0.02$. Very little is known about this cluster's galaxy population despite being part of the well-studied PPS (\citealp{Giovanelli1985}, \citealp{Giovanelli86}, \citealp{Haynes1988}, \citealp{Hudson1997}, \citealp{Hanski2001}). This is due to its location at extremely low Galactic latitudes, $\ell, b \approx 160.52\dg, 0.27\dg$, where severe extinction ($A_{B}$ $\geq$ 5 mag) and stellar confusion in the plane of the Milky Way (MW) make locating even major overdensities in the galaxy distribution inherently difficult and unreliable (\citealp{Kraan-Korteweg2000}, \citealp{Roman2000}).

This galaxy cluster was first identified with the Uhuru satellite \citep{Forman1978} in the 2$-$6 keV band as the X-ray source 4U\,0446$+$44. It was subsequently observed with the ROentgen SATellite (ROSAT)\footnote{https://heasarc.gsfc.nasa.gov/docs/rosat/rosat3.html}, the European Space Agency X-ray Observatory (EXOSAT)\footnote{https://heasarc.gsfc.nasa.gov/docs/exosat/exosat.html}, and the Einstein\footnote{https://heasarc.gsfc.nasa.gov/docs/einstein/heao2\_about.html} satellite. It was also detected and is listed in the CIZA catalogue as CIZA J0450.0$+$4501 \citep{Ebeling2002}. From analyses of these data, \citet{Leahy2000} reported an X-ray temperature of kT=5.5 keV, a total luminosity of $2.7 \times 10^{44}$ erg s$^{-1}$ in the 0.2 $-$ 10 keV, and a total cluster mass of $4.7 \times 10^{14}$\MSUN. For comparison purposes, the Coma cluster has an X-ray temperature of 8.0 keV, a total luminosity of $7.26 \times 10^{44}$ erg s$^{-1}$ \citep{Ebeling1998}, and a total cluster mass of $13 \pm 2 \times 10^{14}$\MSUN\ \citep{Hughes1989}.

\begin{figure*}
   \centering
 \includegraphics[width=175mm, height=80mm]{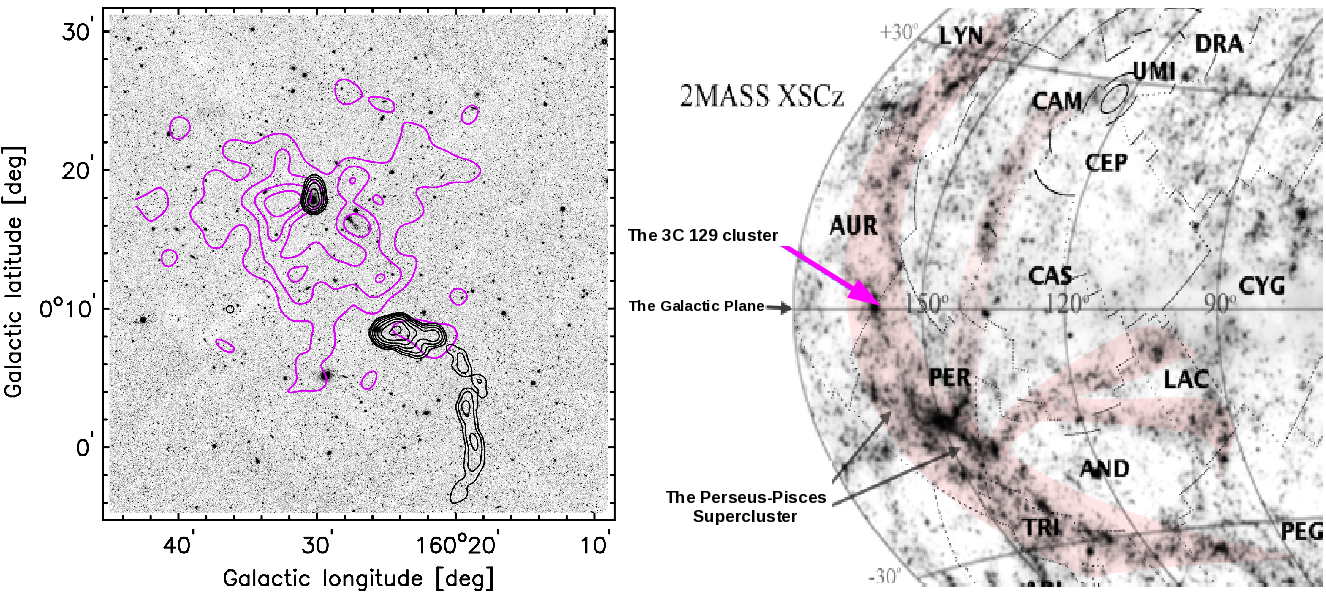}
    \caption{Left panel: The central region of the 3C\,129 cluster ($\sim 37' \times 37'$) showing the $K$-band UKIDSS image. The VLA 1424 MHz contours of the radio sources (NVSS; \citealp{Condon1998}) are overlaid in black. 3C\,129 is the head-tailed source with a curved tail on the right and the wide-angle tailed radio source on the left is 3C\,129.1. The X-ray 0.1 - 2.4 keV emission from ROSAT is overlaid in magenta. Right panel: An Aitoff Galactic projection of the galaxy distribution extracted from the 2MASS Galaxy Catalog (XSCz) in the Local Universe shown from the northern hemisphere \citep{Jarrett2004}. The image highlights the large filamentary structure of the PPS shaded in pink. The 3C\,129 cluster lies in the GP crossing at $\ell \approx 161\dg$. }\label{clusterCore}
 \end{figure*}

The cluster contains two radio galaxies, 3C\,129 and 3C\,129.1 ($cz = 6236$ and $6655$ \kms, respectively \citealp{Spinrad1975}) with jets extending far into the intra-cluster medium (ICM), as is shown in the left panel of Fig.\,\ref{clusterCore}. The radio galaxy 3C\,129 is a well-studied head-tail radio source (\citealp{Owen1979}, \citealp{Jaegers1983}, \citealp{Taylor2001}, \citealp{Harris2002}, \citealp{Lal2004}, \citealp{Murgia2016}) with a curved tail that extends over 427 kpc \citep{Krawczynski2003} and a total flux density of $\sim5.3$ Jy at 1400 MHz \citep{White1992}, while 3C\,129.1 is a wide-angle tail radio source with a total flux density of $\sim1.9$ Jy at 1400 MHz \citep{Condon1998}.  

Due to the high extinction layer of the Milky Way, most of what is known about the cluster is based on its X-ray and radio source properties; its constituent galaxies had not been observed before. For this reason we decided to perform a census of the galaxy population of this cluster by conducting a deep, blind, \HI\ imaging survey with the Westerbork Synthesis Radio Telescope\footnote{http://www.astron.nl/radio-observatory/} (WSRT). This was done using a mosaic of about $3\dg \times 3\dg$ covering the cluster and its immediate surrounding regions of the PPS (see \citealp{Ramatsoku2016} hereafter paper\,\textsc{i}).

Our aims are to investigate the large-scale and sub-structure associated with the 3C\,129 galaxy cluster and to determine its relevance to flow fields around the larger PPS environment. As evident in the 2MASS survey of extended sources, this large-scale structure envelops the cluster as it crosses from the southern to northern Galactic hemispheres \citep{Jarrett2004} (see the right panel of Fig.~\ref{clusterCore}). Our data will also aid in efforts of the 2MASS Tully-Fisher survey (2MTF; \citealp{Masters2008}) since we cover the inner ZoA regions ($b \approx |5\dg|$) which are excluded by this survey. Moreover, since we covered a wide area of the cluster with our \HI\ imaging survey, we will be able to characterise the various environments in and around the cluster to study the environmental effects on the population of galaxies therein.

\HI\ is an ideal diagnostic tool for studying processes in clusters because of its sensitivity to the environment. It has also proved most effective at detecting galaxies at low Galactic latitudes (\citealp{kraan1994}, \citealp{vandriel09}, \citealp{Henning2010}, \citealp{McIntyre2015}, \citealp{StaveleySmith2016}, \citealp{SanchezBarrantes2019}). However, it is unsuitable for tracing the dominant galaxy population of clusters which primarily comprises gas-poor, early-type galaxies. These early-type galaxies stand out at near-infrared wavelengths, particularly in the $J$ and $K$ bands. Combining the \HI\ and NIR data provides a versatile and robust tool for conducting an inventory of the galaxy population in clusters, particularly behind the Galactic Plane (GP) where dust extinction presents a challenge for optical wavelengths. 

In this paper we explore the NIR colour-magnitude properties of galaxies within the observed WSRT \HI\ mosaic to identify cluster members of the 3C\,129 cluster, and to complement our \HI\ detections at NIR wavelengths. This is achieved using data from the UKIDSS Galactic Plane Survey \citep{Lucas2008}. 

The layout of the paper is as follows; in Sect.\,\ref{SampleSect} we give a summary of the \HI\ sample and discuss the near-infrared photometry of the UKIDSS images. The colour-magnitude relation and cluster membership selection are discussed in Sect.\,\ref{CMDSect}. A subset of the catalogue of the cluster galaxies and images are described and presented in Sect.\,\ref{CatSect}, with the full catalogue and images given in Appendix~B. The near-infrared properties of the 3C129 cluster are discussed in Sect.~\ref{nirSect} and are compared with those of other galaxy clusters nearby. We describe the structure of the cluster in Sect.~\ref{substructureP2} and give a summary of the main results in Sect.~\ref{CMDSumDiscuss}. We assume a $\Lambda$ cold dark matter cosmology with $\Omega_{\rm M} = 0.3, \Lambda_{\Omega} = 0.7$ and a Hubble constant H$_{0}$ = 70 \kms\ Mpc$^{-1}$ throughout this paper.\\

%fig 1

\section{Data samples}\label{SampleSect}
\subsection{The HI data}
\begin{figure}
   \centering
   \hspace*{-0.6cm}  
 \includegraphics[width=88mm, height=80mm]{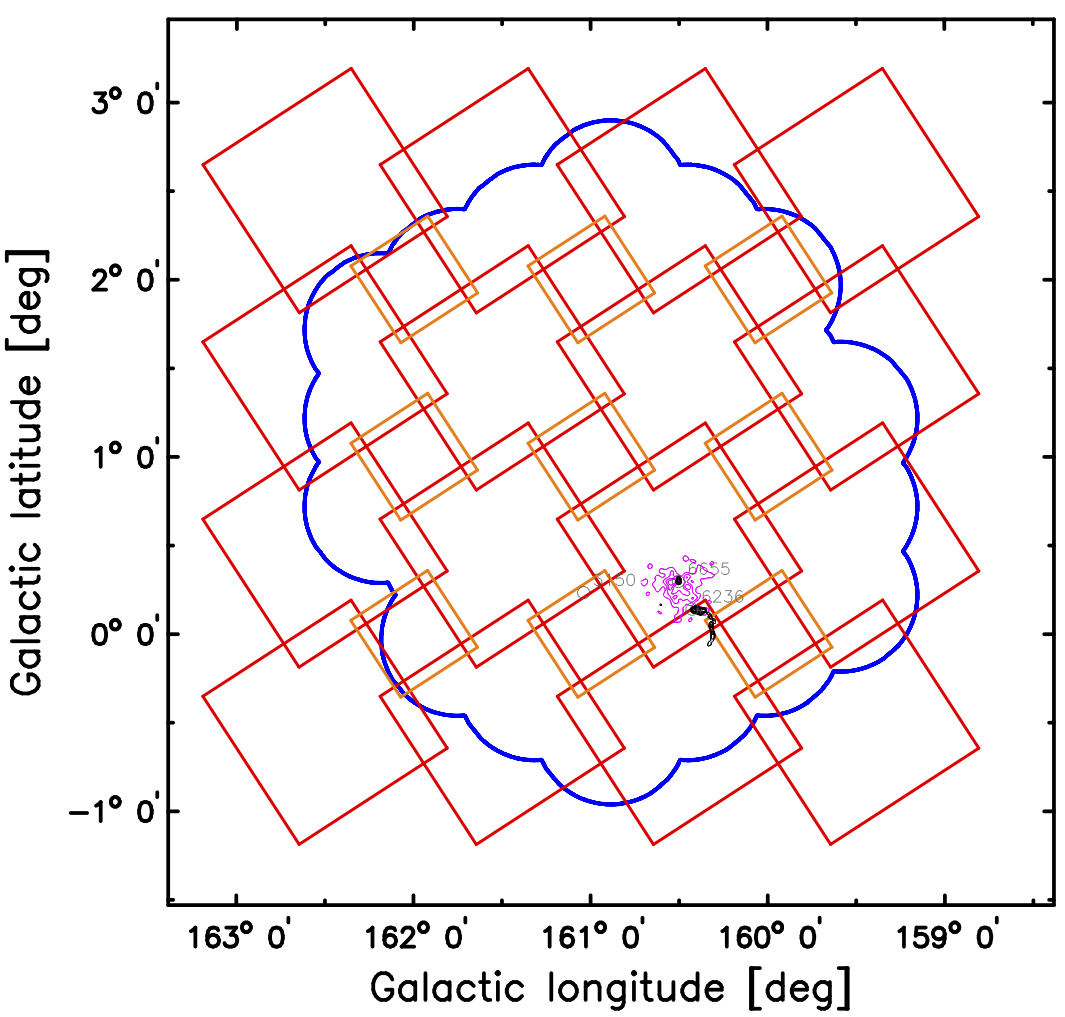}
\caption{The projected sky distribution of the observed WSRT hexagonal mosaic in blue. Coverage of the near-infrared tiles from the UKIDSS - GPS data is outlined in red. The X-ray emission is shown in magenta and the position of the contours of the radio sources 3C\,129 and 3C\,129.1 are shown in black.}\label{footprint}
 \end{figure}

For a detailed description of the WSRT \HI\ survey in and around the 3C\,129 cluster, we refer the reader to paper\,\textsc{i}. Here, only a brief summary is provided below:

The 21\,cm \HI-line imaging survey was conducted using the WSRT comprising 35 pointings each with an integration time of 12 hours, in a hexagonal mosaic centred at $\ell,b  \approx 160\dg, 0.5\dg$ (see Fig.\,\ref{footprint}) which is where the Perseus-Pisces supercluster crosses the zone of avoidance. The total sky area covered was $\sim$9.6 deg$^2$, with a survey rms sensitivity of 0.36 mJy/beam over the radial velocity range of $cz = 2400-16600$ \kms, with 16.5 \kms\  and 23\arcsec\ $\times$ 16\arcsec\ velocity and angular resolution respectively. 
We detected a total of 214 \HI\ galaxies over the entire radial velocity range, with \HI\ masses between $5 \times 10^7 - 2 \times 10^{10}$ \MSUN\ of which 80 were spatially resolved. In total 87 were detected at a redshift of the 3C\,129 cluster ($cz \sim$ $4000 - 8000$ \kms) with 24 within its radius (see Sec.\,\ref{fitRS} for how we define the radius of the cluster). The rest were found in the foreground and background of the cluster, at redshifts of, $cz = 2400-4000$ \kms\ and  $cz \sim$ $10000 - 16600$ \kms, respectively. The footprint of our \HI\ observations is shown in blue in Fig.~\ref{footprint}.

%fig 2

\subsection{The near-infrared data}\label{UKsrcDet}
The cluster is prominent in the 2MASX catalogue as illustrated in the right panel of Fig.~\ref{clusterCore}. However, we extracted our near-infrared sample of galaxies from images provided by the UKIRT Infrared Deep Sky Survey (UKIDSS: \citealp{Lawrence2007}) because of its improved depth and spatial resolution. We used Data Release 10 (DR10) of a sub-survey of UKIDSS, the Galactic Plane Survey (GPS), which covers an area of 1800 deg$^{2}$ to a 5$\sigma$ $K (2.2 \mu m)$-band depth of 18.8 mag (Vega)\citep{Lucas2008}. The UKIDSS-GPS with a pixel scale of 0.2\arcsec/pix (after microstepping) and an average seeing of 0.8\arcsec\ is highly successful at separating stars from galaxies further from the Galactic mid-plane ($|b| \gtrsim 5\dg$). However, it suffers some confusion at lower Galactic latitudes. At these latitudes visual inspection is still a superior method for identifying galaxies \citep{Lucas2008}. 

Given the large area covered by our WSRT mosaic it was not possible to visually inspect the entire region. For this reason we opted to employ intermediary steps prior to visual inspection. We started by conducting source detection using the SExtractor software (\textsc{sextractor} version 2.8.6; \citealp{Bertin1996}). The detection and extraction of sources was conducted over sixteen overlapping UKIDSS-GPS images of size $1\dg \times 1\dg$, supplemented by smaller images ($\sim30\arcmin \times 30\arcmin$) which together cover the entire \HI\ mosaic survey footprint of the WSRT (see Fig.~\ref{footprint}). We used the $K$-band for source detection since it is least affected by Galactic dust extinction, and provides higher signal-to-noise ratios compared to the $J (1.2 \mu m)$- or $H (1.6 \mu m)$-bands. 

The source detection performed by \textsc{sextractor} uses an algorithm that registers a detection when a set number of pixels are connected as specified by the detection parameters, \textsc{detect{\_}minarea} and \textsc{detect{\_}thresh}. The \textsc{global} background noise was estimated using a small mesh set by \textsc{back{\_}size} $= 64$. This size was chosen to prevent an overestimation of the background due to wings of very bright sources. To set the detection parameters, a two-fold approach was adopted; for sources with pixel values above $9\sigma$, a minimum of 10 adjacent pixels was required, and sources between $3\sigma - 9\sigma$ were only accepted if they had a minimum of 50 adjacent pixels.
\newpage
Once the sources were detected, we conducted a deblending procedure to separate objects that were recognised as connected on the sky using the deblending parameters with \textsc{deblend{\_}mincont} $= 0.001$ and \textsc{deblend{\_}nthresh} $= 32$. Values for these parameters were chosen after performing tests with a wide range of values and verifying results by visual inspection to ensure that spurious detections were minimised and that no obvious sources were missed. 
In Table\, A.1, we provide a summary of the \textsc{sextractor} parameters we adopted for detecting sources in the UKIDSS-GPS $K$-band images.

\subsection{Star-galaxy separation}\label{StarGalSep}
Although \textsc{sextractor} parameters described in section \ref{UKsrcDet} were optimised to limit the number of stars and spurious detections, the high stellar density in the GP necessitated further processing to separate the stars from galaxies. In this section we describe the three-step process that was conducted for this purpose. 

\subsubsection{\textsc{sextractor} stellarity index}
\begin{figure}
   \centering
   \hspace*{-0.6cm} 
 \includegraphics[width=90mm, height=60mm]{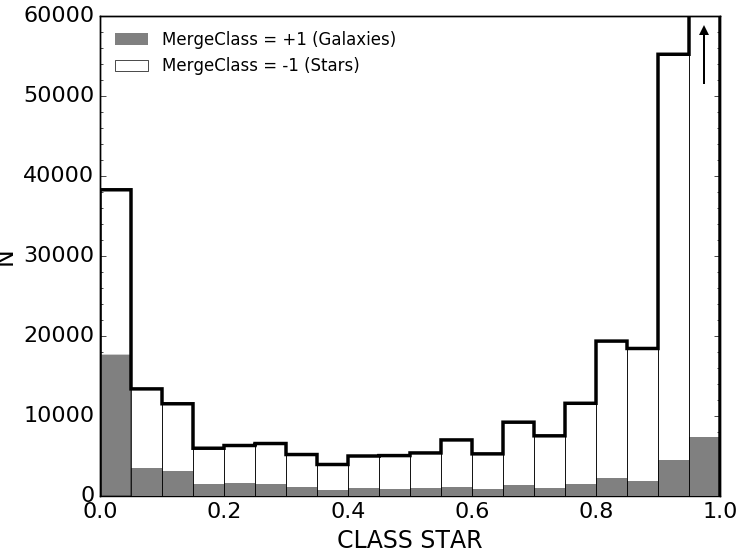}
    \caption{The distribution of \textsc{sextractor} stellarity index, \textsc{class{\_}star} of all objects found in the UKIDSS-GPS searched images. All objects with \textsc{class{\_}star} $>$ 0.8 are stars, those with \textsc{class{\_}star} $<$ 0.35, galaxies and 0.35 $<$\textsc{class{\_}star} $<$ 0.8 are ambiguous objects. The open histogram shows the distribution of objected labelled as stars in the UKIDSS catalogue (\textsc{mergeClass}=-1) and the filled histogram are those labelled as galaxies (\textsc{mergeClass}=+1). The horizontal arrow points to a higher number count of $\sim10^5$ in that bin that is not shown in the plot.}\label{stellarity}
 \end{figure}

%\textbf{\textsc{sextractor} stellarity index:} 
Firstly, we performed a preliminary star-galaxy classification using \textsc{sextractor}'s stellarity index (\textsc{class{\_}star}). This is a dimensionless parameter that characterises objects as either point-like or extended. This parameter is based on a neural-network analysis approach that compares the point spread function to the object scale and then provides a confidence level estimate of the classification which ranges from 0 for galaxies to 1 for stars. 
Figure~\ref{stellarity} shows the distribution of the \textsc{class{\_}star} values of all sources which were detected using procedures described in Sect.~\ref{UKsrcDet}. From this plot two distinct populations with \textsc{class{\_}star} values of 0 and 1 are evident. Moreover, the distribution of sources classified as galaxies shows a sharp decline from \textsc{class{\_}star} = 0.0 to 0.35, it then flattens and rises steeply from \textsc{class{\_}star} = 0.8 to 1.0. Based on the distribution of the stellarity index, \textsc{class{\_}star}, we classified sources as:

\begin{description}
\item{Galaxies; \textsc{class{\_}star} $\leq$ 0.35.}
\item{Stars or galaxies; 0.35 $<$\textsc{class{\_}star} $<$ 0.8.}
\item{Stars; \textsc{class{\_}star} $\geq$ 0.8.}
\end{description}

~\\
\subsubsection{UKIDSS-GPS star-galaxy separator}
%\textbf{UKIDSS-GPS star-galaxy separator:} 
Secondly, we validated objects classified as galaxies by \textsc{class{\_}star} by cross-correlating them with the UKIDSS-GPS catalogue which also provides a star-galaxy classification parameter, \textsc{mergedClass}. This parameter is based on quantitative measurements in the $J, H$ and $K$ passbands of the observed radial profiles by the UKIDSS data reduction pipeline \citep{Dye2006}. The profiles are characterised by the `mergeClassStat` parameter and identified as either point-like, resolved or artefacts. Resolved sources are assigned large ``mergeClassStat`` values, while smaller values are associated with point-like sources. Moreover, the UKIDSS-GPS further separates galaxies from stars by making cuts based on the difference between their Petrosian (resolved) magnitude and the point source magnitude \citep{Lucas2008}. In this catalogue galaxies are assigned \textsc{mergeClass}$=+1$ and stars are assigned \textsc{mergeClass}$=-1$.\\

\subsubsection{Visual inspection}
%\textbf{Visual inspection:} 
We found that the \textsc{class{\_}star} and \textsc{mergeClass} parameters are not reliable under all circumstances. For instance, out of all sources with \textsc{class{\_}star} = 0.0 only about 47\% were classified as galaxies by \textsc{mergeClass} (see Fig.~\ref{stellarity}). Because of this, we needed to ensure that galaxies were not missed based on these parameters, particularly in the ambiguous 0.35 $<$ \textsc{class{\_}star} $<$ 0.8 range. This necessitated a further visual inspection step to improve the classification of each object, and to remove any spurious detections. This was done for all objects with \textsc{class{\_}star} $<$ 0.8 and  \textsc{mergeClass} = +1 and \textsc{mergeClass} = -1, those with \textsc{class{\_}star} $>$ 0.8 were not inspected visually since only 0.05\% of the objects in these bins had \textsc{mergeClass} = +1. We conducted the visual inspection using \textsc{ds9} on the $J, H$ and $K$ bands three-colour (RGB) composite images from UKIDSS. The extended nature of galaxies and their colour made them readily distinguishable in this manner. \\ 

After removing artefacts, stars and duplicate detections in overlapping regions of the UKIDSS images, the final sample consisted of 9737 unique galaxies spatially distributed over the full WSRT mosaic. In section~\ref{fitRS}, we give a full description of how we selected galaxies belonging to the 3C\,129 cluster. \\

\subsection{Star subtraction}
Due to the high stellar density at low Galactic latitudes, images required additional steps to remove stars superimposed on the galaxies to obtain accurate photometry. We conducted this by fitting a point-spread function (PSF) to stars and subsequently removed them from the galaxy. For this purpose we used a four-step script based on the \textsc{iraf} routine \textsc{killall} \citep{Buta1999}, which in turn is based on \textsc{daophot} tasks \citep{Stetson1987}. 
The four steps were as follows: (a) We used \textsc{imstat} to determine the rms and sky background in the image; (b) modelled the galaxy using \textsc{ellipse} and \textsc{bmodel}, and removed this model from the image. It should however, be noted that galaxy structures such as spiral arms may be not be modelled fully and might result in residuals in the galaxy-subtracted image; (c) used \textsc{sextractor} to detect bright stars that are above 3.5$\sigma$ in the image in which the galaxy has been subtracted. The PSF photometry of these sources was then determined using \textsc{daophot} (tasks \textsc{allstar} and \textsc{phot}) and removed from the image. This step was repeated for fainter stars above 1.8$\sigma$. To mitigate against misidentifying residual structures as stars, we required these stars to be detected by  \textsc{daofind} at the 2$\sigma$ threshold;  (d) the two lists with the bright and faint stars were combined and removed from the original image using the \textsc{substar} task; (e) residuals resulting from imperfect PSF fits to stars in the image were found and removed. We repeated steps (b) to (e) four times, each time improving the galaxy model, which resulted in a more reliable star removal. 

\begin{figure}
   \centering
 \includegraphics[width=90mm, height=115mm]{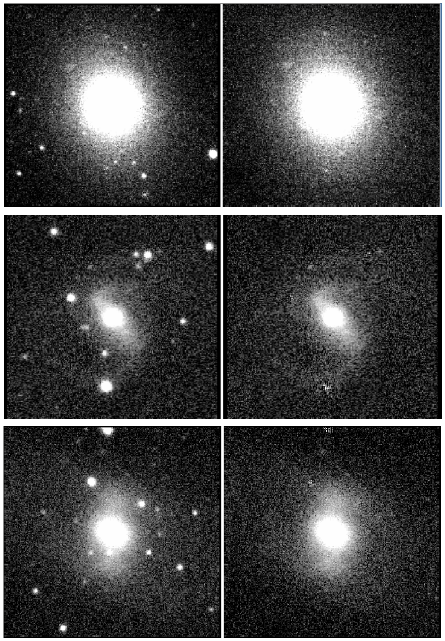}
    \caption{Postage stamp examples of $K$-band images of galaxies before (left panels) and after star-subtraction (right panels).}\label{starsub}
 \end{figure}

Figure~\ref{starsub} shows example images for three galaxies in our sample before (left panel) and after (right panel) star-subtraction. It illustrates that this star removal procedure performs quite well. In fact, it has been shown by \citet{Woudt2005} and \citet{Mutabazi2014}, through simulations of subtracting artificial stars added to non-crowded fields, that removing stars this way has a minimal effect (less than 0.01 mag on average) on the galaxy photometry.\\

\subsection{The NIR photometry}
After subtracting all the stars we extracted the galaxy photometric parameters. Since UKIDSS photometric parameters are only measured within circular apertures which are inadequate for elongated galaxies, we used customised python scripts developed by K. Said\footnote{khaled.said@anu.edu.au} and W. Williams\footnote{w.williams@strw.leidenuniv.nl} based on a combination of \textsc{sextractor} and \textsc{iraf} tasks. 

\subsubsection{Astrometry of positions}
%\textbf{Astrometry of positions:} 
First, the central positions of galaxies were determined by adding the three images from the $J$+$H$+$K$ bands and measuring the intensity-weighted centroid. This method of determining positions is more accurate as it measures the centre of the galaxy from higher signal-to-noise images \citep{Jarrett2000}. 

\subsubsection{Ellipticities and geometries}
%\textbf{Ellipticities and geometries}: 
Next, ellipticities and position angles were determined individually from the $J$, $H$ and $K$ band images. The \textsc{iraf} task \textsc{ellipse} was used to fit ellipses to the galaxy image. Ellipticities, $\epsilon = (1 - b/a)$  and position angles ($\phi$; counter clockwise from north) were then fitted at semi-major axis intervals, while keeping the central coordinates, X and Y fixed. The \textsc{ellipse} task produced tables containing the intensity (in counts) within the ellipse and the ellipticity and position angle at each semi-major axis interval including their errors. Then the galaxy's position angle and ellipticity were determined to be the average value in the outer disc between 1$\sigma$ - 2$\sigma$ isophotes, where $\sigma$ is the sky background rms. 

\subsubsection{Magnitudes}
%\textbf{Magnitudes:} 
Finally, the isophotal magnitudes were measured within elliptical apertures in all bands ($J$, $H$, $K$). The $K$-band surface brightness $\mu$ = 20 mag arcsec$^{-2}$ isophote was adopted to determine the isophotal radius ($r_{K_{20}}$). Fiducial magnitudes were measured within this same aperture of radius $r_{K_{20}}$ for the $J$, $H$, $K$-bands. All the magnitudes are based on the Vega calibration. 

\subsection{Photometric checks}
To check our photometric consistency, we derived from the UKIDSS images galaxy magnitudes within a 7\arcsec\ radius circular aperture and compared these with their counterparts in the 2MASX catalogue \citep{Jarrett2000} in the same 7\arcsec\ circular aperture. Sources in both catalogues were matched if they were separated by less than 1.5\arcsec. We used this small correlation radius of 1.5\arcsec\ because the UKIDSS WFCAM astrometric calibration was derived from 2MASS \citep{Hodgkin2009}, which has a positional accuracy of about 0.5\arcsec. Only galaxies with reliable photometry (as flagged in the 2MASX catalogue) were compared. We found 363 counterparts in the 2MASX catalogue. \\

\begin{figure}[h]
   \centering
 \includegraphics[width=85mm, height=120mm]{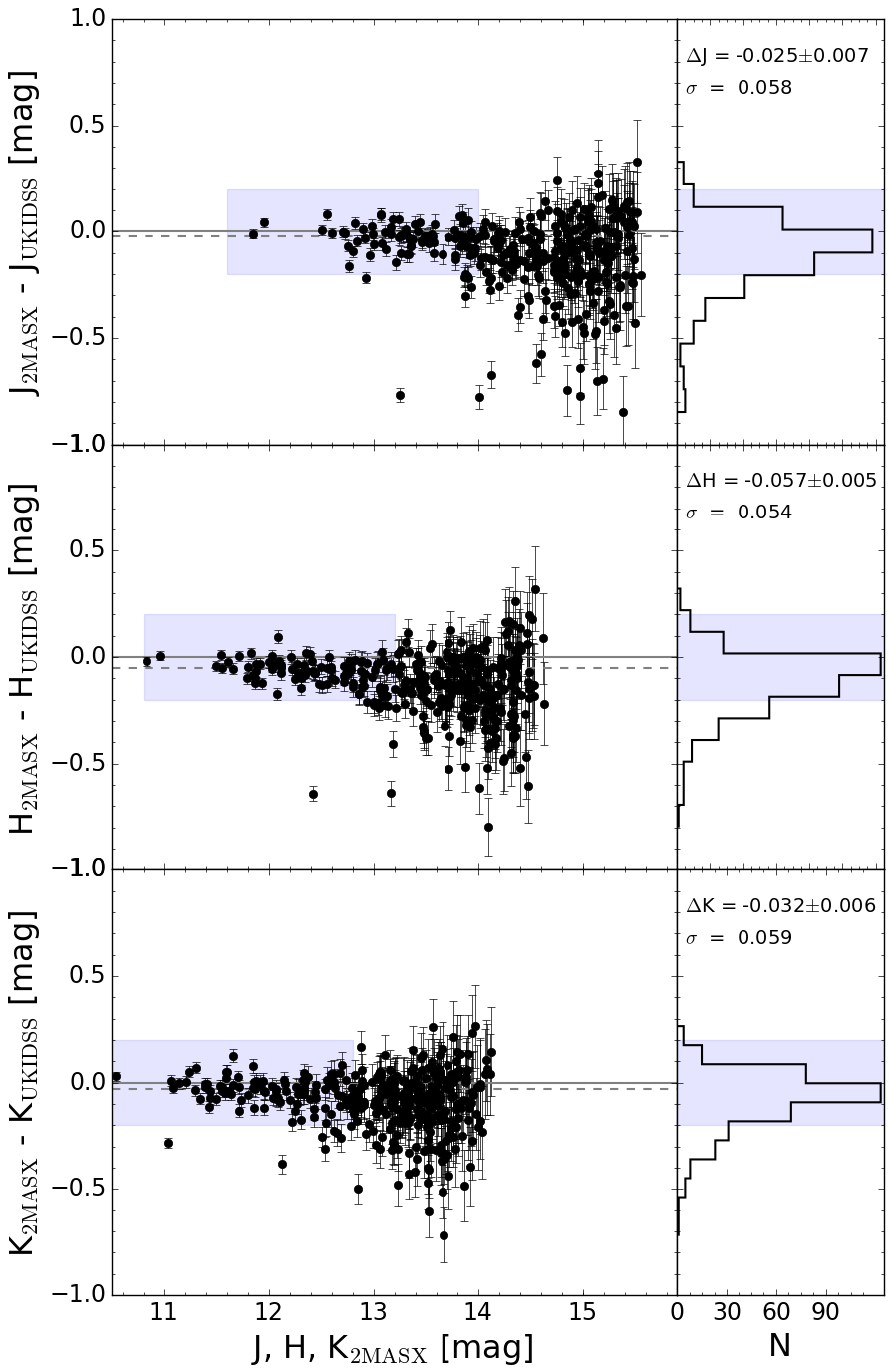}
    \caption{Comparison between the 2MASX and UKIDSS circular aperture magnitudes measured within a 7\arcsec\ radius for the $J$ (top), $H$ (middle) and $K$ (bottom) bands. The median offset between the two surveys is indicated by the dashed line, it is printed in the top right legend. This median was determined within areas marked by the blue shades also marked in the histograms.}\label{comparison2mass}
 \end{figure}

In Fig.~\ref{comparison2mass}, we show a comparison of our UKIDSS photometry with that of the 2MASX. The comparison relation used is:
\begin{equation}
\Delta m = m_{2MASX} - m_{UKIDSS}.
\end{equation}

\noindent
By this equation, a positive $\Delta$m indicates that 2MASX magnitudes are fainter than our UKIDSS magnitudes.  
The offsets given in the top right corners of Fig.~\ref{comparison2mass} indicate that the 2MASX galaxies are slightly brighter than their UKIDSS counterparts. These offsets can be attributed to the 2MASX low angular resolution ($\sim$ 2.0\arcsec), that cannot resolve the sources. The improved resolution of the UKIDSS ($\sim$ 0.8\arcsec) on the other hand allows for the detection and subtraction of fainter foregrounds stars that would otherwise contribute to the galaxy's brightness within the measurement aperture. Moreover the different filters may also have contributed to slight offsets because the UKIDSS uses the Mauna Kea Observatory (MKO; \citealp{Hodgkin2009}) filter set which is slightly different from the 2MASS filter set \citep{Cohen2003}.\\

\subsection{Extinction}\label{extinction}
Galaxies in the ZoA tend to have their brightness reduced due to the dust obscuration in the Milky Way. This effect is less prominent in the near-infrared compared to the optical, but it is nonetheless not negligible. Within our WSRT field the Galactic extinction values range from $A_{K} = 0.16 - 0.56$ mag (see  Fig.~7 of paper\,\textsc{i}).

We correct for this effect using the extinction law; 

\begin{equation}
\frac{A_{\lambda}}{E(B - V)} = R_{\rm v},
\end{equation}
~\\ 
where A$_{\lambda}$ is the extinction in a given band and $E(B - V)$ is the colour reddening from \citet{Schlegel1998} with the \citet{Schlafly2011} correction factor of 0.86 applied. The value for $R_{V}$ = 3.1 is given by \citet{Schlegel1998}. 
Using the parametrisation by \citet{Fitzpatrick1999} we derive the extinction in the $J$, $H$ and $K$ bands to be,
 
 \begin{equation}
 A_{J} = 0.741E(B - V), 
\end{equation}
  \begin{equation}
 A_{H} = 0.456E(B - V), 
\end{equation}
 \begin{equation}
 A_{K} = 0.310E(B - V).
\end{equation}

The effect of extinction on our galaxies is low given that our field has a median extinction of  $A_{K} = 0.29 \pm 0.05$ mag, however this is not negligible. This extinction also translates into a reduction of the isophotal radius of the galaxy due to the loss of the fainter, low surface brightness outer regions. According to \citet{Riad2010} extinction at for example, $A_{K}$ = 1.0 mag can reduce the isophotal radius by $\sim28$\%.  %We thus applied further corrections to the magnitudes for changes in the isophotal size due to dust extinction following the optimised correction based on central surface brightness as prescribed by \citet{Riad2010}. 
  
\section{The colour-magnitude relation}\label{CMDSect}
The colour-magnitude relation (CMR) has been shown to be a powerful technique for selecting galaxies in clusters since they form a distinct linear feature in the colour-magnitude diagram \citep{Visvanathan1977}. This feature, known as the red sequence is typically composed of passive early-type galaxies with old stellar populations. It is not yet clear when in the history of the Universe the red sequence emerged, however, several studies have shown that it is present at all redshifts up to $z \sim 1$ (\citealp{Bell2004}, \citealp{Menci2005}, \citealp{Scarlata2007}, \citealp{Mei2009}). Great efforts have been taken with spectroscopic and photometric surveys to push the redshift limit further. Several massive red galaxies have been detected at high redshifts through these studies, however, they are often constrained by low sample numbers in the case of the former, while the latter suffers uncertainties in the rest-frame colour measurements at high redshifts resulting in a large scatter on the CMR and less-defined red sequence (\citealp{Cirasuolo2007}, \citealp{Franzetti2007}, \citealp{Cassata2008}, \citealp{Taylor2009}, \citealp{Tanaka2010}, \citealp{Spitler2012}). Regardless of these limitations, when the red sequence has been clearly observed, it has a well-defined slope that evolves with redshift and a small ($< 0.1$ mag) intrinsic scatter (\citealp{Bower1992}, \citealp{Gladders1998}, \citealp{Lopez2004}, \citealp{Stott2009}). 

In addition to the tight CMR, galaxies in regular clusters have a radial distribution that is typically centrally concentrated \citep{Dressler1980}. They also form the dominant population at the bright end of the galaxy luminosity function and create a high contrast against the background, thus making them easily recognisable in surveys \citep{Gladders1998}. 

Given the scarcity of redshift information of galaxies in the 3C\,129 cluster we use the CMR in the near-infrared to identify cluster members because most galaxies are invisible at the optical wavelength due to extinction in this region. We use the red sequence feature in the CMR as well as the angular on-sky projection distribution to detect clustering. The advantage of using this technique is that it is not affected by projection effects since any random distribution of galaxies will not form a coherent red sequence signature in the colour-magnitude diagram.\\

\begin{figure}
   \centering
  %\hspace*{-0.6cm}  
 \includegraphics[width=90mm, height=57mm]{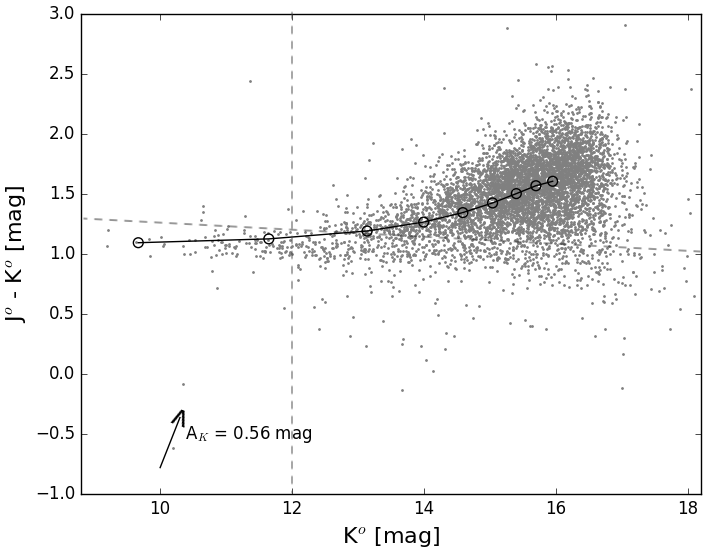}
    \caption{The colour-magnitude diagram of 9737 sources from the UKIDSS-GPS within the WSRT mosaic (see Sect.~\ref{UKsrcDet}). The reddening curve computed from redshift $z =  0.02$ to $z = 0.4$ is shown by the black line with open circles representing the redshift bin and the Galactic dust reddening vector for a K-band extinction of 0.56 mag is indicated with the arrow. Dashed lines represent the boundaries of red sequence cluster members compared to blue outliers (vertical line) or background galaxies (slanted line).}\label{cmdall}
 \end{figure}

In Fig.~\ref{cmdall} we present the NIR colour-magnitude diagram (CMD) of all 9737 galaxies within the observed WSRT mosaic (see Fig.~\ref{footprint}). Colours were measured inside the $\mu_{k}$ = 20 mag arcsec$^{-2}$ fiducial isophotal aperture defined in the $K$-band and corrected for foreground extinction based on the DIRBE/IRAS maps as described in Sect.~\ref{extinction}. \\

~\\ 

Two well-defined populations are apparent in this plot. The dominant population ($K^{o} >$ 12 mag, $J^{\rm o} - K^{\rm o} \gtrsim 1.3$ mag) consists of redshifted background galaxies. The overlaid reddening curve\footnote{http://kcor.sai.msu.ru} \citep{Chilingarian2010} supports this claim. The other identifiable population is a slightly slanted band of galaxies that stretches between $8 \lesssim K \lesssim 16$ mag with colours, $J^{\rm o} - K^{\rm o} \approx 1.1$ mag. This second feature contains within it the red sequence of the 3C\,129 cluster. Below the red sequence is a small population of blue outliers which were visually inspected and found to be close binary stars and/or artefacts that were missed during the star-galaxy separation process in the UKIDSS images (see Sect.~\ref{StarGalSep}).

 \subsection{The red sequence and cluster membership}\label{fitRS}
To determine the galaxy cluster membership we used a three step process;\\

Firstly, a region of the CMD containing the red sequence was selected and an estimate of the initial fit was made by visual inspection. All points deviating by more than 3$\sigma$ from this initial fit were removed from the full sample. We then performed an iterative linear fit of the form, ${J-K} = \alpha K + c$ where the slope of the red sequence is denoted by $\alpha$ to points that were within $\pm3\sigma$ of the initial fit. 
The final fit gives $(J-K)^{o} = -(0.023\pm0.002)K^{o} + 1.25$ with a $1\sigma$ dispersion of 0.039 mag. The slope matches that of the Coma cluster of $\alpha = -0.017 \pm 0.009$ (measured by \citealp{Stott2009}), which is at a similar redshift of $z \approx 0.02$ as the 3C\,129 cluster. The $J-K$ colours of the red sequences are also comparable to that of galaxy clusters in the WIde-field Nearby Galaxy-cluster Survey (WINGS: \citealp{Valentinuzzi2011}). Galaxies within $\pm3\sigma$ of the final fit formed the initial list of the cluster member candidates. \\

Secondly, we limited the relative background contamination in the initial sample list of red sequence galaxies by only selecting galaxies with $r_{K_{20}} > $3\arcsec. For galaxies with $r_{K_{20}} > $3\arcsec\ the red sequence becomes increasingly contaminated by background galaxies which are hard to discriminate from true cluster members. Moreover, at these sizes cluster members will fall in the dwarf regime and are of no interest to this study, and fall below the cluster completeness limit of $K^{o} \approx 15.0$~mag or ($M_{\rm K} \approx -19.7$~mag; see Sect.\,\ref{CatSect}) .\\

%This was motivated by the fact that the red sequence of the cluster becomes less well-defined \textbf{and heavily contaminated for galaxies with $r_{K_{20}} < $ 3\arcsec\ since these comprise 82\% of all galaxies in the sample with magnitudes of $K^{o} \gtrsim 15.5$~mag and therefore probably background sources}. 

Thirdly, we defined a radius $r_{cl}$ of the cluster centred on the ROSAT X-ray emission. This cluster radius was chosen based on the spatial distribution (cf. Fig.~\ref{spatialScatter}) of the red sequence galaxies which have radii larger than $r_{K_{20}} > $ 3\arcsec\ out to where they became sparsely distributed. Assuming a redshift of $z = 0.02$ the radius was found to be $r_{cl} \approx 1.7$~Mpc ($\sim 0.8r_{A}$)\footnote{$r_{A}$ is the Abell radius}. It corresponds to about $1.34R_{200}$, where $R_{200} $ is the radius at which the average interior density of the cluster is $200 \times \rho_{c}$, and $\rho_{c}$ is the critical density of the Universe. The $R_{200}$ of the 3C\,129 cluster is $\sim 1.24$~Mpc based on the 0.1 $-$ 2.4~keV band measurements by ROSAT \citep{Piffaretti2011}. We therefore only selected red sequence galaxies within $r_{cl} < 1.34R_{200}$ as potential cluster members.\\ 

\begin{figure}[h]
   \centering
 %\hspace*{-0.6cm}   
 \includegraphics[width=88mm, height=60mm]{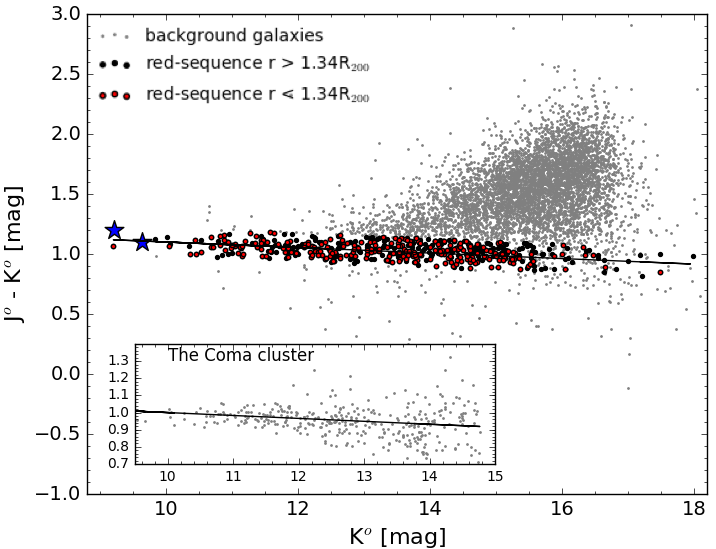}
    \caption{The colour-magnitude diagram showing the red sequence signature of the 3C\,129 cluster in red with the two blue stars identifying the bright radio sources, 3C\,129 and 3C\,129.1. The fitted slope, $\alpha = -0.023 \pm 0.002$ mag of the cluster is illustrated by the black solid line and compared to that of the Coma cluster using 2MASX data in the inserted CMD. Black points are galaxies outside the cluster radius of $r_{cl} < 1.34R_{200}$ and grey points are mostly background galaxies.}\label{cmdRS}
 \end{figure}

After applying the steps outlined above, we obtained a sample of 250 galaxy members of the 3C\,129 cluster which lie on the red-sequence. Of these galaxies, 13 were already detected in our \HI\ survey \citep{Ramatsoku2016}. The CMD of all galaxies on the red sequence is shown in Fig~\ref{cmdRS}. In addition, we included 11 more galaxies with $r_{K_{20}} < 3$\arcsec\ that were detected in \HI\  within the velocity window and the defined radius of the cluster on the sky regardless of their location on the red-sequence. Thus the final sample of the 3C\,129 cluster members comprises 261 galaxies. 

We note that we cannot measure the contamination of the final sample by background galaxies without redshift measurements. However, it has been previously shown that this method of finding galaxy cluster members in both the infrared and optical wavelengths results in a background contamination of less than 5\% \citep{Gladders2000}.\\

\subsection{Morphologies of the cluster members}
\begin{figure}
   \centering
  %\hspace*{-0.6cm}  
 \includegraphics[width=88mm, height=60mm]{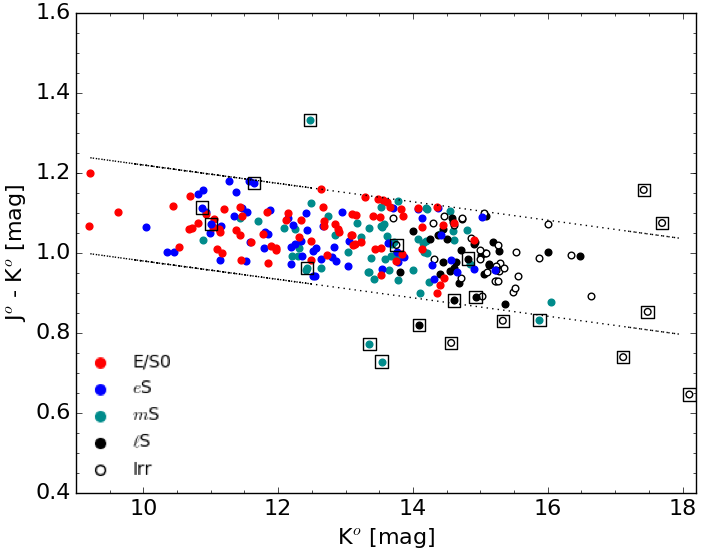}
    \caption{The CMD of the 3C\,129 cluster identified as members. Colours indicate morphological type. Circles are the red-sequence galaxies, and those enclosed in squares are H\textsc{i} detected galaxies within the radius and velocity range of the cluster. The dotted lines show the $\pm3\sigma$ rms dispersion (0.12 mag) from the fitted red sequence relation:  ${J-K} = -0.023K + 1.25$.}\label{cmdMorph}
 \end{figure}

Figure~\ref{cmdMorph} shows the CMD of the cluster with galaxies separated into approximate morphologies based on a visual inspection and a subjective estimate of their bulge-to-disc ratios, their ellipticities and compactness. The morphological classification was independently conducted and adjudicated by co-authors. This classification is only meant to distinguish large-scale trends and to give a general description of galaxy morphologies in the cluster. Galaxies were classified into five broad groups of early-types (E/S0), early spirals ($e$S) such as Sa/Sb, medium spirals ($m$S) which were typically of Sc/Sd type, the late-type spirals ($\ell$S) which were mostly Sdm, and Irregular (Irr) galaxies without any discernible structure. In this plot we also indicate with black squares, the galaxies detected in \HI\ (with UKIDSS counterparts) within the velocity range and radius of the cluster.  

From Fig.\,\ref{cmdMorph} we find that early-type galaxies (E/S0) and the early-spirals are mostly located at the bright end of the red sequence. This is to be expected given that we are evaluating a potentially rich cluster which tends to have a higher fraction of these bright early-type galaxies than star forming late-type galaxies. 

The mid-range spiral galaxies ($m$S) show no obvious trends in the CMD except that they are predominantly at fainter magnitudes compared to the early-type galaxies. There is a number of these type of galaxies ($\sim$ 23\%) detected near the centre of the cluster (see also Figs.~\ref{spatialScatter} and \ref{morphR} in Sect.~\ref{SpaSect}). Interestingly, they were not detected in \HI, suggesting they might be stripped of their gas by ram-pressure from the ICM upon infall, and are \HI\ deficient. Judging by their morphologies alone, these should have otherwise been detected in our \HI\ observations. 

Late-type spirals and irregular galaxies are mostly found at the faint end of the CMD, that is, at $K^{o} \gtrsim 14$ mag. These are typically LSB galaxies which could have lost their gas through tidal stripping or interaction with the ICM \citep{Conselice2003}. On the other hand, most of the \HI\ detected galaxies also populate this end of the CMD, but tend to be slightly bluer ($J^{\rm o} - K^{\rm o} \lesssim 1.0$ mag) than their LSB counterparts. They are located in the cluster outskirts where we expect less interaction and where we expect galaxies to have retained their gas.\\

Analyses of these populations of galaxies as well as the effect of the various environments in which they reside, requires  a more comprehensive analysis than presented in this paper. Here we only present a first description. We will study and discuss the detailed nature of these galaxies in the forthcoming paper\,\textsc{iii}.\\

\subsubsection{The catalogue of candidate cluster galaxies}\label{CatSect}
\begin{figure*}
\centering

\begin{subfigure}[b]{0.19\textwidth}\includegraphics[width=3.4cm, height=3.4cm]{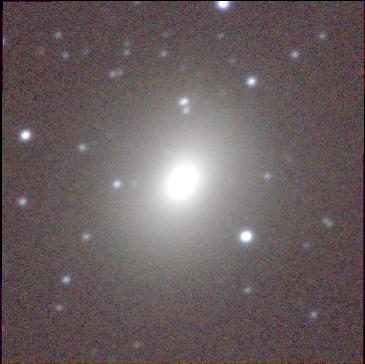}\caption*{1. E/S0}\end{subfigure}
\begin{subfigure}[b]{0.19\textwidth}\includegraphics[width=3.4cm, height=3.4cm]{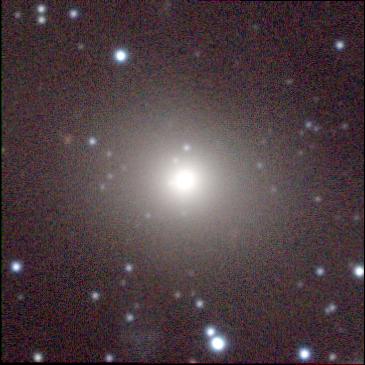}\caption*{2. E/S0}\end{subfigure}
\begin{subfigure}[b]{0.19\textwidth}\includegraphics[width=3.4cm, height=3.4cm]{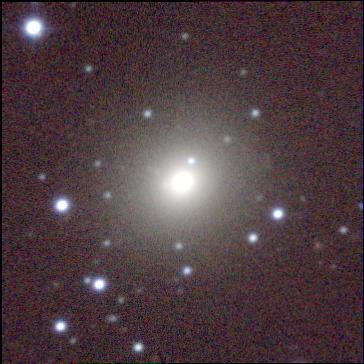}\caption*{3. E/S0}\end{subfigure}
\begin{subfigure}[b]{0.19\textwidth}\includegraphics[width=3.4cm, height=3.3cm,cfbox=cyan 1.5pt 0.0pt]{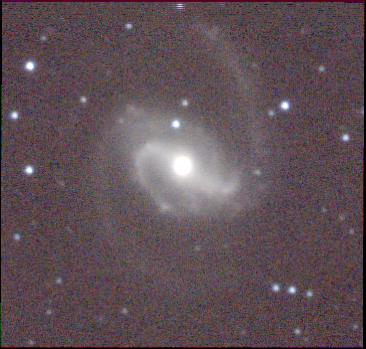}\caption*{4. $e$S}\end{subfigure}
\begin{subfigure}[b]{0.19\textwidth}\includegraphics[width=3.4cm, height=3.4cm]{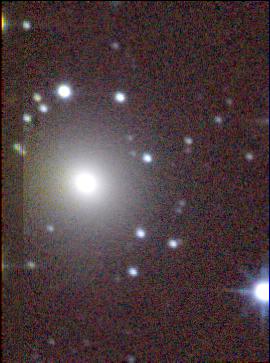}\caption*{5. E/S0}\end{subfigure}

\begin{subfigure}[b]{0.19\textwidth}\includegraphics[width=3.4cm, height=3.4cm]{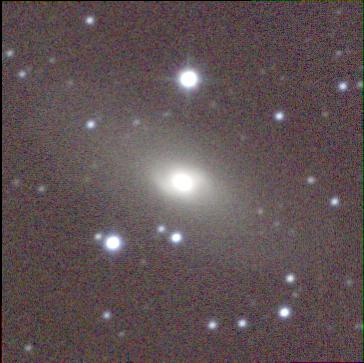}\caption*{6. $e$S}\end{subfigure}
\begin{subfigure}[b]{0.19\textwidth}\includegraphics[width=3.4cm, height=3.4cm]{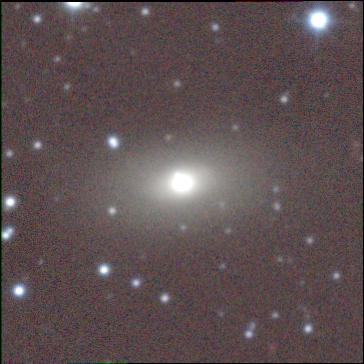}\caption*{7. E/S0}\end{subfigure}
\begin{subfigure}[b]{0.19\textwidth}\includegraphics[width=3.4cm, height=3.4cm]{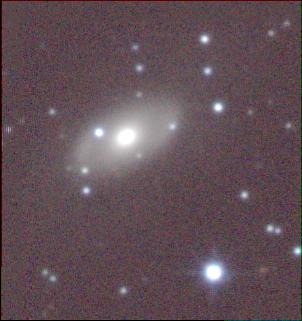}\caption*{8. $e$S}\end{subfigure}
\begin{subfigure}[b]{0.19\textwidth}\includegraphics[width=3.4cm, height=3.4cm]{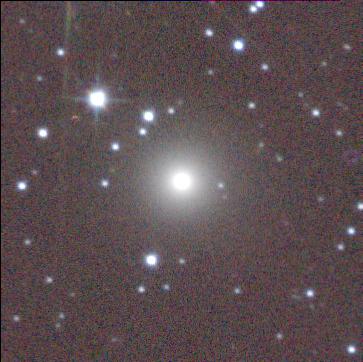}\caption*{9. E/S0}\end{subfigure}
\begin{subfigure}[b]{0.19\textwidth}\includegraphics[width=3.4cm, height=3.4cm]{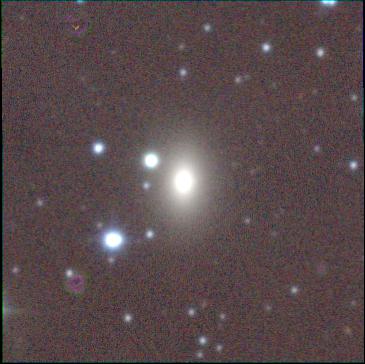}\caption*{10. E/S0}\end{subfigure}

\begin{subfigure}[b]{0.19\textwidth}\includegraphics[width=3.4cm,height=3.3cm,cfbox=cyan 1.5pt 0.0pt]{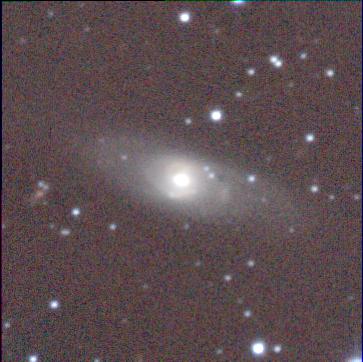}\caption*{11. $e$S}\end{subfigure}
\begin{subfigure}[b]{0.19\textwidth}\includegraphics[width=3.4cm, height=3.4cm]{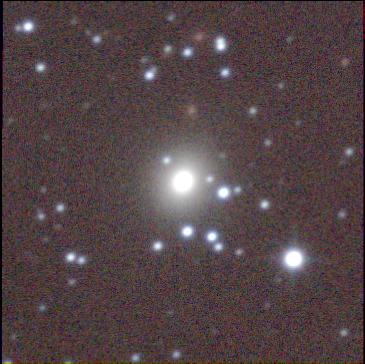}\caption*{12. E/S0}\end{subfigure}
\begin{subfigure}[b]{0.19\textwidth}\includegraphics[width=3.4cm, height=3.4cm]{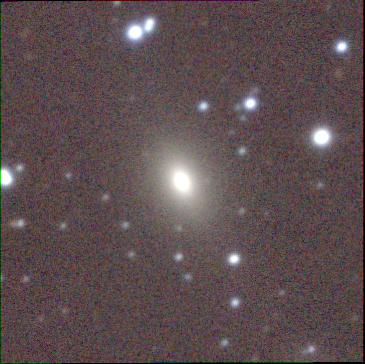}\caption*{13. E/S0}\end{subfigure}
\begin{subfigure}[b]{0.19\textwidth}\includegraphics[width=3.4cm, height=3.4cm]{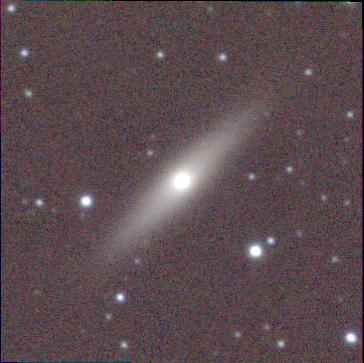}\caption*{14. $e$S}\end{subfigure}
\begin{subfigure}[b]{0.19\textwidth}\includegraphics[width=3.4cm, height=3.4cm]{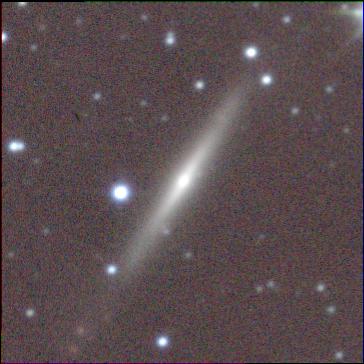}\caption*{15. M - S}\end{subfigure}

\begin{subfigure}[b]{0.19\textwidth}\includegraphics[width=3.4cm, height=3.3cm,cfbox=cyan 1.5pt 0.0pt]{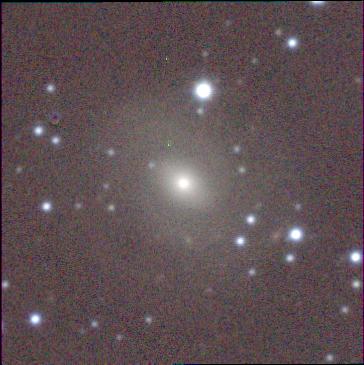}\caption*{16. $e$S}\end{subfigure}
\begin{subfigure}[b]{0.19\textwidth}\includegraphics[width=3.4cm, height=3.4cm]{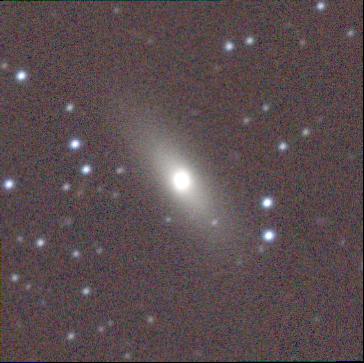}\caption*{17. E/S0}\end{subfigure}
\begin{subfigure}[b]{0.19\textwidth}\includegraphics[width=3.4cm, height=3.4cm]{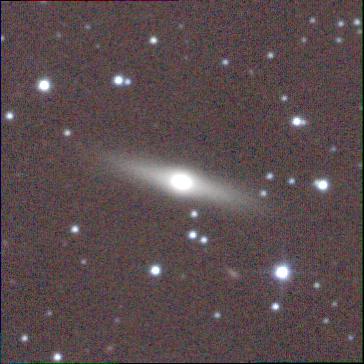}\caption*{18. E/S0}\end{subfigure}
\begin{subfigure}[b]{0.19\textwidth}\includegraphics[width=3.4cm, height=3.4cm]{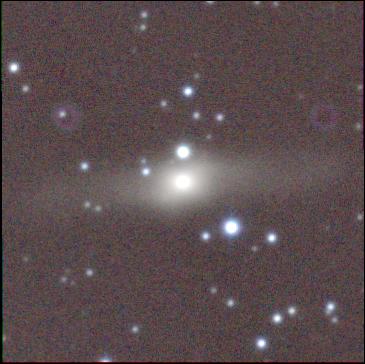}\caption*{19. $e$S}\end{subfigure}
\begin{subfigure}[b]{0.19\textwidth}\includegraphics[width=3.4cm, height=3.4cm]{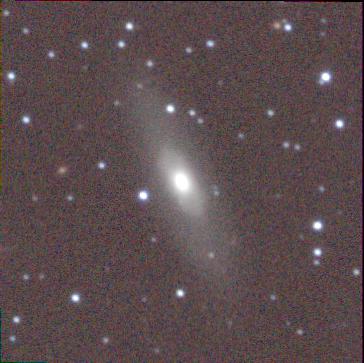}\caption*{20. $e$S}\end{subfigure}

\begin{subfigure}[b]{0.19\textwidth}\includegraphics[width=3.4cm, height=3.4cm]{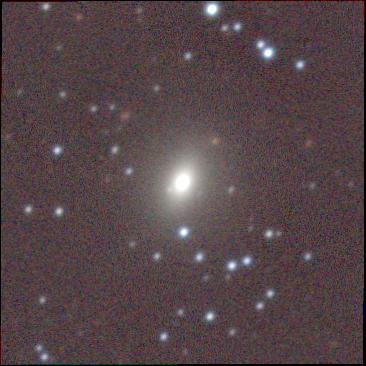}\caption*{21. E/S0}\end{subfigure}
\begin{subfigure}[b]{0.19\textwidth}\includegraphics[width=3.4cm, height=3.4cm]{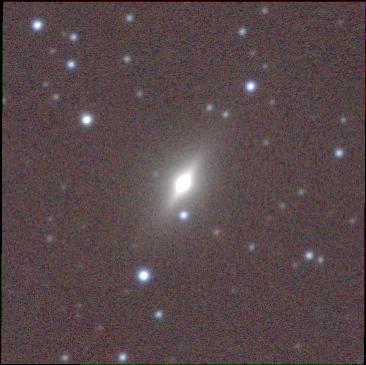}\caption*{22. E/S0}\end{subfigure}
\begin{subfigure}[b]{0.19\textwidth}\includegraphics[width=3.4cm, height=3.4cm]{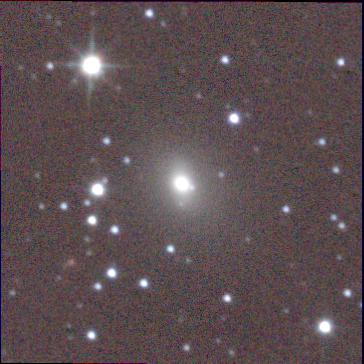}\caption*{23. E/S0}\end{subfigure}
\begin{subfigure}[b]{0.19\textwidth}\includegraphics[width=3.4cm, height=3.4cm]{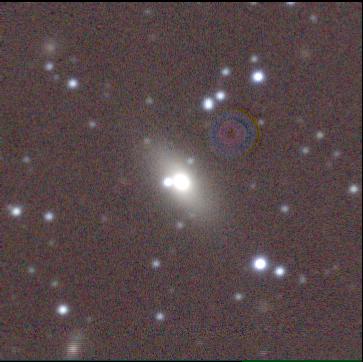}\caption*{24. E/S0}\end{subfigure}
\begin{subfigure}[b]{0.19\textwidth}\includegraphics[width=3.4cm, height=3.4cm]{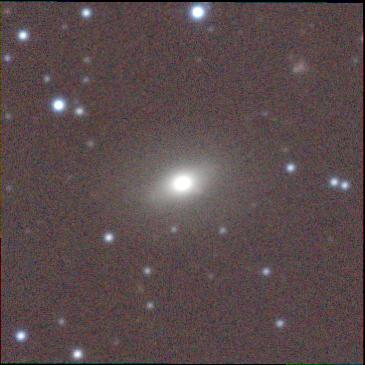}\caption*{25. E/S0}\end{subfigure}

\begin{subfigure}[b]{0.19\textwidth}\includegraphics[width=3.4cm, height=3.4cm]{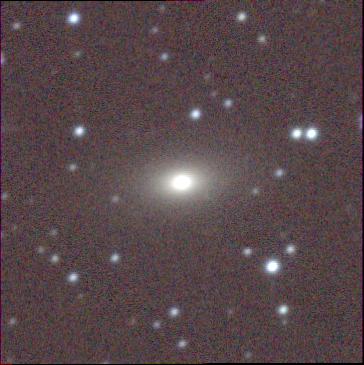}\caption*{26. E/S0}\end{subfigure}
\begin{subfigure}[b]{0.19\textwidth}\includegraphics[width=3.4cm, height=3.4cm]{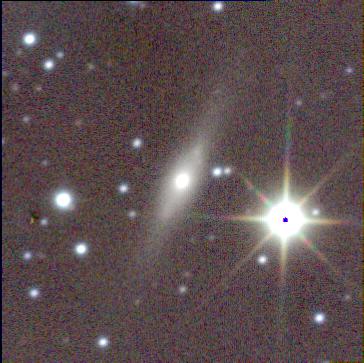}\caption*{27. $e$S}\end{subfigure}
\begin{subfigure}[b]{0.19\textwidth}\includegraphics[width=3.4cm, height=3.4cm]{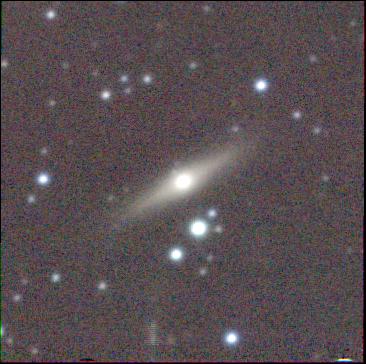}\caption*{28. $e$S}\end{subfigure}
\begin{subfigure}[b]{0.19\textwidth}\includegraphics[width=3.4cm, height=3.4cm]{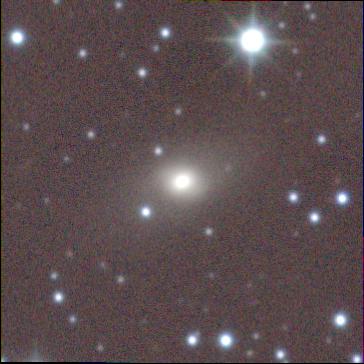}\caption*{29. $e$S}\end{subfigure}
\begin{subfigure}[b]{0.19\textwidth}\includegraphics[width=3.4cm, height=3.4cm]{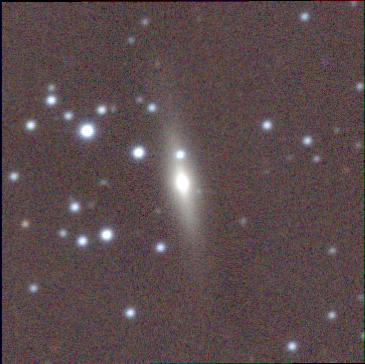}\caption*{30. $e$S}\end{subfigure}
\caption{The false-colour $-$ $J$ (blue), $H$ (green) and $K$ (red) representation (1.2\arcmin\ $\times$ 1.2\arcmin) of the 30 brightest galaxy candidates in the 3C\,129 cluster. The cyan frames indicate H\textsc{i} detection in the WSRT H\textsc{i}-survey.}\label{30postage}
\end{figure*}

A full catalogue of the 261 cluster members is presented in Appendix~B, Table~B. Galaxies are listed in ascending order of their $K$-band magnitudes (from the brightest to the faintest). For each galaxy a 1.2\arcmin\ $\times$ 1.2\arcmin\ image in the $J, H$ and $K$ band was extracted. The false colour (RGB) postage stamps of these images are shown in Fig.~B in the same order as the catalogue. Galaxy morphologies are printed at the bottom of the images. The galaxies that were detected in \HI\ are framed in cyan. The first 30 entries of the catalogue and their corresponding images are shown in Table.~\ref{30Candidates} and Fig~\ref{30postage}, respectively.\\

%The columns are as follows:\\
%\begin{flushleft}
%Column\,1 - The galaxy identifying number.\\
%\vspace{5.5pt}
%Column\,2- Unique identifiers based on the Right Ascension (RA) and Declination (Dec) ZoAJhhmmss.ss$\pm$ddmmss.s.\\
%\vspace{5.5pt}
%Columns\,3 \& 4  - Equatorial coordinates, RA and Dec (J2000), respectively in degrees. \\
%\vspace{5.5pt}
%Columns\, 5 \& 6  - Galactic longitude and latitude in degrees.\\
%\vspace{5.5pt}
%Column\,7 - Ellipticity in the $K$-band at the  20 mag arcsec$^{-2}$ isophote.\\ 
%\vspace{5.5pt}
%Column\,8 - The $K$-band position angle in degrees, measured from celestial north to east.\\
%\vspace{5.5pt}
%Column\,9 - The $K_{20}$ isophotal major-axis radius in arcseconds.\\
%\vspace{5.5pt}
%Columns\,10 \& 11 \& 12 - $J, H$ and $K$ band fiducial isophotal magnitudes (not corrected for foreground extinction).\\
%\vspace{5.5pt}
%Column\,13 - The line-of-sight Galactic reddening in magnitudes.\\
%\vspace{5.5pt}
%Column\,14 - $J^{\rm o} - K^{\rm o}$ colours, corrected for foreground extinction.\\
%\vspace{5.5pt}
%Column\,15  - The estimated morphology of galaxies.\\
%\vspace{5.5pt}
%Column\,16  - The radial velocities from the WSRT %\HI\ measurements, and two optical velocities for the radio galaxies 3C\,129 and 3C\,129.1.
%\end{flushleft}

\onecolumn
\begin{centering}
\begin{landscape}
%\scriptsize 
%\onecolumn
% \footnote \noindentsize
\setlength{\tabcolsep}{2.0pt}
 \hspace*{100cm}
\begin{longtable}{crrrrrrrrcrccclc}

 %\centering
 
\caption[Parameters of cluster member candidates]{Near$-$infrared parameters of the first 30 entries of 3C\,129 candidate cluster galaxies. The full catalogue is available in Appendix B}\label{30Candidates}\\

% \multicolumn{9}{c}{\small{\bf{\tablename} \thetable{}.} \textsc{Hi} parameters of reliable
% detections.} \\[0.5ex]
\hline \\[-1.8ex]
\thead{ID no.}       &\thead{Unique ID}   &  \thead{RA}    &  \thead{Dec}    & \thead{$\ell$} &  \thead{$b$}  &  \thead{$\epsilon_{K}$}  &  \thead{$\phi_{K}$}  & \thead{$r_{K20}$} & \thead{$J_{K20}$}  & \thead{$H_{K20}$}  & \thead{$K_{20}$}   &  \thead{$E(B $-$ V)$} & \thead{J$^{\rm o}$$-$K$^{\rm o}$} &  \thead{Type} & \thead{v$_{rad}$} \\
\\
                     & \thead{ZoA}        &   \thead{deg}  &  \thead{deg}    &  \thead{deg}   &  \thead{deg}  &  \thead{}                &  \thead{deg}         &  \thead{\arcsec}    &  \thead{mag}         &  \thead{mag}         &  \thead{mag}        &    \thead{mag}         & \thead{mag}         &                 & \thead{\kms}\\
\\
\thead{(1)}          & \thead{(2)}        &   \thead{(3)}  &  \thead{(4)}    &  \thead{(5)}   &  \thead{(6)}  &  \thead{(7)}             &   \thead{(8)}        &   \thead{(9)}       &   \thead{(10)}       &  \thead{(11)}        &  \thead{(12)}     &  \thead{(13)}            & \thead{(14)}        &  \thead{(15)}   &  \thead{(16)}       \\

\hline \\[-1.8ex]
\endfirsthead

\multicolumn{9}{c}{{\tablename} \thetable{} -- Continued} \\[0.5ex]
\hline \\[-1.8ex]

\thead{ID no.}       &\thead{Unique ID}   &  \thead{RA}    &  \thead{Dec}    & \thead{$\ell$} &  \thead{$b$}  &  \thead{$\epsilon_{K}$}  &  \thead{$\phi_{K}$}  & \thead{$r_{K20}$} & \thead{$J_{K20}$}  & \thead{$H_{K20}$}  & \thead{$K_{K20}$}   &  \thead{$E(B $-$ V)$}& \thead{J$^{\rm o}$$-$K$^{\rm o}$} &  \thead{Type} & \thead{v$_{rad}$} \\
\\
                     & \thead{ZoA}        &   \thead{deg}  &  \thead{deg}    &  \thead{deg}   &  \thead{deg}  &  \thead{}                &  \thead{deg}         &  \thead{\arcsec}    &  \thead{mag}         &  \thead{mag}         &  \thead{mag}        &    \thead{mag}         & \thead{mag}         &                 & \thead{\kms}\\
\\
\thead{(1)}          & \thead{(2)}        &   \thead{(3)}  &  \thead{(4)}    &  \thead{(5)}   &  \thead{(6)}  &  \thead{(7)}             &   \thead{(8)}        &   \thead{(9)}       &   \thead{(10)}       &  \thead{(11)}        &  \thead{(12)}     &  \thead{(13)}            & \thead{(14)}        &  \thead{(15)}   &  \thead{(16)}       \\

\hline \\[-1.8ex]
\endhead

\hline \\[-1.8ex]
\\
\endfoot

\endlastfoot
1	&	J044908.26$+$445540.3	&	72.284	&	44.928	&	160.490	&	0.086	&	0.17	&	24.71	    &	28.10	&	10.96	$\pm$	0.02	&	~9.94	$\pm$	0.02	&	~9.51	$\pm$	0.02	&	0.89	&	1.07	&	E/S0	&--- \\
2	&	J045006.67$+$450305.8	&	72.528	&	45.052	&	160.505	&	0.298	&	0.20	&	65.00	    &	37.30	&	11.21	$\pm$	0.04	&	10.10	$\pm$	0.04	&	~9.59	$\pm$	0.04	&	0.98	&	1.20	&	E/S0	&	6655$^{*}$\\
3	&	J044909.06$+$450039.4	&	72.288	&	45.011	&	160.427	&	0.142	&	0.10	&	$-$60.00	&	25.60	&	11.50	$\pm$	0.03	&	10.46	$\pm$	0.04	&	10.00	$\pm$	0.04	&	0.94	&	1.10	&	E/S0	&	6236$^{\times}$ \\
4	&	J045145.56$+$443602.6	&	72.940	&	44.601	&	161.039	&	0.235	&	0.40	&	60.00		&	22.80	&	11.75	$\pm$	0.03	&	10.82	$\pm$	0.03	&	10.35	$\pm$	0.04	&	0.79	&	1.07	&	$e$S	&	5086$^{+}$ \\
5	&	J044939.78$+$440922.1	&	72.416	&	44.156	&	161.141	&	$-$0.337&	0.09	&	$-$55.01	&	12.79	&	12.05	$\pm$	0.02	&	11.16	$\pm$	0.02	&	10.76	$\pm$	0.02	&	0.65	&	1.01	&	E/S0	&--- \\	
6	&	J044842.39$+$454818.5	&	72.177	&	45.805	&	159.769	&	0.593	&	0.40	&	$-$61.15	&	21.62	&	12.25	$\pm$	0.02	&	11.23	$\pm$	0.02	&	10.77	$\pm$	0.02	&	1.12	&	1.00	&	$e$S	&--- \\	
7	&	J045245.69$+$450106.2	&	73.190	&	45.018	&	160.829	&	0.638	&	0.44	&	85.11		&	19.63	&	12.31	$\pm$	0.02	&	11.23	$\pm$	0.02	&	10.78	$\pm$	0.02	&	0.94	&	1.12	&	E/S0	&--- \\
8	&	J045219.88$+$451546.1	&	73.083	&	45.263	&	160.592	&	0.734	&	0.50	&	$-$60.00	&	25.90	&	12.23	$\pm$	0.05	&	11.26	$\pm$	0.05	&	10.83	$\pm$	0.06	&	0.91	&	1.00	&	$e$S	&--- \\
9	&	J045326.75$+$441900.7	&	73.361	&	44.317	&	161.449	&	0.288	&	0.06	&	26.77		&	14.31	&	12.32	$\pm$	0.02	&	11.38	$\pm$	0.02	&	10.96	$\pm$	0.02	&	0.69	&	1.06	&	E/S0	&--- \\
10	&	J045045.92$+$450659.7	&	72.691	&	45.117	&	160.529	&	0.428	&	0.27	&	0.36		&	13.39	&	12.62	$\pm$	0.02	&	11.56	$\pm$	0.02	&	11.05	$\pm$	0.02	&	1.00	&	1.14	&	E/S0	&--- \\
11	&	J044719.31$+$441701.6	&	71.830	&	44.284	&	160.774	&	$-$0.576&	0.47	&	$-$73.54	&	16.96	&	12.50	$\pm$	0.02	&	11.56	$\pm$	0.02	&	11.06	$\pm$	0.02	&	0.69	&	1.15	&	$e$S	&	4993\\
12	&	J044843.25$+$445216.0	&	72.180	&	44.871	&	160.486	&	$-$0.007&	0.11	&	32.33	    &	9.98	&	12.54	$\pm$	0.02	&	11.54	$\pm$	0.02	&	11.11	$\pm$	0.02	&	0.84	&	1.07	&	E/S0	&--- \\
13	&	J044459.45$+$453344.1	&	71.248	&	45.562	&	159.534	&	$-$0.059&	0.30	&	$-$23.53	&	14.62	&	12.83	$\pm$	0.02	&	11.73	$\pm$	0.02	&	11.19	$\pm$	0.02	&	1.35	&	1.06	&	E/S0	&--- \\
14	&	J044429.89$+$442914.8	&	71.125	&	44.487	&	160.291	&	$-$0.827&	0.31	&	44.80	    &	14.65	&	12.81	$\pm$	0.02	&	11.78	$\pm$	0.02	&	11.23	$\pm$	0.02	&	0.98	&	1.16	&	$e$S	&--- \\
15	&	J044724.21$+$445928.3	&	71.851	&	44.991	&	160.244	&	$-$0.107&	0.80	&	$-$35.00	&	31.50	&	12.64	$\pm$	0.05	&	11.67	$\pm$	0.05	&	11.25	$\pm$	0.07	&	0.84	&	1.03	&	$m$S	&--- \\
16	&	J045414.90$+$450315.1	&	73.562	&	45.054	&	160.967	&	0.864	&	0.17	&	$-$41.28	&	19.00	&	12.79	$\pm$	0.02	&	11.75	$\pm$	0.02	&	11.29	$\pm$	0.02	&	0.93	&	1.10	&	$e$S	&	6269\\
17	&	J045129.31$+$451852.0	&	72.872	&	45.314	&	160.458	&	0.652	&	0.57	&	$-$35.96	&	17.04	&	12.86	$\pm$	0.02	&	11.84	$\pm$	0.02	&	11.31	$\pm$	0.02	&	1.04	&	1.10	&	E/S0	&--- \\
18	&	J045028.01$+$443407.6	&	72.617	&	44.569	&	160.916	&	0.037	&	0.70	&	$-$74.56	&	15.98	&	12.74	$\pm$	0.02	&	11.79	$\pm$	0.02	&	11.32	$\pm$	0.02	&	0.80	&	1.07	&	E/S0	&--- \\
19	&	J045156.57$+$445815.0	&	72.986	&	44.971	&	160.774	&	0.496	&	0.47	&	$-$12.60	&	14.73	&	12.79	$\pm$	0.02	&	11.79	$\pm$	0.02	&	11.33	$\pm$	0.02	&	0.94	&	1.05	&	$e$S	&--- \\
20	&	J045251.94$+$444122.5	&	73.216	&	44.690	&	161.095	&	0.443	&	0.56	&	63.37	    &	16.94	&	12.90	$\pm$	0.02	&	11.92	$\pm$	0.02	&	11.47	$\pm$	0.02	&	0.85	&	1.07	&	$e$S	&--- \\
21	&	J045018.46$+$454152.2	&	72.577	&	45.698	&	160.031	&	0.738	&	0.27	&	24.97	    &	12.06	&	13.08	$\pm$	0.02	&	11.99	$\pm$	0.02	&	11.49	$\pm$	0.02	&	1.17	&	1.08	&	E/S0	&--- \\
22	&	J045324.45$+$451127.6	&	73.352	&	45.191	&	160.768	&	0.835	&	0.40	&	29.03	    &	12.08	&	12.97	$\pm$	0.02	&	11.93	$\pm$	0.02	&	11.50	$\pm$	0.02	&	0.98	&	1.05	&	E/S0	&--- \\
23	&	J044753.50$+$443250.9	&	71.973	&	44.547	&	160.638	&	$-$0.328&	0.10	&	25.00	    &	13.10	&	12.88	$\pm$	0.04	&	11.87	$\pm$	0.05	&	11.52	$\pm$	0.06	&	0.84	&	1.00	&	E/S0	&--- \\	
24	&	J044734.77$+$452912.6	&	71.895	&	45.487	&	159.885	&	0.237	&	0.41	&	50.35	    &	8.87	&	13.14	$\pm$	0.02	&	12.05	$\pm$	0.02	&	11.56	$\pm$	0.02	&	1.19	&	1.06	&	E/S0	&--- \\
25	&	J045332.55$+$453232.6	&	73.386	&	45.542	&	160.510	&	1.076	&	0.39	&	$-$23.08	&	14.16	&	13.09	$\pm$	0.02	&	12.03	$\pm$	0.02	&	11.56	$\pm$	0.02	&	0.98	&	1.11	&	E/S0	&--- \\	
26	&	J044639.43$+$454052.2	&	71.664	&	45.681	&	159.633	&	0.240	&	0.41	&	85.72	    &	12.30	&	13.12	$\pm$	0.02	&	12.08	$\pm$	0.02	&	11.57	$\pm$	0.02	&	1.25	&	1.01	&	E/S0	&--- \\
27	&	J044730.43$+$454548.7	&	71.877	&	45.764	&	159.666	&	0.406	&	0.20	&	47.01	    &	10.49	&	13.14	$\pm$	0.02	&	12.11	$\pm$	0.02	&	11.62	$\pm$	0.02	&	1.25	&	0.98	&	$e$S	&--- \\
28	&	J044953.41$+$451613.5	&	72.473	&	45.270	&	160.312	&	0.408	&	0.60	&	$-$60.00	&	15.10	&	13.33	$\pm$	0.05	&	12.16	$\pm$	0.04	&	11.71	$\pm$	0.06	&	1.01	&	1.18	&	$e$S	&--- \\	
29	&	J045032.79$+$445411.7	&	72.637	&	44.903	&	160.668	&	0.262	&	0.29	&	$-$10.36	&	12.04	&	13.26	$\pm$	0.02	&	12.21	$\pm$	0.02	&	11.71	$\pm$	0.02	&	0.92	&	1.15	&	$e$S	&--- \\
30	&	J044940.04$+$451119.0	&	72.417	&	45.189	&	160.350	&	0.325	&	0.62	&	$-$10.81	&	13.76	&	13.23	$\pm$	0.02	&	12.22	$\pm$	0.02	&	11.72	$\pm$	0.02	&	0.98	&	1.09	&	$e$S	&--- \\
\hline
%\multicolumn{1}{|p{\columnwidth}|} 
 
\caption*{\textbf{Notes.} The columns are: (1) the galaxy identifying number; (2) unique identifiers based on the right ascension (RA) and declination (Dec) ZoAJhhmmss.ss$\pm$ddmmss.s.; (3) \& (4) equatorial coordinates, RA and Dec (J2000), respectively in degrees; (5) and (6) galactic longitude and latitude in degrees; (7) ellipticity in the $K$-band at the  20 mag arcsec$^{-2}$ isophote; (8) the $K$-band position angle in degrees, measured from celestial north to east; (9) the $K_{20}$ isophotal major-axis radius in arcseconds; (10),  (11) and (12) $J, H$ and $K$ band fiducial isophotal magnitudes (not corrected for foreground extinction); (13) the line-of-sight Galactic reddening in magnitudes; (14) $J^{\rm o} - K^{\rm o}$ colours, corrected for foreground extinction; (15) the estimated morphology of galaxies; (16) the radial velocities from the WSRT \HI\ measurements, and two optical velocities for the radio galaxies 3C\,129 and 3C\,129.1. v$_{opt}$: [*] \citealp{Spinrad1975}, [$\times$] \citealp{Spinrad1975}, [+] v$_{opt}$ = 5150 \citep{Takata1994}.}

\end{longtable}

%\begin{flushleft}
%\small{[*] v$_{opt}$: \citealp{Spinrad1975}}\\
%\small{[$\times$] v$_{opt}$: \citealp{Spinrad1975}}\\
%\small{[+] v$_{opt}$ = 5150 %\citep{Takata1994}}\\
%\end{flushleft}
\end{landscape}
\end{centering}

\twocolumn
%\documentclass{article}
%\usepackage{graphicx}
%\usepackage{epsfig}
%\usepackage{color}
%\usepackage{amssymb}
%\usepackage{txfonts}
%\usepackage{lscape}
%\usepackage{float}
%\usepackage{amsmath}
%\usepackage{natbib}
%\usepackage{longtable}
%\usepackage{journal_names}
%\usepackage{rotating}
%\usepackage{enumitem}
%\usepackage{subcaption}
%\usepackage{xcolor}
%\usepackage[export]{adjustbox}
%\usepackage{caption}
%\DeclareCaptionLabelFormat{continued}{Figure \ref{postage}}
%\captionsetup[ContinuedFloat]{labelformat=continued}

%\begin{document}

%ref version
%\input{Ref30PostageStamp.tex}

\clearpage

\subsubsection{Catalogue completeness}
\begin{figure}[h!]
   \centering
 %\hspace*{-0.6cm}  
 \includegraphics[width=87mm, height=70mm]{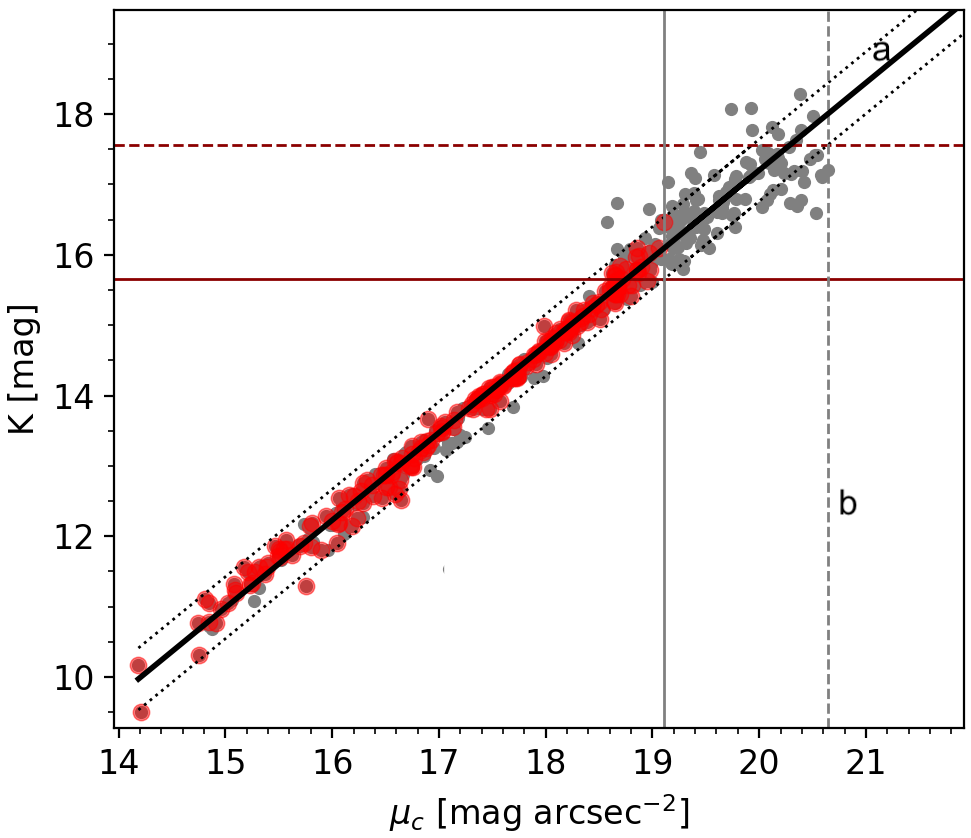}
    \caption{An estimation of the galaxy magnitude completeness limit. The grey points are all galaxies on red sequence in the cluster radius. Red points are those with $r_{K_{20}} > 3$\arcsec\ (that is, the cluster sample). Line $a$ is the relation fitted through the grey points with the $1\sigma$ and $-1\sigma$ deviation shown by the grey dotted lines. The surface brightness threshold for all 3C\,129 galaxies on the red sequence is illustrated by line $b$. The completeness limit is given by the intercept of line $b$ and the $-1\sigma$ line. The corresponding surface brightness threshold and completeness limit lines for the bonafide cluster sample with $r_{K_{20}} > 3$\arcsec\ are denoted by the solid grey vertical and red line, respectively.}\label{Compl}
 \end{figure}

Given the severity and unknown small-scale patchiness of the Galactic extinction and the high stellar density in
this Galactic field, it is non-trivial to perform a robust and quantitative completeness analysis. This type of analysis would involve measuring the detectability of model galaxies (as a function of magnitudes) which imitate the spatial surface brightness distribution of real galaxies. Since it is not possible to conduct such a quantitative completeness analysis in this ZoA region we estimate the completeness of the selected sample of the 3C129 cluster members using the technique described in \citet{Garilli1999} and \citet{Skelton2009}. With this technique galaxies are not detectable when their central surface brightness falls below the detection limit at a given magnitude. 
Figure\,\ref{Compl} shows the relation between the fiducial $K$-band aperture magnitudes and the central surface brightness of all galaxies (grey points) on the red sequence within the cluster radius. We performed a linear fit through these data points. This fit is shown by the black solid line ($a$) with the 1$\sigma$ and $-1\sigma$ deviation indicated by the dotted lines. The surface brightness limit of this sample is illustrated by the vertical grey dashed line ($b$). If we take the intercept of line $a$ and $b$ to be the completeness limit, then galaxies to the right of the line $b$ and below line $a$ are not detected (That is, this is  completeness limit in K-band magnitude, based on the faintest surface brightness that is detected). However, there is a scatter in the distribution around the fitted line $a$, thus the completeness limit is at a magnitude where the line $b$ intersects the $-1\sigma$ line. Using this technique we derive a completeness limit of $K = 17.5$ mag for all 3C\,129 galaxies on the red sequence.

However, for this study we only considered galaxies with  $r_{K_{20}} > 3$\arcsec\ (red points) as cluster members (see Sect.\,\ref{fitRS}), this resulted in a change in the surface brightness limit but only modest changes in the linear fit and its scatter. Therefore, for the sample of galaxies on the red-sequence that we formally define as 3C\,129 cluster members, we estimate a magnitude complete limit of $K$ = 15.6 mag (or extinction corrected $K^{o} = 15.0$ mag). 

\begin{figure}[h!]
   \centering
 %\vspace*{-0.6cm}  
 \hspace*{0.05cm}  
    \includegraphics[width=86mm, height=35mm]{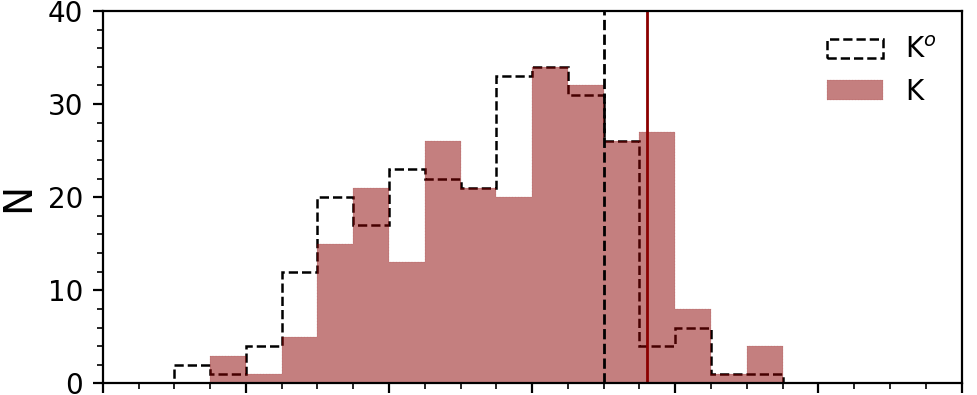} 
  \includegraphics[width=88mm, height=40mm]{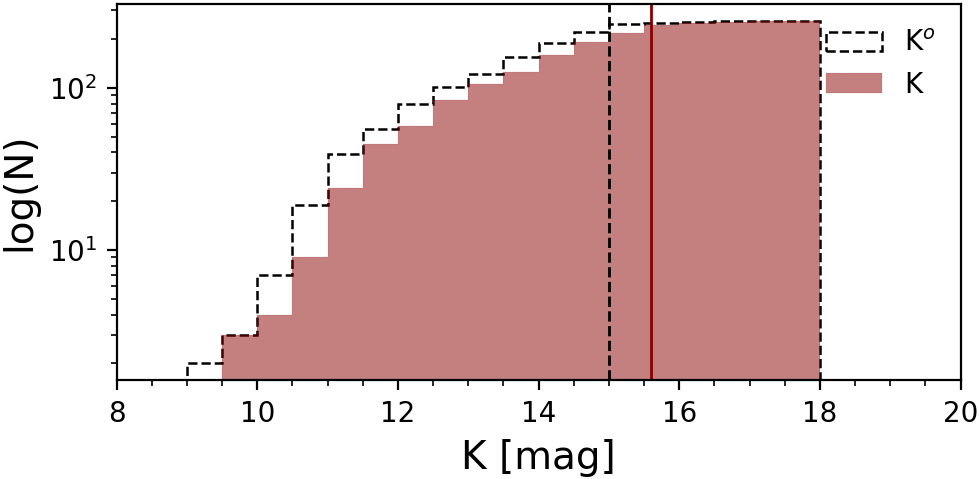}

    \caption{The top panel is the histogram of the fiducial isophotal magnitudes for the $K$-band plotted in 0.5 magnitude bins and the bottom panel is the cumulative galaxy number count distribution as a function of the $K$-band magnitudes. In both panels the solid and dashed distributions represents the observed and extinction-corrected magnitudes, respectively. The estimated completeness magnitude limit is shown by the red vertical line for the observed magnitudes and dashed vertical line for the extinction corrected magnitudes.}\label{nirDistribution}
 \end{figure}

To test the validity of our completeness estimates we use the cumulative galaxy counts shown in the bottom panel of Fig.~\ref{nirDistribution} to derive the completeness limit. Since a linear increase is expected in these counts as a function of the $K$-band magnitude (for example, \citealp{Kochanek2001}) a deviation from this trend indicates incompleteness. Therefore, using this method the completeness magnitude of the 3C\,129 galaxy sample is defined where the cumulative galaxy distribution flattens. By this technique we find that the logN distribution flattens at similar magnitudes as those derived from Fig.\,\ref{Compl}, that is,, $\sim K$ = 15.6 mag or extinction corrected $\sim K^{o} = 15.0$ mag. This also corresponds with the magnitude where the number counts start to decrease significantly as a function of the magnitude as shown in the histogram (bottom panel). 
Moreover, based on our visual inspection of the UKIDSS-GPS images, we are confident that we have identified all galaxies brighter than $K \approx 15.6$ mag ($K^{o} = 15.0$ mag).  We refer to Table~B  and Fig.~B (Appendix~B) for a visual impression of galaxies with these magnitudes.

\subsubsection{The 3C\,129 cluster galaxy density}\label{nirSect}
In this subsection we assess the density properties of the 3C\,129 cluster. We conduct this by comparing the $K$-band magnitude distribution of its galaxies with those of three well-known galaxy clusters at similar distances. Our comparison sample comprises the Perseus, Coma and Norma galaxy clusters. All three represent well-studied and rich clusters nearby ($z \approx 0.02$). Parameters of these clusters are summarised in Table~\ref{CCSummary}.

\begin{table}[h!]
\caption {A summary of parameters of comparison galaxy clusters.}\label{CCSummary}
\begin{tabular}{lcl}
\hline
Cluster & Distance          & Cluster members            \\  
                  &Mpc             & catalogue                        \\  
~~~(1)              &(2)               &    ~~~~(3)                                \\         

\hline
3C\,129     & 89.1            &This work                           \\
Perseus     &76.6             & \citealp{Brunzendorf1999}   \\
Norma       &70.0             & \citealp{Skelton2009}          \\
Coma        &98.9             & \citealp{Eisenhardt2007}    \\
\hline \\
\end{tabular}
\end{table}

The comparison samples for Perseus, Coma and Norma were selected from catalogues of confirmed galaxy cluster members by \citet{Brunzendorf1999}, \citet{Eisenhardt2007} and \citet{Skelton2009}, respectively. We used 2MASX to extract the $K$-band elliptical aperture magnitudes of their galaxies at the 20 mag arcsec$^{-2}$ isophote by cross-correlating their positions with 2MASX sources. We only considered galaxies as matched if they were separated by less than 1.5\arcsec\ in both catalogues. This offset corresponds to the maximum positional uncertain between 2MASX and other galaxy catalogues \citep{KraanKorteweg2005}.

\begin{figure}
   \centering
% \hspace*{-0.6cm}   
 \includegraphics[width=90mm, height=35mm]{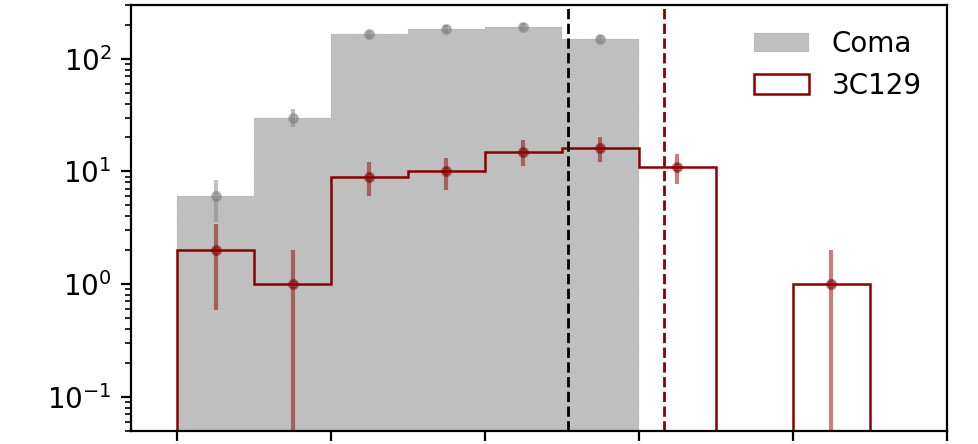}\\
  \includegraphics[width=90mm, height=35mm]{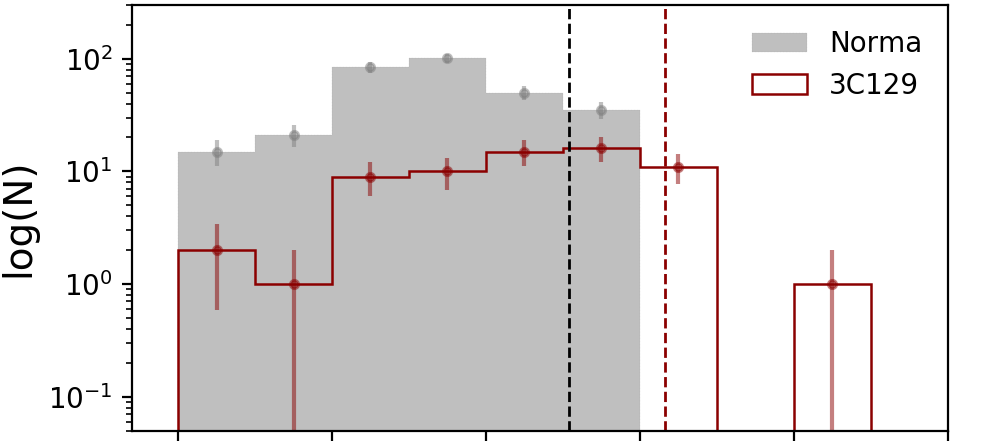}\\
   %\hspace*{0.16cm} 
  \includegraphics[width=90mm, height=45mm]{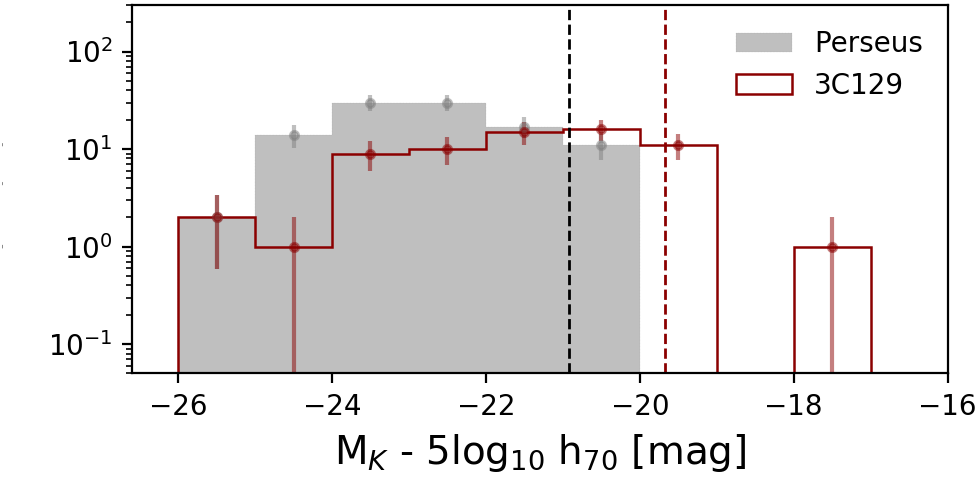}\\
    \caption{Comparison of the extinction corrected $K$-band magnitude distribution (with Poisson error bars) per 1.0 mag bin of the 3C\,129 cluster (red) with three well known clusters at similar distances, the Coma and Norma and Perseus clusters indicated by the solid histograms. The solid black and red lines are the respective 2MASX and 3C129 UKIDSS sample completeness limits.}\label{Kdistcomp}
 \end{figure}

Figure~\ref{Kdistcomp} shows the extinction corrected $K$-band magnitude distribution spanning the magnitude range $-16.5 \geq  M_{K^{\rm o}} \geq -26$ mag of the three clusters compared with the 3C\,129 cluster. The galaxy number counts were only compared within the cores of the clusters over an area of radius, $r_{cl}$ = 0.5~Mpc. The size of this area was chosen based on the confirmed galaxy members' sample with the smallest area coverage. 
The Coma distribution shows a steep increase in the galaxy counts from the bright end, $M_{K^{\rm o}} \geq -26$ mag until it flattens at  $M_{K^{\rm o}} \geq -20.5$ mag reaching its maximum number counts. It only shows a slight downturn just before $M_{K^{\rm o}} \approx -20$ mag because of the incompleteness of 2MASX at this magnitude. The Norma distribution shows a steady increase from the bright end with a downturn at $\sim M_{K^{\rm o}} \approx -22$ mag, before the 2MASX completeness limit, while the Perseus distribution on the other hand does not show this number density and increase as seen in the Coma and Norma clusters. We find that compared to these three distributions, the 3C\,129 cluster, shows the same number density on bright side and similar increase towards the faint end as the Perseus cluster. Based on this analysis we deduce that within the magnitude regime of, $-21.5  \leq M_{K^{\rm o}} \leq -26.0$ mag (limited by 2MASX completeness), the central galaxy density within $r_{cl} = 0.5$~Mpc in the Coma and Norma cluster are comparable, while the 3C\,129 cluster appears to be poorer than the aforementioned, but similar to the Perseus cluster.

%\newpage
\section{The spatial distribution of the 3C\,129 cluster galaxies}\label{SpaSect}
 
\begin{figure*}
   \centering
 \includegraphics[width=130mm, height=115mm]{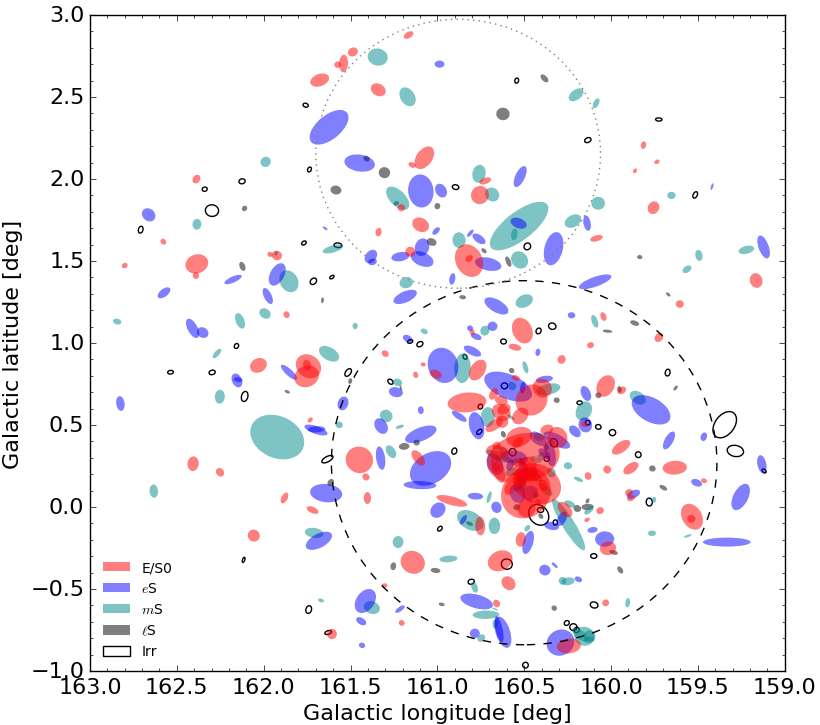}
    \caption{Spatial distribution of galaxies on the red sequence. Points sizes and shapes are shown relative to their radii, ellipticities and positions angles. The defined projected radius of the cluster, $r_{cl}$ = 1.7 Mpc is shown by the black dashed line. The substructure of the cluster as described in Sect.~\ref{substructureP2} is denoted by the smaller dotted grey circle. The different colours illustrate the morphologies of the galaxies. }\label{spatialScatter}
 \end{figure*}

The spatial distribution of all galaxies on our defined red-sequence is shown in Fig.\,\ref{spatialScatter}. The plot highlights the different morphologies and sizes of these galaxies. Members of the 3C\,129 cluster are enclosed within the large dashed circle of radius of 1.7~Mpc centred on its X-ray emission ($\ell, b \approx 160.52\dg, 0.27\dg$). The smaller dotted circle encloses a prominent substructure to be discussed in Sect.\,\ref{substructureP2}. In this section we only discuss galaxies within the defined extent of the cluster (dashed circle).\\

~\\
Galaxies within the inner region of the cluster of $r_{core}$ =  0.8 Mpc, corresponding to its $R_{500}$ \citep{Piffaretti2011}, are highly concentrated in the centre and are mostly early-type galaxies (E/S0). At larger projected radii the number of spirals ($e$S to Irr) increases steadily while the number of E/S0 galaxies decreases. 

The spatial distribution in Fig.~\ref{spatialScatter} is essentially a demonstration of the classical morphology-density relation \citep{Dressler1980}. Most studies of the morphology-density (T - $\Sigma$) or morphology-radius (T - R) relation in galaxy clusters often use a composite of a large number of clusters to obtain an adequately high number of galaxies for each morphological type (\citealp{Dressler1980}, \citealp{Goto2003}, \citealp{Thomas2006}). However, these composite (T - $\Sigma$) or (T - R)  relations often lose information on the individual clusters. They also introduce large uncertainties given the large and sometimes inhomogeneous samples of clusters at different redshifts that are combined to make one cluster (\citealp{Andreon1996}, \citealp{vanDokkum2000}).  

\begin{figure}
   \centering
 \hspace*{-0.6cm}  
 \includegraphics[width=90mm, height=57mm]{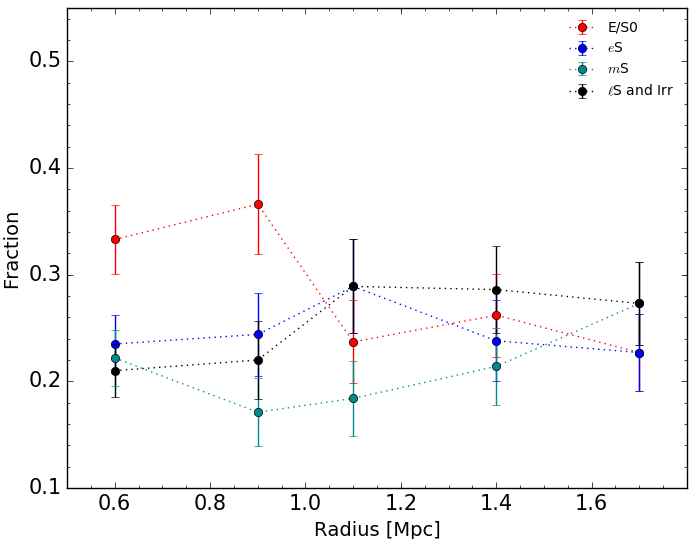}
    \caption{The morphology-radius relation for the E/S0 (red), $e$S (blue), $m$S (darkcyan), $\ell$S \& Irr (black) members of the cluster.}\label{morphR}
 \end{figure}

To study the (T - R) relation specific to the 3C\,129 cluster, we use a sample of all our classified galaxies (N=261). We show the results in Fig.~\ref{morphR} where the azimuthally averaged fractions of the morphological types are plotted as function of the projected distance from the core of the cluster to the outskirts. The fraction of E/S0 galaxies decreases from $\sim$36\% in the core to almost 22\% in the outskirts of the cluster. The early spirals ($e$S) fraction increases from 24\% and peaks to 29\% where the E/S0 population is lowest at $r_{cl} = 1.1$ Mpc and then decreases rapidly to 22\% similar to the E/S0 fraction. On the other hand, $\ell$S and Irr galaxies show the opposite trend with an increasing fraction from the core towards the cluster outskirts. The morphology-radius relation of the 3C\,129 cluster shows an obvious morphological segregation that is qualitatively consistent with those of the centrally concentrated rich clusters \citep{Dressler1997}. 

\subsection{Substructure in the 3C\,129 cluster}\label{substructureP2}
\subsubsection{Projected density distribution}
To investigate the possible presence of substructure we analyse the projected galaxy number densities which is apparent already in Fig.\,\ref{spatialScatter}. Although this method ignores the redshift information it is optimal at identifying substructures with enhanced densities that are relatively large in separation \citep{Pinkney1996}. For the galaxies on the red sequence we study the projected galaxy number density using the bivariate kernel density estimator \citep{Feigelson2012},   

\begin{equation}\label{kde}
f_{kern}(\textbf{x},\textbf{h}) = {\frac{1}{n\prod_{j=1}^p h_{j}}} \sum_{i=1}^{n} \Bigg[\prod_{j=1}^p h_{j} Gk \bigg(\frac{\textbf{x}_{i} - \textbf{X}_{ij}}{h_{ij}}\bigg) \Bigg]
\end{equation}

\noindent
where $p$ = 2 is the number of variables, which are Galactic coordinates ($\ell, b$) in our case represented by $\textbf{X}_{i}=\textbf{X}_{i1}, \textbf{X}_{i2}$, with the smoothing length (bandwidth), \textbf{h} = $(h_1, h_2)$.  The number of galaxies is denoted by $n$ and the bivariate Gaussian kernel function by $Gk$.  

\begin{figure}
   \centering
 \hspace*{-0.6cm} 
 \includegraphics[width=95mm, height=80mm]{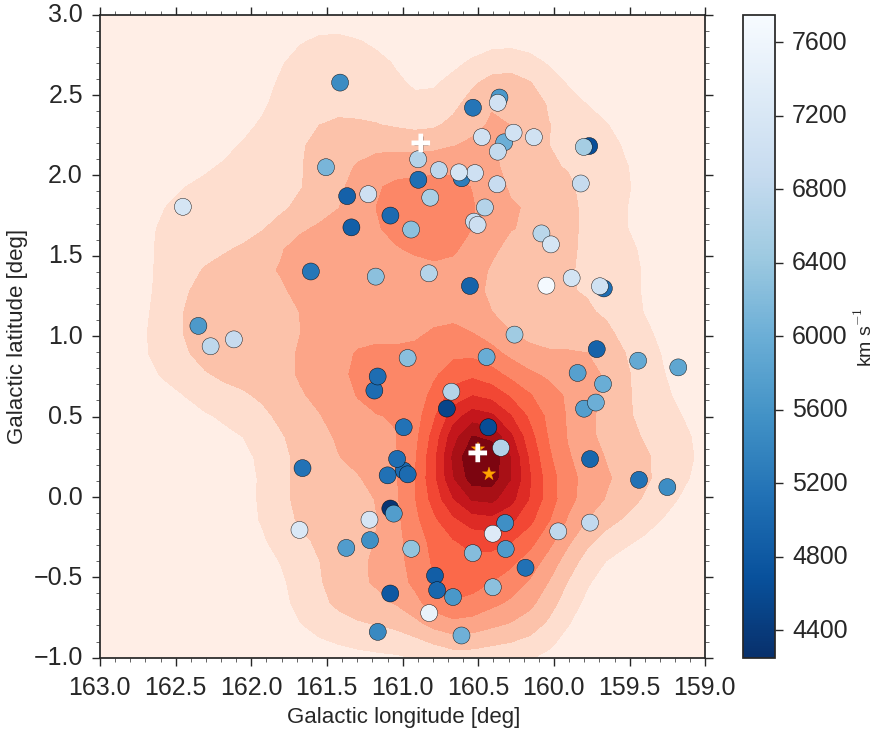}
    \caption{A two-dimensional kernel density map of galaxies on the red-sequence (red) in position space. Dots in gradient blues are galaxies detected in H\textsc{i} over the radial velocity range of the 3C\,129 cluster (4000 - 8000 \kms). The displayed colour bar is the velocity range of the H\textsc{i} galaxies. The two white plus markers indicate the position of the X-ray emission over which the cluster is centred and the location near which the 3C129-A group lies and the orange stars are positions of the two radio sources, 3C129 and 3C129.1.}\label{spatialDensity}
 \end{figure}

The resulting density map constructed with the red sequence galaxies is plotted in Fig.~\ref{spatialDensity}. It shows a distribution that is slightly elongated north-east to south-west in Galactic coordinates. This is possibly related to an intrinsically more elongated mass distribution of the large-scale filamentary structure (PPS) within which this cluster is embedded. Upon closer inspection two distinct structures become apparent. A dominant density peak corresponds to the 3C\,129 cluster centre, located at ($\ell, b$) $\approx$ ($160.52\dg, 0.27\dg$). The other, a more diffuse density peak about 2.9 Mpc (assuming $z = 0.02$) north from the cluster core located near ($\ell, b$) $\approx$ ($160.88\dg, 2.20\dg$), indicates a substructure, henceforth called the 3C129-A group.

A comparison with the X-ray emission map in Fig.~\ref{clusterCore} reveals that the core of the cluster is closely aligned with the X-ray gas emission. The X-ray contours are irregular and are not associated with any X-ray point source in the cluster. Thus supporting what was inferred by \citet{Leahy2000} that the 3C\,129 cluster is under assembly through a merger and it is possibly growing along the filament. \\

Furthermore, we compared this galaxy projected density distribution with the spatial distribution of galaxies detected in \HI\ in the velocity range of the cluster, marked by gradient blue dots. The latter shows the 3C129-A group to be more populated by \HI-rich galaxies and also shows galaxies to have distinctly higher velocities than those around the core of the 3C\,129 cluster. Furthermore, a relatively smaller number of \HI\ detections was found in the core of the cluster, which is indicative of a gas removal mechanism taking place in this region. The implication of this is beyond the scope of this paper and will be discussed in detail in a paper that follows where we will be analysing the environmental effects on the \HI\ properties of the galaxies.\\

\subsubsection{Velocity distribution}
\begin{figure}
%   \centering
 \hspace*{-0.5cm}
 \includegraphics[width=95mm, height=65mm]{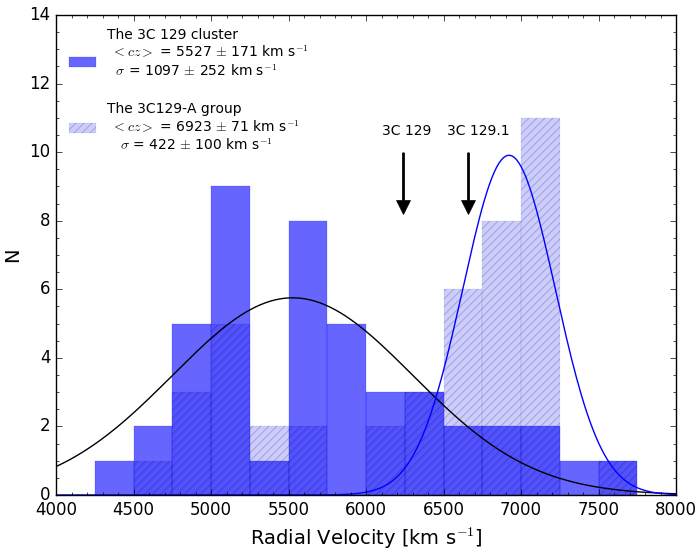}
    \caption{The distribution of radial velocities of the 3C\,129 cluster (dark blue) and its northern substructure; 3C\,129-A (hatched light blue). Velocities bins are 250 \kms\ wide. The black line is a fitted Gaussian profile of the 3C\,129 cluster and blue line for the 3C\,129-A group. The radial velocities of the radio sources, 3C\,129 and 3C\,129.1 are indicated by arrows.}\label{velodist}
 \end{figure}

Having established the existence of the projected 3C129-A substructure in association with the 3C\,129 cluster, we now investigate the radial velocity distribution of their constituent galaxies. We do this by using the 87 galaxies with measured \HI\ radial velocities within the velocity range of the cluster's parent supercluster (PPS) of 4000 - 8000 \kms\ with and without a NIR-counterpart \citep{Ramatsoku2016}. A visual inspection of Fig.\,\ref{velodist} indicates a bimodal velocity distribution of the \HI\ detections. To check whether the velocity distribution is indeed statistically different from a Gaussian distribution we applied a Lilliefors test (see \citealp{Feigelson2012}). The resulting $p$-value is $p$ $<$ 0.01 which indicates that the null hypothesis, which states that the full distribution is Gaussian, can be rejected, thus consistent with the presence of substructure in the radial velocity distribution. 

We estimate the mean radial velocities of the two structures by selecting 43 \HI-detected galaxies within a radius of $1.1\dg$ centred on the cluster's centre, and 44 within a radius of $0.8\dg$ centred on the 3C129-A group (see Fig.\,\ref{spatialScatter}). The latter's radius was chosen to be as large as possible while maintaining mutual exclusivity with the 3C\,129 cluster. The velocity distributions of the galaxies selected in the two structures are shown in Fig.~\ref{velodist}. We fit two Gaussian profiles and obtained $cz = 5227 \pm 171$~\kms\ and  $\sigma = 1097 \pm 252$~\kms\ for the 3C\,129 cluster and  $cz = 6923 \pm 71$~\kms\ and $\sigma = 422 \pm 100$~\kms\ for the 3C129-A group. We note that the velocity dispersion of the cluster is higher than that which was determined from the $\beta$-model of the cluster of $\sigma = 765$~\kms\ \citep{Leahy2000} but also note that our measurement is subject to large uncertainties given the lack of optical spectroscopy for gas-poor galaxies in the cluster core.

Combining the results from the velocity and density distribution analysis we find a distinct substructure to the north of the 3C\,129 cluster with higher velocity and more gas-rich spirals. Although the line-of-sight velocity difference of $cz_{gr} - cz_{cl} = 1696 \pm 185$~\kms\ is quite large, it is possible that the 3C129-A group is falling into the cluster. In paper\,\textsc{iii} we will also perform a Tully-Fisher analysis to get hints of the infall properties of the group. A more detailed and robust dynamical analysis can only be conducted with more redshifts of the gas-poor galaxies in the cluster and the immediate surrounding regions.

\section{Summary}\label{CMDSumDiscuss}
We combined near-infrared images and colours from the UKIDSS Galactic Plane Survey with \HI\ data from the Westerbork Synthesis Radio Telescope, in and around the 3C\,129 cluster to identify its galaxy members. This was done by analysing the properties of the $J - K$ versus $K$ magnitude of galaxies to define and fit the red-sequence of this cluster. The slope of its red sequence was found to be $\alpha\ = -0.023 \pm 0.002$ mag, similar to the Coma cluster ($\alpha = -0.017 \pm 0.009$). We determined the extent of the cluster, centred on the X-ray emission, out to a radius of 1.7~Mpc ($1.34R_{200}$), and obtained a sample of 261 identified 3C\,129 cluster galaxies with $r_{K_{20}} > $ 3\arcsec. Of these, 26 have redshift measurements, 24 from our WSRT \HI\ observations and 3 from previous optical observations. A catalogue and colour images of the galaxy cluster members is presented with the photometric parameters. We note that the way the catalogue was defined and the lack of redshifts might lead to contamination of $\leq$5\% by background galaxies.\\

The galaxies within the core radius of the 3C\,129 cluster were used to assess its galaxy density, which we compared with those of three well-known clusters at similar redshift, namely the Coma, Norma and Perseus clusters. Results show that the galaxy density at the core of the 3C\,129 cluster is lower than in the Coma and Norma cluster, but comparable to that in the Perseus. This places the 3C\,129 cluster among the richest clusters in the Perseus-Pisces filament.\\ 

We visually determined the approximate morphologies of all the 261 cluster galaxies using UKIDSS images and derived the morphology-radius relation of the cluster. A clear morphological segregation occurs at 0.6 - 0.9 Mpc from the cluster core, where the fraction of early-type galaxies increases inwards to more than 35\%, while the fraction of late-type galaxies drops to about 16\%. The segregation continues towards the cluster outskirts where the fraction of early-type galaxies drops to 22\% and the fraction of late-types increases to almost 30\%. In paper\,\textsc{iii} we will assess how the environment may have affected the morphologies of these galaxies.\\

The peak of the spatial distribution of the galaxies in the 3C\,129 cluster core shows no \HI\ detections implying a \HI\ deficient cluster centre, consistent with rich cluster environments. 
The core galaxy distribution also shows a slight asymmetry which is aligned with the irregular X-ray emission and not associated with any X-ray point source or any galaxy in the cluster.  These features strengthen claims by \citet{Leahy2000} that the cluster has undergone (or is undergoing) a merger and has not yet reached a dynamically relaxed state. The \HI-detected galaxies around the cluster core were measured to have, $cz =  5227 \pm 171$~\kms\ with $\sigma = 1097 \pm 252$~\kms. We identified a substructure to the north of the main central region with a radial velocity distribution of $cz = 6923 \pm 71$~\kms\ and $\sigma = 422 \pm 100$~\kms. We suspect that this substructure is infalling, which suggests the cluster is still growing by accreting from its parent large-scale structure that forms the Perseus-Pisces filament. However, this remains unconfirmed as more detailed analyses are required. These will be performed as part of the analysis of the 3C\,129 cluster in paper\,\textsc{iii}.\\

\begin{acknowledgements}
%\begin{scriptsize}
The authors would like to thank Hans Boehringer and Gayoung Chon for assisting with the X-ray data reduction for this study. We thank the anonymous reviewer for the helpful comments and suggestions. MR acknowledges financial support from the Ubbo Emmius Fund of the University of Groningen and the financial support provided by SKA South-Africa and the South African National Research Foundation. The SA and NL authors of this collaboration all benefitted tremendously from collaborative exchanges support by the NRF/NWO bilateral agreement for Astronomy and Astronomy Enabling Technologies. MR's research is supported by the SARAO HCD programme via the "New Scientific Frontiers with Precision Radio Interferometry" research group grant. This work is based upon research supported by the South African Research Chairs Initiative of the Department of Science and Technology and National Research Foundation. MR also acknowledges INAF-OAC for their support. We would also like to thank W. Williams for the photometric contributions to the project. This research has also made use of the UKIRT Infrared Deep Sky Survey database. We acknowledge the UKIDSS support team, including M. Reads, S Hodgkin and P. Lucas for their various support on assessing the UKIDSS images. This work is part of MV's Vici research programme "The Panoramic Perspective on Gas and Galaxy Evolution" with project number 016.130.338, and (partly) financed by the Netherlands Organisation for Scientific Research (NWO). 
\end{acknowledgements}
%\end{scriptsize}

%----------------------------------------------------------------------------------------
%    REFERENCE LIST
%----------------------------------------------------------------------------------------

\bibliographystyle{aa} % mn2e.bst for references
\bibliography{ReferencesPaper2} % your file with extension .bib containing references

% the Appendix should come directly after the References

%-----------------------------APPENDIX--------------------------------------------------------
\onecolumn
\appendix
%\pagenumbering{arabic}
\section{Source extractor parameters}\label{appendixa}
\begin{scriptsize}
\begin{table}[h!]
\caption {Parameters for \textsc{sextractor} 2.8.6.}\label{sextractorparam}
%\footnotesize
\centering
\begin{tabular}{ll}
\hline
 Parameter& Value\\     
\hline\hline

DETECT\_MINAREA  & 50 ( 10 for sources above 9$\sigma$)\\   %        
DETECT\_THRESH   & 3   (9 for min. of 50 adjacent pixels) \\ %        
ANALYSIS\_THRESH & 3 \\          
 FILTER          & Y   \\           
FILTER\_NAME     &default.conv  \\ 
DEBLEND\_NTHRESH & 32  \\          
DEBLEND\_MINCONT & 0.001  \\      
CLEAN           & Y   \\           
CLEAN\_PARAM     & 1.0\\            
MASK\_TYPE       & CORRECT  \\           
PHOT\_APERTURES  & 5  \\            
PHOT\_AUTOPARAMS & 2.5, 3.5  \\     
PHOT\_PETROPARAMS& 2.0, 3.5 \\     
SATUR\_LEVEL     & 422784/ADU \\   
MAG\_ZEROPOINT   & 24.105 \\        
GAIN            & 4.5/e-/ADU  \\          
GAIN\_KEY        & GAIN  \\         
PIXEL\_SCALE     & 0.4/arcsec  \\  
SEEING\_FWHM     & 0.8/arcsec  \\
STARNNW\_NAME    & default.nnw \\  
BACK\_SIZE       & 64 \\            
BACK\_FILTERSIZE & 3  \\           
BACKPHOTO\_TYPE   &GLOBAL \\       
 
\hline\hline\\
\end{tabular}
\end{table}
\end{scriptsize}

\section{The near-infrared images}\label{appendixb}
\begin{centering}
\begin{landscape}
%\scriptsize 
%\onecolumn
% \footnote \noindentsize

\setlength{\tabcolsep}{2.0pt}
\begin{longtable}{crrrrrrrrcrccclc}
\caption*{\textbf{Table B.} Near$-$infrared parameters of the 3C\,129 candidate cluster galaxies.}\label{Candidates}\\

% \multicolumn{9}{c}{\small{\bf{\tablename} \thetable{}.} \textsc{Hi} parameters of reliable
% detections.} \\[0.5ex]
\hline \\[-1.8ex]
\thead{ID no.}       &\thead{Unique ID}   &  \thead{RA}    &  \thead{Dec}    & \thead{$\ell$} &  \thead{$b$}  &  \thead{$\epsilon_{K}$}  &  \thead{$\phi_{K}$}  & \thead{$r_{K20}$} & \thead{$J_{K20}$}  & \thead{$H_{K20}$}  & \thead{$K_{20}$}   &  \thead{$E(B $-$ V)$} & J$^{\rm o}$ $-$ K$^{\rm o}$ &  \thead{Type} & v$_{rad}$ \\
\\
                     & \thead{ZoA}        &   \thead{deg}  &  \thead{deg}    &  \thead{deg}   &  \thead{deg}  &  \thead{}                &  \thead{deg}         &  \thead{\arcsec}    &  \thead{mag}         &  \thead{mag}         &  \thead{mag}        &    \thead{mag}         & \thead{mag}         &                 & \thead{\kms}\\
\\
\thead{(1)}          & \thead{(2)}        &   \thead{(3)}  &  \thead{(4)}    &  \thead{(5)}   &  \thead{(6)}  &  \thead{(7)}             &   \thead{(8)}        &   \thead{(9)}       &   \thead{(10)}       &  \thead{(11)}        &  \thead{(12)}     &  \thead{(13)}            & \thead{(14)}        &  \thead{(15)}   &  \thead{(16)}       \\

\hline \\[-1.8ex]
\endfirsthead

\multicolumn{9}{c}{{\textbf{Table B}} -- Continued} \\[0.5ex]
\hline \\[-1.8ex]

\thead{ID no.}       &\thead{Unique ID}   &  \thead{RA}    &  \thead{Dec}    & \thead{$\ell$} &  \thead{$b$}  &  \thead{$\epsilon_{K}$}  &  \thead{$\phi_{K}$}  & \thead{$r_{K20}$} & \thead{$J_{K20}$}  & \thead{$H_{K20}$}  & \thead{$K_{K20}$}   &  \thead{$E(B $-$ V)$}& \thead{J$^{\rm o}$$-$K$^{\rm o}$} &  \thead{Type} & \thead{v$_{rad}$} \\
\\
                     & \thead{ZoA}        &   \thead{deg}  &  \thead{deg}    &  \thead{deg}   &  \thead{deg}  &  \thead{}                &  \thead{deg}         &  \thead{\arcsec}    &  \thead{mag}         &  \thead{mag}         &  \thead{mag}        &    \thead{mag}         & \thead{mag}         &                 & \thead{\kms}\\
\\
\thead{(1)}          & \thead{(2)}        &   \thead{(3)}  &  \thead{(4)}    &  \thead{(5)}   &  \thead{(6)}  &  \thead{(7)}             &   \thead{(8)}        &   \thead{(9)}       &   \thead{(10)}       &  \thead{(11)}        &  \thead{(12)}     &  \thead{(13)}            & \thead{(14)}        &  \thead{(15)}   &  \thead{(16)}       \\

\hline \\[-1.8ex]
\endhead

\hline \\[-1.8ex]
\\
\endfoot

\endlastfoot
1	&	J044908.26$+$445540.3	&	72.284	&	44.928	&	160.490	&	0.086	&	0.17	&	24.71	    &	28.10	&	10.96	$\pm$	0.02	&	~9.94	$\pm$	0.02	&	~9.51	$\pm$	0.02	&	0.89	&	1.07	&	E/S0	&--- \\
2	&	J045006.67$+$450305.8	&	72.528	&	45.052	&	160.505	&	0.298	&	0.20	&	65.00	    &	37.30	&	11.21	$\pm$	0.04	&	10.10	$\pm$	0.04	&	~9.59	$\pm$	0.04	&	0.98	&	1.20	&	E/S0	&	6655$^{*}$\\
3	&	J044909.06$+$450039.4	&	72.288	&	45.011	&	160.427	&	0.142	&	0.10	&	$-$60.00	&	25.60	&	11.50	$\pm$	0.03	&	10.46	$\pm$	0.04	&	10.00	$\pm$	0.04	&	0.94	&	1.10	&	E/S0	&	6236$^{\times}$\\
4	&	J045145.56$+$443602.6	&	72.940	&	44.601	&	161.039	&	0.235	&	0.40	&	60.00		&	22.80	&	11.75	$\pm$	0.03	&	10.82	$\pm$	0.03	&	10.35	$\pm$	0.04	&	0.79	&	1.07	&	$e$S	&	5086$^{+}$\\
5	&	J044939.78$+$440922.1	&	72.416	&	44.156	&	161.141	&	$-$0.337&	0.09	&	$-$55.01	&	12.79	&	12.05	$\pm$	0.02	&	11.16	$\pm$	0.02	&	10.76	$\pm$	0.02	&	0.65	&	1.01	&	E/S0	&--- \\	
6	&	J044842.39$+$454818.5	&	72.177	&	45.805	&	159.769	&	0.593	&	0.40	&	$-$61.15	&	21.62	&	12.25	$\pm$	0.02	&	11.23	$\pm$	0.02	&	10.77	$\pm$	0.02	&	1.12	&	1.00	&	$e$S	&--- \\	
7	&	J045245.69$+$450106.2	&	73.190	&	45.018	&	160.829	&	0.638	&	0.44	&	85.11		&	19.63	&	12.31	$\pm$	0.02	&	11.23	$\pm$	0.02	&	10.78	$\pm$	0.02	&	0.94	&	1.12	&	E/S0	&--- \\
8	&	J045219.88$+$451546.1	&	73.083	&	45.263	&	160.592	&	0.734	&	0.50	&	$-$60.00	&	25.90	&	12.23	$\pm$	0.05	&	11.26	$\pm$	0.05	&	10.83	$\pm$	0.06	&	0.91	&	1.00	&	$e$S	&--- \\
9	&	J045326.75$+$441900.7	&	73.361	&	44.317	&	161.449	&	0.288	&	0.06	&	26.77		&	14.31	&	12.32	$\pm$	0.02	&	11.38	$\pm$	0.02	&	10.96	$\pm$	0.02	&	0.69	&	1.06	&	E/S0	&--- \\
10	&	J045045.92$+$450659.7	&	72.691	&	45.117	&	160.529	&	0.428	&	0.27	&	0.36		&	13.39	&	12.62	$\pm$	0.02	&	11.56	$\pm$	0.02	&	11.05	$\pm$	0.02	&	1.00	&	1.14	&	E/S0	&--- \\
11	&	J044719.31$+$441701.6	&	71.830	&	44.284	&	160.774	&	$-$0.576&	0.47	&	$-$73.54	&	16.96	&	12.50	$\pm$	0.02	&	11.56	$\pm$	0.02	&	11.06	$\pm$	0.02	&	0.69	&	1.15	&	$e$S	&	4993\\
12	&	J044843.25$+$445216.0	&	72.180	&	44.871	&	160.486	&	$-$0.007&	0.11	&	32.33	    &	9.98	&	12.54	$\pm$	0.02	&	11.54	$\pm$	0.02	&	11.11	$\pm$	0.02	&	0.84	&	1.07	&	E/S0	&--- \\
13	&	J044459.45$+$453344.1	&	71.248	&	45.562	&	159.534	&	$-$0.059&	0.30	&	$-$23.53	&	14.62	&	12.83	$\pm$	0.02	&	11.73	$\pm$	0.02	&	11.19	$\pm$	0.02	&	1.35	&	1.06	&	E/S0	&--- \\
14	&	J044429.89$+$442914.8	&	71.125	&	44.487	&	160.291	&	$-$0.827&	0.31	&	44.80	    &	14.65	&	12.81	$\pm$	0.02	&	11.78	$\pm$	0.02	&	11.23	$\pm$	0.02	&	0.98	&	1.16	&	$e$S	&--- \\
15	&	J044724.21$+$445928.3	&	71.851	&	44.991	&	160.244	&	$-$0.107&	0.80	&	$-$35.00	&	31.50	&	12.64	$\pm$	0.05	&	11.67	$\pm$	0.05	&	11.25	$\pm$	0.07	&	0.84	&	1.03	&	$m$S	&--- \\
16	&	J045414.90$+$450315.1	&	73.562	&	45.054	&	160.967	&	0.864	&	0.17	&	$-$41.28	&	19.00	&	12.79	$\pm$	0.02	&	11.75	$\pm$	0.02	&	11.29	$\pm$	0.02	&	0.93	&	1.10	&	$e$S	&	6269\\
17	&	J045129.31$+$451852.0	&	72.872	&	45.314	&	160.458	&	0.652	&	0.57	&	$-$35.96	&	17.04	&	12.86	$\pm$	0.02	&	11.84	$\pm$	0.02	&	11.31	$\pm$	0.02	&	1.04	&	1.10	&	E/S0	&--- \\
18	&	J045028.01$+$443407.6	&	72.617	&	44.569	&	160.916	&	0.037	&	0.70	&	$-$74.56	&	15.98	&	12.74	$\pm$	0.02	&	11.79	$\pm$	0.02	&	11.32	$\pm$	0.02	&	0.80	&	1.07	&	E/S0	&--- \\
19	&	J045156.57$+$445815.0	&	72.986	&	44.971	&	160.774	&	0.496	&	0.47	&	$-$12.60	&	14.73	&	12.79	$\pm$	0.02	&	11.79	$\pm$	0.02	&	11.33	$\pm$	0.02	&	0.94	&	1.05	&	$e$S	&--- \\
20	&	J045251.94$+$444122.5	&	73.216	&	44.690	&	161.095	&	0.443	&	0.56	&	63.37	    &	16.94	&	12.90	$\pm$	0.02	&	11.92	$\pm$	0.02	&	11.47	$\pm$	0.02	&	0.85	&	1.07	&	$e$S	&--- \\
21	&	J045018.46$+$454152.2	&	72.577	&	45.698	&	160.031	&	0.738	&	0.27	&	24.97	    &	12.06	&	13.08	$\pm$	0.02	&	11.99	$\pm$	0.02	&	11.49	$\pm$	0.02	&	1.17	&	1.08	&	E/S0	&--- \\
22	&	J045324.45$+$451127.6	&	73.352	&	45.191	&	160.768	&	0.835	&	0.40	&	29.03	    &	12.08	&	12.97	$\pm$	0.02	&	11.93	$\pm$	0.02	&	11.50	$\pm$	0.02	&	0.98	&	1.05	&	E/S0	&--- \\
23	&	J044753.50$+$443250.9	&	71.973	&	44.547	&	160.638	&	$-$0.328&	0.10	&	25.00	    &	13.10	&	12.88	$\pm$	0.04	&	11.87	$\pm$	0.05	&	11.52	$\pm$	0.06	&	0.84	&	1.00	&	E/S0	&--- \\	
24	&	J044734.77$+$452912.6	&	71.895	&	45.487	&	159.885	&	0.237	&	0.41	&	50.35	    &	8.87	&	13.14	$\pm$	0.02	&	12.05	$\pm$	0.02	&	11.56	$\pm$	0.02	&	1.19	&	1.06	&	E/S0	&--- \\
25	&	J045332.55$+$453232.6	&	73.386	&	45.542	&	160.510	&	1.076	&	0.39	&	$-$23.08	&	14.16	&	13.09	$\pm$	0.02	&	12.03	$\pm$	0.02	&	11.56	$\pm$	0.02	&	0.98	&	1.11	&	E/S0	&--- \\	
26	&	J044639.43$+$454052.2	&	71.664	&	45.681	&	159.633	&	0.240	&	0.41	&	85.72	    &	12.30	&	13.12	$\pm$	0.02	&	12.08	$\pm$	0.02	&	11.57	$\pm$	0.02	&	1.25	&	1.01	&	E/S0	&--- \\
27	&	J044730.43$+$454548.7	&	71.877	&	45.764	&	159.666	&	0.406	&	0.20	&	47.01	    &	10.49	&	13.14	$\pm$	0.02	&	12.11	$\pm$	0.02	&	11.62	$\pm$	0.02	&	1.25	&	0.98	&	$e$S	&--- \\
28	&	J044953.41$+$451613.5	&	72.473	&	45.270	&	160.312	&	0.408	&	0.60	&	$-$60.00	&	15.10	&	13.33	$\pm$	0.05	&	12.16	$\pm$	0.04	&	11.71	$\pm$	0.06	&	1.01	&	1.18	&	$e$S	&--- \\	
29	&	J045032.79$+$445411.7	&	72.637	&	44.903	&	160.668	&	0.262	&	0.29	&	$-$10.36	&	12.04	&	13.26	$\pm$	0.02	&	12.21	$\pm$	0.02	&	11.71	$\pm$	0.02	&	0.92	&	1.15	&	$e$S	&--- \\
30	&	J044940.04$+$451119.0	&	72.417	&	45.189	&	160.350	&	0.325	&	0.62	&	$-$10.81	&	13.76	&	13.23	$\pm$	0.02	&	12.22	$\pm$	0.02	&	11.72	$\pm$	0.02	&	0.98	&	1.09	&	$e$S	&--- \\
31	&	J045020.02$+$445348.1	&	72.583	&	44.897	&	160.649	&	0.229	&	0.58	&	$-$15.99	&	14.81	&	13.20	$\pm$	0.02	&	12.22	$\pm$	0.02	&	11.73	$\pm$	0.02	&	0.95	&	1.06	&	E/S0	&--- \\
32	&	J045049.32$+$445742.3	&	72.706	&	44.962	&	160.655	&	0.337	&	0.36	&	$-$76.66	&	11.22	&	13.28	$\pm$	0.02	&	12.26	$\pm$	0.02	&	11.80	$\pm$	0.02	&	0.91	&	1.09	&	E/S0	&--- \\
33	&	J045346.20$+$450747.8	&	73.443	&	45.130	&	160.855	&	0.846	&	0.41	&	7.89	    &	15.79	&	13.31	$\pm$	0.02	&	12.22	$\pm$	0.02	&	11.80	$\pm$	0.02	&	0.96	&	1.09	&	$m$S	&--- \\
34	&	J044414.05$+$443036.7	&	71.059	&	44.510	&	160.243	&	$-$0.848&	0.39	&	28.13		&	12.21	&	13.30	$\pm$	0.02	&	12.30	$\pm$	0.02	&	11.81	$\pm$	0.02	&	1.01	&	1.05	&	E/S0	&--- \\	
35	&	J045147.22$+$450757.0	&	72.947	&	45.133	&	160.632	&	0.577	&	0.17	&	$-$44.96	&	10.23	&	13.21	$\pm$	0.02	&	12.24	$\pm$	0.02	&	11.81	$\pm$	0.02	&	0.95	&	0.98	&	E/S0	&--- \\
36	&	J045117.73$+$451208.6	&	72.824	&	45.202	&	160.523	&	0.555	&	0.25	&	35.05		&	9.63	&	13.37	$\pm$	0.02	&	12.35	$\pm$	0.02	&	11.82	$\pm$	0.02	&	1.01	&	1.12	&	E/S0	&--- \\
37	&	J045132.32$+$442920.1	&	72.885	&	44.489	&	161.100	&	0.133	&	0.73	&	$-$87.70	&	16.54	&	13.36	$\pm$	0.02	&	12.43	$\pm$	0.02	&	11.85	$\pm$	0.02	&	0.76	&	1.18	&	$e$S	&	5129\\
38	&	J044918.07$+$445318.3	&	72.325	&	44.888	&	160.538	&	0.083	&	0.21	&	$-$18.37	&	10.40	&	13.23	$\pm$	0.02	&	12.26	$\pm$	0.02	&	11.86	$\pm$	0.02	&	0.90	&	0.98	&	$e$S	&--- \\
39	&	J044941.69$+$452719.7	&	72.424	&	45.455	&	160.148	&	0.500	&	0.18	&	$-$13.54	&	8.04	&	13.42	$\pm$	0.02	&	12.42	$\pm$	0.02	&	11.86	$\pm$	0.02	&	1.04	&	1.11	&	$e$S	&--- \\
40	&	J045021.54$+$450047.4	&	72.590	&	45.013	&	160.563	&	0.307	&	0.53	&	$-$84.10	&	13.27	&	13.43	$\pm$	0.02	&	12.39	$\pm$	0.02	&	11.90	$\pm$	0.02	&	0.98	&	1.10	&	$e$S	&--- \\
41	&	J044558.57$+$441637.7	&	71.494	&	44.277	&	160.623	&	$-$0.764&	0.55	&	$-$11.81	&	17.24	&	13.30	$\pm$	0.02	&	12.31	$\pm$	0.02	&	11.91	$\pm$	0.02	&	0.85	&	1.03	&	$e$S	&--- \\	
42	&	J045135.55$+$450302.9	&	72.898	&	45.051	&	160.673	&	0.499   &	0.50	&	54.65		&	10.20	&	13.39	$\pm$	0.02	&	12.43	$\pm$	0.02	&	11.95	$\pm$	0.02	&	0.95	&	1.03	&	E/S0	&--- \\
43	&	J044913.04$+$443545.2	&	72.304	&	44.596	&	160.753	&	$-$0.116&	0.48	&	4.20	    &	9.99	&	13.45	$\pm$	0.02	&	12.55	$\pm$	0.02	&	12.14	$\pm$	0.02	&	0.78	&	0.97	&	E/S0	&--- \\	
44	&	J045005.82$+$453012.6	&	72.524	&	45.503	&	160.156	&	0.585	&	0.23	&	18.87		&	10.69	&	13.69	$\pm$	0.02	&	12.72	$\pm$	0.02	&	12.14	$\pm$	0.02	&	1.09	&	1.08	&	$m$S	&--- \\
45	&	J044940.15$+$445844.4	&	72.417	&	44.979	&	160.511	&	0.191	&	0.64	&	7.97		&	14.62	&	13.63	$\pm$	0.02	&	12.63	$\pm$	0.02	&	12.17	$\pm$	0.02	&	0.94	&	1.05	&	$e$S	&--- \\
46	&	J045134.83$+$452431.0	&	72.895	&	45.409	&	160.395	&	0.725	&	0.12	&	$-$70.79	&	10.37	&	13.68	$\pm$	0.02	&	12.66	$\pm$	0.02	&	12.18	$\pm$	0.02	&	1.05	&	1.05	&	E/S0	&--- \\	
47	&	J045217.59$+$443511.6	&	73.073	&	44.587	&	161.110	&	0.299	&	0.46	&	$-$36.63	&	9.46	&	13.53	$\pm$	0.02	&	12.62	$\pm$	0.02	&	12.19	$\pm$	0.02	&	0.76	&	1.02	&	E/S0	&--- \\
48	&	J044747.17$+$444434.0	&	71.947	&	44.743	&	160.477	&	$-$0.216&	0.50	&	10.00		&	12.70	&	13.59	$\pm$	0.05	&	12.61	$\pm$	0.06	&	12.20	$\pm$	0.07	&	0.88	&	1.01	&	$e$S	&--- \\
49	&	J044935.63$+$450035.3	&	72.398	&	45.010	&	160.479	&	0.201	&	0.40	&	60.58		&	10.79	&	13.67	$\pm$	0.02	&	12.65	$\pm$	0.02	&	12.21	$\pm$	0.02	&	0.95	&	1.05	&	$e$S	&--- \\
50	&	J044956.71$+$451705.5	&	72.486	&	45.285	&	160.307	&	0.425	&	0.17	&	56.59		&	11.24	&	13.81	$\pm$	0.02	&	12.71	$\pm$	0.02	&	12.26	$\pm$	0.02	&	1.04	&	1.10	&	E/S0	&--- \\
51	&	J045018.81$+$451136.0	&	72.578	&	45.193	&	160.419	&	0.416	&	0.47	&	$-$47.84	&	10.04	&	13.83	$\pm$	0.02	&	12.79	$\pm$	0.02	&	12.28	$\pm$	0.02	&	1.02	&	1.11	&	$e$S	&--- \\	
52	&	J044557.04$+$450410.9	&	71.488	&	45.070	&	160.017	&	$-$0.252&	0.09	&	$-$21.67	&	8.27	&	13.78	$\pm$	0.02	&	12.78	$\pm$	0.02	&	12.35	$\pm$	0.02	&	0.98	&	1.01	&	E/S0	&--- \\
53	&	J044820.62$+$453134.0	&	72.086	&	45.526	&	159.942	&	0.364	&	0.51	&	53.15		&	10.89	&	13.84	$\pm$	0.02	&	12.83	$\pm$	0.02	&	12.38	$\pm$	0.02	&	1.04	&	1.01	&	E/S0	&--- \\
54	&	J044608.05$+$441621.5	&	71.534	&	44.273	&	160.644	&	$-$0.745&	0.75	&	$-$11.80	&	14.15	&	13.93	$\pm$	0.02	&	12.96	$\pm$	0.02	&	12.48	$\pm$	0.02	&	0.84	&	1.09	&	$e$S	&--- \\	
55	&	J045303.58$+$442506.5	&	73.265	&	44.418	&	161.327	&	0.298	&	0.68	&	$-$3.61		&	12.52	&	13.80	$\pm$	0.02	&	12.92	$\pm$	0.02	&	12.50	$\pm$	0.02	&	0.78	&	0.97	&	$e$S	&--- \\
56	&	J044934.14$+$443426.7	&	72.392	&	44.574	&	160.810	&	$-$0.082&	0.44	&	35.56		&	14.58	&	13.84	$\pm$	0.02	&	12.91	$\pm$	0.02	&	12.52	$\pm$	0.02	&	0.76	&	0.99	&	$m$S	&--- \\	
57	&	J045038.23$+$450521.6	&	72.659	&	45.089	&	160.535	&	0.393	&	0.36	&	47.90		&	6.84	&	14.07	$\pm$	0.02	&	13.09	$\pm$	0.02	&	12.54	$\pm$	0.02	&	1.00	&	1.10	&	E/S0	&--- \\
58	&	J045125.48$+$452551.4	&	72.856	&	45.431	&	160.361	&	0.718	&	0.52	&	79.07		&	12.39	&	14.08	$\pm$	0.02	&	13.04	$\pm$	0.02	&	12.54	$\pm$	0.02	&	1.09	&	1.06	&	$m$S	&--- \\
59	&	J044949.59$+$450319.7	&	72.457	&	45.055	&	160.470	&	0.262	&	0.20	&	$-$30.00	&	8.50	&	14.07	$\pm$	0.07	&	13.03	$\pm$	0.06	&	12.57	$\pm$	0.08	&	0.96	&	1.08	&	E/S0	&--- \\	
60	&	J045401.51$+$451426.6	&	73.506	&	45.241	&	160.798	&	0.951	&	0.55	&	$-$61.16	&	9.67	&	14.01	$\pm$	0.02	&	13.01	$\pm$	0.02	&	12.57	$\pm$	0.02	&	0.98	&	1.02	&	$e$S	&--- \\
61	&	J045444.87$+$453116.6	&	73.687	&	45.521	&	160.660	&	1.226	&	0.50	&	51.50	&	13.71	&	13.94	$\pm$	0.02	&	12.98	$\pm$	0.02	&	12.57	$\pm$	0.02	&	0.80	&	1.02	&	$e$S	&--- \\
62	&	J045233.08$+$445548.6	&	73.138	&	44.930	&	160.874	&	0.553	&	0.63	&	$-$57.68	&	8.91	&	14.06	$\pm$	0.02	&	13.08	$\pm$	0.02	&	12.59	$\pm$	0.02	&	0.94	&	1.07	&	$e$S	&--- \\	
63	&	J044646.93$+$441620.4	&	71.696	&	44.272	&	160.720	&	$-$0.657&	0.61	&	$-$84.87	&	13.48	&	13.94	$\pm$	0.02	&	12.99	$\pm$	0.02	&	12.60	$\pm$	0.02	&	0.74	&	1.01	&	$m$S	&--- \\
64	&	J044801.07$+$444319.3	&	72.004	&	44.722	&	160.519	&	$-$0.198&	0.26	&	24.21	&	7.83	&	14.05	$\pm$	0.02	&	13.08	$\pm$	0.02	&	12.64	$\pm$	0.02	&	0.86	&	1.04	&	E/S0	&--- \\
65	&	J045040.23$+$445701.0	&	72.668	&	44.950	&	160.646	&	0.309	&	0.70	&	$-$67.23	&	14.57	&	14.07	$\pm$	0.02	&	13.09	$\pm$	0.02	&	12.68	$\pm$	0.02	&	0.92	&	0.99	&	$m$S	&--- \\
66	&	J044409.43$+$443700.4	&	71.039	&	44.617	&	160.153	&	$-$0.788&	0.40	&	$-$51.65	&	10.30	&	14.22	$\pm$	0.02	&	13.18	$\pm$	0.02	&	12.70	$\pm$	0.02	&	1.09	&	1.06	&	$m$S	&--- \\	
67	&	J044902.91$+$445639.9	&	72.262	&	44.944	&	160.467	&	0.085	&	0.07	&	$-$89.11	&	8.41	&	14.11	$\pm$	0.02	&	13.16	$\pm$	0.02	&	12.70	$\pm$	0.02	&	0.90	&	1.03	&	$e$S	&--- \\
68	&	J044927.08$+$450121.8	&	72.363	&	45.023	&	160.452	&	0.190	&	0.29	&	$-$83.90	&	7.96	&	14.22	$\pm$	0.02	&	13.18	$\pm$	0.02	&	12.73	$\pm$	0.02	&	0.95	&	1.08	&	E/S0	&--- \\
69	&	J045429.83$+$444425.9	&	73.624	&	44.741	&	161.239	&	0.700	&	0.38	&	$-$74.39	&	7.18	&	14.23	$\pm$	0.02	&	13.23	$\pm$	0.02	&	12.75	$\pm$	0.02	&	0.87	&	1.10	&	$e$S	&--- \\
70	&	J045156.51$+$450314.5	&	72.985	&	45.054	&	160.710	&	0.548	&	0.40	&	30.50	&	10.67	&	14.16	$\pm$	0.02	&	13.23	$\pm$	0.02	&	12.80	$\pm$	0.02	&	0.94	&	0.96	&	$m$S	&	4523\\
71	&	J044642.74$+$444225.2	&	71.678	&	44.707	&	160.381	&	$-$0.385&	0.02	&	$-$69.65	&	5.85	&	14.25	$\pm$	0.02	&	13.26	$\pm$	0.02	&	12.80	$\pm$	0.02	&	0.92	&	1.06	&	$e$S	&--- \\	
72	&	J045031.44$+$453451.2	&	72.631	&	45.581	&	160.145	&	0.692	&	0.43	&	$-$20.20	&	8.32	&	14.37	$\pm$	0.02	&	13.35	$\pm$	0.02	&	12.86	$\pm$	0.02	&	1.20	&	0.99	&	$e$S	&--- \\
73	&	J044741.94$+$445903.4	&	71.925	&	44.984	&	160.282	&	$-$0.072&	0.36	&	17.96	&	8.27	&	14.24	$\pm$	0.02	&	13.30	$\pm$	0.02	&	12.87	$\pm$	0.02	&	0.86	&	1.01	&	$e$S	&--- \\
74	&	J045030.71$+$442816.0	&	72.628	&	44.471	&	160.997	&	$-$0.019&	0.35	&	45.97	&	8.77	&	14.15	$\pm$	0.02	&	13.25	$\pm$	0.02	&	12.87	$\pm$	0.02	&	0.79	&	0.94	&	$e$S	&--- \\
75	&	J045149.06$+$450924.8	&	72.954	&	45.157	&	160.616	&	0.597	&	0.22	&	44.85	&	7.22	&	14.29	$\pm$	0.02	&	13.33	$\pm$	0.02	&	12.87	$\pm$	0.02	&	0.93	&	1.03	&	E/S0	&--- \\
76	&	J044615.88$+$450530.3	&	71.566	&	45.092	&	160.036	&	$-$0.195&	0.65	&	62.54	&	9.64	&	14.26	$\pm$	0.02	&	13.44	$\pm$	0.02	&	12.91	$\pm$	0.02	&	0.94	&	0.94	&	$e$S	&--- \\	
77	&	J044752.80$+$450202.3	&	71.970	&	45.034	&	160.265	&	$-$0.015&	0.40	&	$-$40.00	&	9.50	&	14.26	$\pm$	0.08	&	13.35	$\pm$	0.08	&	12.91	$\pm$	0.09	&	0.86	&	0.98	&	E/S0	&--- \\	
78	&	J045005.99$+$444129.8	&	72.525	&	44.692	&	160.780	&	0.066	&	0.57	&	$-$84.01	&	7.73	&	14.31	$\pm$	0.02	&	13.41	$\pm$	0.02	&	12.97	$\pm$	0.02	&	0.82	&	0.98	&	$e$S	&--- \\
79	&	J044707.90$+$442944.4	&	71.783	&	44.496	&	160.590	&	$-$0.465&	0.10	&	$-$69.40	&	7.97	&	14.43	$\pm$	0.02	&	13.54	$\pm$	0.02	&	12.98	$\pm$	0.02	&	0.78	&	1.11	&	E/S0	&--- \\	
80	&	J044918.88$+$454737.1	&	72.329	&	45.794	&	159.846	&	0.666	&	0.21	&	$-$51.00	&	9.01	&	14.60	$\pm$	0.02	&	13.56	$\pm$	0.02	&	12.98	$\pm$	0.02	&	1.14	&	1.12	&	$m$S	&--- \\
81	&	J045418.02$+$453951.7	&	73.575	&	45.664	&	160.499	&	1.255	&	0.39	&	49.59	&	9.30	&	14.35	$\pm$	0.02	&	13.41	$\pm$	0.02	&	13.00	$\pm$	0.02	&	0.88	&	0.96	&	$m$S	&--- \\
82	&	J045353.83$+$443238.8	&	73.474	&	44.544	&	161.324	&	0.494	&	0.37	&	37.88	&	8.88	&	14.45	$\pm$	0.02	&	13.51	$\pm$	0.02	&	13.02	$\pm$	0.02	&	0.85	&	1.07	&	$e$S	&--- \\
83	&	J045146.37$+$452706.8	&	72.943	&	45.452	&	160.384	&	0.778	&	0.59	&	79.33	&	8.71	&	14.54	$\pm$	0.02	&	13.53	$\pm$	0.02	&	13.04	$\pm$	0.02	&	1.13	&	1.01	&	$e$S	&--- \\
84	&	J044520.92$+$454430.7	&	71.337	&	45.742	&	159.440	&	0.110	&	0.77	&	24.83	&	14.61	&	14.94	$\pm$	0.02	&	13.68	$\pm$	0.02	&	13.05	$\pm$	0.02	&	1.28	&	1.33	&	$m$S	&	5134\\
85	&	J045206.26$+$451041.6	&	73.026	&	45.178	&	160.632	&	0.649	&	0.23	&	$-$75.26	&	6.58	&	14.51	$\pm$	0.02	&	13.48	$\pm$	0.02	&	13.05	$\pm$	0.02	&	0.90	&	1.07	&	E/S0	&--- \\
86	&	J044859.13$+$450201.1	&	72.246	&	45.034	&	160.391	&	0.134	&	0.22	&	$-$23.15	&	6.11	&	14.55	$\pm$	0.02	&	13.54	$\pm$	0.02	&	13.06	$\pm$	0.02	&	0.93	&	1.08	&	E/S0	&--- \\
87	&	J045022.58$+$451600.9	&	72.594	&	45.267	&	160.370	&	0.472	&	0.26	&	65.62	&	6.78	&	14.66	$\pm$	0.02	&	13.54	$\pm$	0.02	&	13.06	$\pm$	0.02	&	1.03	&	1.16	&	E/S0	&--- \\
88	&	J044848.38$+$450140.8	&	72.202	&	45.028	&	160.375	&	0.106	&	0.48	&	$-$43.50	&	8.56	&	14.60	$\pm$	0.02	&	13.54	$\pm$	0.02	&	13.11	$\pm$	0.02	&	0.90	&	1.09	&	$m$S	&--- \\	
89	&	J044941.44$+$445807.9	&	72.423	&	44.969	&	160.521	&	0.188	&	0.21	&	$-$34.30	&	6.52	&	14.52	$\pm$	0.02	&	13.55	$\pm$	0.02	&	13.12	$\pm$	0.02	&	0.94	&	1.00	&	E/S0	&--- \\
90	&	J044922.34$+$444427.2	&	72.343	&	44.741	&	160.660	&	$-$0.002&	0.08	&	$-$15.60	&	6.15	&	14.55	$\pm$	0.02	&	13.59	$\pm$	0.02	&	13.17	$\pm$	0.02	&	0.84	&	1.01	&	$e$S	&--- \\	
91	&	J044812.49$+$450920.1	&	72.052	&	45.156	&	160.210	&	0.107	&	0.24	&	$-$45.33	&	7.28	&	14.60	$\pm$	0.02	&	13.61	$\pm$	0.02	&	13.20	$\pm$	0.02	&	0.94	&	0.99	&	$e$S	&--- \\
92	&	J044743.67$+$445439.8	&	71.932	&	44.911	&	160.342	&	$-$0.115&	0.41	&	$-$69.81	&	7.60	&	14.56	$\pm$	0.02	&	13.68	$\pm$	0.02	&	13.21	$\pm$	0.02	&	0.84	&	0.98	&	$e$S	&--- \\	
93	&	J045131.32$+$450558.1	&	72.880	&	45.099	&	160.627	&	0.520	&	0.03	&	$-$16.62	&	6.54	&	14.74	$\pm$	0.02	&	13.72	$\pm$	0.02	&	13.25	$\pm$	0.02	&	0.95	&	1.07	&	E/S0	&--- \\
94	&	J044944.65$+$452304.9	&	72.436	&	45.385	&	160.208	&	0.462	&	0.37	&	$-$41.69	&	7.31	&	14.70	$\pm$	0.02	&	13.70	$\pm$	0.02	&	13.26	$\pm$	0.02	&	1.06	&	0.99	&	$e$S	&--- \\
95	&	J044933.20$+$445558.6	&	72.388	&	44.933	&	160.533	&	0.146	&	0.10	&	$-$56.04	&	4.90	&	14.75	$\pm$	0.02	&	13.76	$\pm$	0.02	&	13.28	$\pm$	0.02	&	0.94	&	1.06	&	E/S0	&--- \\
96	&	J044935.99$+$445213.7	&	72.400	&	44.870	&	160.586	&	0.112	&	0.24	&	2.46	&	6.83	&	14.75	$\pm$	0.02	&	13.73	$\pm$	0.02	&	13.30	$\pm$	0.02	&	0.93	&	1.05	&	E/S0	&--- \\
97	&	J044943.68$+$443245.1	&	72.432	&	44.546	&	160.850	&	$-$0.078&	0.16	&	$-$59.70	&	5.72	&	14.63	$\pm$	0.02	&	13.73	$\pm$	0.02	&	13.35	$\pm$	0.02	&	0.75	&	0.97	&	$e$S	&--- \\
98	&	J045216.50$+$441218.2	&	73.069	&	44.205	&	161.403	&	0.055	&	0.27	&	$-$3.60	&	6.37	&	14.72	$\pm$	0.02	&	13.77	$\pm$	0.02	&	13.35	$\pm$	0.02	&	0.63	&	1.10	&	E/S0	&--- \\
99	&	J044947.45$+$450944.5	&	72.448	&	45.162	&	160.384	&	0.325	&	0.43	&	$-$44.49	&	7.29	&	14.89	$\pm$	0.02	&	13.88	$\pm$	0.02	&	13.37	$\pm$	0.02	&	0.98	&	1.10	&	$e$S	&--- \\	
100	&	J044417.15$+$443615.4	&	71.071	&	44.604	&	160.178	&	$-$0.779&	0.50	&	$-$50.00	&	9.40	&	14.98	$\pm$	0.11	&	13.80	$\pm$	0.09	&	13.46	$\pm$	0.12	&	1.09	&	1.04	&	$m$S	&--- \\
101	&	J045226.36$+$451011.7	&	73.110	&	45.170	&	160.676	&	0.690	&	0.39	&	21.47	&	6.53	&	14.89	$\pm$	0.02	&	13.93	$\pm$	0.02	&	13.47	$\pm$	0.02	&	0.87	&	1.05	&	E/S0	&--- \\
102	&	J045216.79$+$455402.2	&	73.070	&	45.901	&	160.093	&	1.132	&	0.52	&	$-$17.24	&	7.05	&	14.93	$\pm$	0.02	&	13.96	$\pm$	0.02	&	13.49	$\pm$	0.02	&	1.02	&	1.00	&	$m$S	&--- \\	
103	&	J045301.01$+$442921.2	&	73.254	&	44.489	&	161.267	&	0.337	&	0.39	&	34.50	&	7.43	&	14.92	$\pm$	0.02	&	13.97	$\pm$	0.02	&	13.50	$\pm$	0.02	&	0.80	&	1.08	&	$m$S	&--- \\
104	&	J045314.91$+$452642.6	&	73.312	&	45.445	&	160.553	&	0.974	&	0.49	&	$-$74.74	&	6.44	&	14.98	$\pm$	0.02	&	13.98	$\pm$	0.02	&	13.54	$\pm$	0.02	&	0.98	&	1.02	&	E/S0	&--- \\	
105	&	J044650.13$+$452342.0	&	71.709	&	45.395	&	159.871	&	0.078	&	0.45	&	65.44	&	7.23	&	15.07	$\pm$	0.02	&	14.04	$\pm$	0.02	&	13.55	$\pm$	0.02	&	1.13	&	1.03	&	$e$S	&--- \\
106	&	J045004.69$+$450036.1	&	72.520	&	45.010	&	160.533	&	0.267	&	0.34	&	31.33	&	4.94	&	15.09	$\pm$	0.02	&	14.08	$\pm$	0.02	&	13.57	$\pm$	0.02	&	0.97	&	1.10	&	E/S0	&--- \\
107	&	J045100.26$+$445855.4	&	72.751	&	44.982	&	160.659	&	0.375	&	0.24	&	36.21	&	4.72	&	15.04	$\pm$	0.02	&	14.05	$\pm$	0.02	&	13.62	$\pm$	0.02	&	0.91	&	1.03	&	E/S0	&--- \\
108	&	J044959.35$+$445647.7	&	72.497	&	44.947	&	160.572	&	0.214	&	0.33	&	$-$53.30	&	6.69	&	15.11	$\pm$	0.02	&	14.13	$\pm$	0.02	&	13.65	$\pm$	0.02	&	0.97	&	1.04	&	$m$S	&--- \\
109	&	J045432.63$+$451846.8	&	73.636	&	45.313	&	160.799	&	1.067	&	0.16	&	$-$0.60	&	3.03	&	15.19	$\pm$	0.02	&	14.23	$\pm$	0.02	&	13.67	$\pm$	0.02	&	0.90	&	1.14	&	E/S0	&--- \\
110	&	J045445.42$+$453911.3	&	73.689	&	45.653	&	160.560	&	1.310	&	0.48	&	$-$39.85	&	9.48	&	14.87	$\pm$	0.02	&	14.01	$\pm$	0.02	&	13.73	$\pm$	0.03	&	0.84	&	0.77	&	$m$S	&	4941\\
111	&	J045143.42$+$454722.1	&	72.931	&	45.789	&	160.118	&	0.987	&	0.26	&	$-$66.54	&	3.51	&	15.37	$\pm$	0.02	&	14.40	$\pm$	0.02	&	13.77	$\pm$	0.02	&	1.34	&	1.02	&	E/S0	&--- \\
112	&	J044900.93$+$441937.0	&	72.254	&	44.327	&	160.936	&	$-$0.316&	0.69	&	$-$85.32	&	9.15	&	15.20	$\pm$	0.02	&	14.27	$\pm$	0.02	&	13.81	$\pm$	0.03	&	0.64	&	1.11	&	$m$S	&	6332\\
113	&	J044555.95$+$444549.4	&	71.483	&	44.764	&	160.248	&	$-$0.453&	0.24	&	70.62	&	6.78	&	15.19	$\pm$	0.02	&	14.23	$\pm$	0.02	&	13.81	$\pm$	0.03	&	0.99	&	0.95	&	$m$S	&--- \\
114	&	J045403.78$+$443621.7	&	73.516	&	44.606	&	161.294	&	0.556	&	0.49	&	$-$43.64	&	6.81	&	15.12	$\pm$	0.02	&	14.20	$\pm$	0.02	&	13.81	$\pm$	0.02	&	0.87	&	0.94	&	$m$S	&--- \\
115	&	J044823.57$+$451244.2	&	72.098	&	45.212	&	160.187	&	0.169	&	0.03	&	41.98	&	5.03	&	15.23	$\pm$	0.02	&	14.28	$\pm$	0.02	&	13.82	$\pm$	0.02	&	0.98	&	0.99	&	$m$S	&--- \\
116	&	J045105.62$+$443545.0	&	72.773	&	44.596	&	160.970	&	0.140	&	0.05	&	38.79	&	5.25	&	14.93	$\pm$	0.02	&	14.18	$\pm$	0.02	&	13.87	$\pm$	0.03	&	0.77	&	0.73	&	$m$S	&	5044\\
117	&	J045156.82$+$453619.8	&	72.987	&	45.606	&	160.284	&	0.900	&	0.12	&	$-$0.68	&	4.67	&	15.62	$\pm$	0.02	&	14.53	$\pm$	0.02	&	13.88	$\pm$	0.02	&	1.62	&	1.04	&	E/S0	&--- \\
118	&	J044920.79$+$450159.4	&	72.337	&	45.033	&	160.433	&	0.183	&	0.50	&	$-$60.00	&	7.60	&	15.25	$\pm$	0.13	&	14.00	$\pm$	0.09	&	13.89	$\pm$	0.14	&	0.95	&	0.95	&	$m$S	&--- \\
119	&	J045029.63$+$441012.4	&	72.623	&	44.170	&	161.226	&	$-$0.214&	0.15	&	2.40	&	6.40	&	15.15	$\pm$	0.02	&	14.31	$\pm$	0.02	&	13.92	$\pm$	0.03	&	0.64	&	0.96	&	$m$S	&--- \\
120	&	J044706.52$+$451102.1	&	71.777	&	45.184	&	160.063	&	$-$0.022&	0.07	&	$-$35.59	&	3.95	&	15.36	$\pm$	0.02	&	14.40	$\pm$	0.02	&	13.94	$\pm$	0.02	&	0.94	&	1.01	&	E/S0	&--- \\
121	&	J044950.94$+$454506.7	&	72.462	&	45.752	&	159.938	&	0.711	&	0.14	&	$-$35.38	&	4.73	&	15.53	$\pm$	0.02	&	14.48	$\pm$	0.02	&	13.94	$\pm$	0.02	&	1.15	&	1.09	&	E/S0	&--- \\
122	&	J044434.42$+$445034.2	&	71.143	&	44.843	&	160.031	&	$-$0.584&	0.10	&	9.12	&	3.65	&	15.44	$\pm$	0.02	&	14.40	$\pm$	0.02	&	13.96	$\pm$	0.02	&	1.11	&	1.01	&	E/S0	&--- \\
123	&	J044946.26$+$450747.6	&	72.443	&	45.130	&	160.406	&	0.302	&	0.49	&	24.10	&	6.81	&	15.45	$\pm$	0.02	&	14.37	$\pm$	0.02	&	13.97	$\pm$	0.03	&	0.98	&	1.07	&	$m$S	&--- \\
124	&	J044658.40$+$452255.0	&	71.743	&	45.382	&	159.896	&	0.088	&	0.07	&	$-$35.86	&	4.02	&	15.61	$\pm$	0.02	&	14.62	$\pm$	0.02	&	13.99	$\pm$	0.02	&	1.11	&	1.14	&	E/S0	&--- \\
125	&	J044816.56$+$451604.5	&	72.069	&	45.268	&	160.131	&	0.189	&	0.01	&	$-$39.80	&	4.12	&	15.38	$\pm$	0.02	&	14.43	$\pm$	0.02	&	14.00	$\pm$	0.02	&	1.02	&	0.94	&	E/S0	&--- \\
126	&	J045042.89$+$445937.4	&	72.679	&	44.994	&	160.618	&	0.343	&	0.32	&	$-$63.47	&	5.81	&	15.48	$\pm$	0.02	&	14.43	$\pm$	0.02	&	14.00	$\pm$	0.02	&	0.93	&	1.07	&	$m$S	&--- \\
127	&	J044403.24$+$443827.5	&	71.014	&	44.641	&	160.123	&	$-$0.786&	0.00	&	90.00	&	5.90	&	15.56	$\pm$	0.16	&	14.36	$\pm$	0.13	&	14.02	$\pm$	0.17	&	1.10	&	1.06	&	$m$S	&--- \\
128	&	J045021.00$+$451216.6	&	72.587	&	45.205	&	160.414	&	0.428	&	0.26	&	$-$72.94	&	5.67	&	15.53	$\pm$	0.02	&	14.49	$\pm$	0.02	&	14.02	$\pm$	0.02	&	1.02	&	1.07	&	$m$S	&--- \\
129	&	J045029.29$+$460723.4	&	72.622	&	46.123	&	159.724	&	1.034	&	0.28	&	49.67	&	5.21	&	15.57	$\pm$	0.02	&	14.62	$\pm$	0.02	&	14.02	$\pm$	0.02	&	0.99	&	1.13	&	E/S0	&--- \\
130	&	J044646.85$+$452209.9	&	71.695	&	45.369	&	159.884	&	0.054	&	0.09	&	$-$26.10	&	4.14	&	15.59	$\pm$	0.02	&	14.60	$\pm$	0.02	&	14.03	$\pm$	0.02	&	1.10	&	1.09	&	E/S0	&--- \\
131	&	J045419.04$+$452229.0	&	73.579	&	45.375	&	160.726	&	1.075	&	0.20	&	72.14	&	5.28	&	15.48	$\pm$	0.02	&	14.53	$\pm$	0.02	&	14.03	$\pm$	0.02	&	0.93	&	1.04	&	$m$S	&--- \\
132	&	J045416.62$+$452530.8	&	73.569	&	45.425	&	160.682	&	1.102	&	0.00	&	90.00	&	5.00	&	15.42	$\pm$	0.14	&	14.49	$\pm$	0.14	&	14.07	$\pm$	0.16	&	0.90	&	0.96	&	$e$S	&--- \\
133	&	J044721.08$+$450806.7	&	71.838	&	45.135	&	160.128	&	$-$0.021&	0.25	&	$-$58.10	&	5.70	&	15.47	$\pm$	0.02	&	14.50	$\pm$	0.02	&	14.09	$\pm$	0.03	&	0.90	&	0.99	&	$m$S	&--- \\
134	&	J045002.01$+$450215.3	&	72.508	&	45.038	&	160.507	&	0.278	&	0.48	&	55.19	&	6.12	&	15.54	$\pm$	0.02	&	14.54	$\pm$	0.02	&	14.09	$\pm$	0.02	&	0.98	&	1.03	&	$e$S	&--- \\
135	&	J045541.23$+$445953.2	&	73.922	&	44.998	&	161.171	&	1.026	&	0.33	&	$-$54.60	&	5.30	&	15.60	$\pm$	0.02	&	14.64	$\pm$	0.02	&	14.11	$\pm$	0.02	&	0.90	&	1.11	&	$e$S	&--- \\
136	&	J044802.83$+$452237.9	&	72.012	&	45.377	&	160.022	&	0.229	&	0.17	&	22.56	&	4.32	&	15.71	$\pm$	0.02	&	14.76	$\pm$	0.02	&	14.12	$\pm$	0.02	&	1.07	&	1.13	&	E/S0	&--- \\
137	&	J044524.63$+$451918.2	&	71.353	&	45.322	&	159.764	&	$-$0.160	&	0.02	&	$-$32.18	&	4.01	&	15.60	$\pm$	0.02	&	14.71	$\pm$	0.02	&	14.13	$\pm$	0.03	&	1.14	&	0.98	&	$m$S	&	6830\\
138	&	J044908.53$+$450705.9	&	72.286	&	45.118	&	160.344	&	0.209	&	0.12	&	26.00	&	5.19	&	15.57	$\pm$	0.02	&	14.58	$\pm$	0.02	&	14.14	$\pm$	0.03	&	0.96	&	1.02	&	$m$S	&--- \\
139	&	J044755.58$+$450438.2	&	71.982	&	45.077	&	160.237	&	0.019	&	0.43	&	$-$62.81	&	5.82	&	15.52	$\pm$	0.02	&	14.55	$\pm$	0.02	&	14.15	$\pm$	0.03	&	0.88	&	1.00	&	$m$S	&--- \\
140	&	J045029.48$+$450139.1	&	72.623	&	45.028	&	160.567	&	0.334	&	0.15	&	$-$24.88	&	3.98	&	15.66	$\pm$	0.02	&	14.65	$\pm$	0.02	&	14.15	$\pm$	0.02	&	0.97	&	1.09	&	Irr	&--- \\
141	&	J045026.25$+$445346.6	&	72.609	&	44.896	&	160.661	&	0.243	&	0.04	&	42.27	&	4.30	&	15.56	$\pm$	0.02	&	14.64	$\pm$	0.02	&	14.18	$\pm$	0.02	&	0.93	&	0.98	&	E/S0	&--- \\
142	&	J044616.70$+$440932.4	&	71.570	&	44.159	&	160.748	&	$-$0.799	&	0.17	&	26.77	&	4.30	&	15.52	$\pm$	0.02	&	14.62	$\pm$	0.02	&	14.19	$\pm$	0.03	&	0.79	&	0.99	&	$m$S	&--- \\	
143	&	J044632.01$+$440858.6	&	71.633	&	44.150	&	160.784	&	$-$0.770	&	0.00	&	90.00	&	5.00	&	15.53	$\pm$	0.14	&	14.43	$\pm$	0.04	&	14.21	$\pm$	0.17	&	0.75	&	1.00	&	$e$S	&--- \\
144	&	J044603.60$+$444430.9	&	71.515	&	44.742	&	160.279	&	$-$0.450	&	0.09	&	$-$5.20	&	4.19	&	15.61	$\pm$	0.02	&	14.62	$\pm$	0.02	&	14.24	$\pm$	0.03	&	0.99	&	0.93	&	$m$S	&--- \\
145	&	J044357.81$+$443721.5	&	70.991	&	44.623	&	160.126	&	$-$0.810	&	0.48	&	81.34	&	5.71	&	15.74	$\pm$	0.02	&	14.74	$\pm$	0.02	&	14.25	$\pm$	0.03	&	1.09	&	1.03	&	$m$S	&--- \\
146	&	J045159.93$+$451925.2	&	73.000	&	45.324	&	160.508	&	0.728	&	0.58	&	8.91	&	6.35	&	15.66	$\pm$	0.02	&	14.69	$\pm$	0.02	&	14.25	$\pm$	0.03	&	0.99	&	0.98	&	$e$S	&--- \\
147	&	J045229.45$+$452451.7	&	73.123	&	45.414	&	160.493	&	0.852	&	0.59	&	$-$19.33	&	6.61	&	15.66	$\pm$	0.02	&	14.72	$\pm$	0.02	&	14.25	$\pm$	0.02	&	0.98	&	0.99	&	$m$S	&--- \\	
148	&	J044728.19$+$450836.8	&	71.867	&	45.144	&	160.135	&	0.000	&	0.49	&	88.14	&	6.20	&	15.63	$\pm$	0.02	&	14.71	$\pm$	0.02	&	14.26	$\pm$	0.03	&	0.95	&	0.95	&	$\ell$S	&--- \\
149	&	J045409.86$+$445915.6	&	73.541	&	44.988	&	161.010	&	0.810	&	0.42	&	$-$78.78	&	6.35	&	15.66	$\pm$	0.02	&	14.73	$\pm$	0.02	&	14.28	$\pm$	0.03	&	0.91	&	1.00	&	E/S0	&--- \\	
150	&	J044944.04$+$452717.6	&	72.434	&	45.455	&	160.153	&	0.505	&	0.28	&	$-$69.33	&	3.95	&	15.88	$\pm$	0.02	&	14.92	$\pm$	0.02	&	14.32	$\pm$	0.02	&	1.04	&	1.11	&	E/S0	&--- \\
151	&	J044456.88$+$453304.0	&	71.237	&	45.551	&	159.537	&	$-$0.072	&	0.04	&	$-$58.90	&	4.28	&	16.03	$\pm$	0.03	&	14.95	$\pm$	0.02	&	14.33	$\pm$	0.03	&	1.35	&	1.12	&	E/S0	&--- \\	
152	&	J045213.66$+$455728.1	&	73.057	&	45.958	&	160.043	&	1.161	&	0.38	&	$-$41.73	&	4.79	&	15.89	$\pm$	0.02	&	14.96	$\pm$	0.02	&	14.35	$\pm$	0.03	&	1.04	&	1.09	&	E/S0	&--- \\
153	&	J045253.13$+$443406.9	&	73.221	&	44.569	&	161.191	&	0.369	&	0.36	&	78.70	&	5.33	&	15.76	$\pm$	0.02	&	14.84	$\pm$	0.02	&	14.36	$\pm$	0.03	&	0.78	&	1.05	&	$\ell$S	&--- \\
154	&	J044744.15$+$441950.0	&	71.934	&	44.331	&	160.786	&	$-$0.489	&	0.60	&	$-$18.30	&	6.04	&	15.63	$\pm$	0.02	&	14.83	$\pm$	0.02	&	14.42	$\pm$	0.03	&	0.70	&	0.90	&	$\ell$S	&	4878\\
155	&	J044434.53$+$452801.4	&	71.144	&	45.467	&	159.558	&	$-$0.176	&	0.19	&	$-$44.80	&	3.98	&	16.13	$\pm$	0.02	&	15.09	$\pm$	0.02	&	14.44	$\pm$	0.03	&	1.30	&	1.13	&	$m$S	&--- \\
156	&	J045238.00$+$453139.1	&	73.158	&	45.528	&	160.421	&	0.943	&	0.30	&	16.90	&	3.91	&	15.94	$\pm$	0.02	&	14.93	$\pm$	0.02	&	14.44	$\pm$	0.02	&	1.17	&	0.99	&	$e$S	&--- \\
157	&	J044926.79$+$445532.2	&	72.362	&	44.926	&	160.526	&	0.127	&	0.67	&	$-$51.10	&	7.25	&	15.81	$\pm$	0.02	&	14.92	$\pm$	0.02	&	14.45	$\pm$	0.03	&	0.92	&	0.96	&	$m$S	&--- \\	
158	&	J044651.22$+$442147.3	&	71.713	&	44.363	&	160.659	&	$-$0.588	&	0.06	&	$-$23.49	&	4.07	&	15.86	$\pm$	0.03	&	15.03	$\pm$	0.02	&	14.48	$\pm$	0.03	&	0.73	&	1.07	&	E/S0	&--- \\
159	&	J045216.28$+$451657.3	&	73.068	&	45.283	&	160.570	&	0.738	&	0.20	&	34.80	&	4.58	&	15.96	$\pm$	0.03	&	15.02	$\pm$	0.02	&	14.52	$\pm$	0.03	&	0.92	&	1.04	&	$m$S	&--- \\
160	&	J044858.75$+$441329.9	&	72.245	&	44.225	&	161.010	&	$-$0.387	&	0.47	&	$-$82.16	&	4.88	&	15.89	$\pm$	0.02	&	14.98	$\pm$	0.02	&	14.57	$\pm$	0.03	&	0.67	&	1.03	&	$\ell$S	&--- \\
161	&	J044947.40$+$444454.6	&	72.447	&	44.749	&	160.701	&	0.060	&	0.67	&	16.72	&	6.49	&	15.96	$\pm$	0.03	&	14.99	$\pm$	0.02	&	14.59	$\pm$	0.03	&	0.86	&	1.00	&	$m$S	&--- \\
162	&	J045224.39$+$451948.4	&	73.102	&	45.330	&	160.548	&	0.787	&	0.39	&	$-$64.60	&	4.15	&	16.01	$\pm$	0.02	&	15.11	$\pm$	0.02	&	14.60	$\pm$	0.03	&	0.94	&	1.01	&	E/S0	&--- \\	
163	&	J045012.16$+$445802.4	&	72.551	&	44.967	&	160.580	&	0.256	&	0.57	&	18.53	&	6.47	&	16.12	$\pm$	0.03	&	15.08	$\pm$	0.02	&	14.61	$\pm$	0.03	&	0.97	&	1.09	&	$e$S	&--- \\
164	&	J044621.16$+$441634.4	&	71.588	&	44.276	&	160.667	&	$-$0.713	&	0.49	&	$-$5.56	&	4.98	&	15.90	$\pm$	0.02	&	15.00	$\pm$	0.02	&	14.62	$\pm$	0.03	&	0.81	&	0.93	&	$m$S	&--- \\	
165	&	J045532.82$+$450602.8	&	73.887	&	45.101	&	161.075	&	1.071	&	0.24	&	$-$51.61	&	3.82	&	16.05	$\pm$	0.02	&	15.09	$\pm$	0.02	&	14.63	$\pm$	0.03	&	0.91	&	1.03	&	$m$S	&--- \\
166	&	J044858.41$+$451309.8	&	72.243	&	45.219	&	160.248	&	0.252	&	0.10	&	$-$46.62	&	4.23	&	16.21	$\pm$	0.03	&	15.16	$\pm$	0.02	&	14.69	$\pm$	0.03	&	0.96	&	1.11	&	$m$S	&--- \\
167	&	J045011.73$+$445955.1	&	72.549	&	44.999	&	160.555	&	0.275	&	0.41	&	51.90	&	4.58	&	16.12	$\pm$	0.02	&	15.12	$\pm$	0.02	&	14.70	$\pm$	0.03	&	0.98	&	1.00	&	$m$S	&--- \\
168	&	J044651.60$+$455551.6	&	71.715	&	45.931	&	159.465	&	0.428	&	0.30	&	$-$7.58	&	4.66	&	16.34	$\pm$	0.03	&	15.33	$\pm$	0.03	&	14.71	$\pm$	0.03	&	1.22	&	1.11	&	$e$S	&--- \\
169	&	J044739.96$+$443348.7	&	71.917	&	44.564	&	160.600	&	$-$0.348	&	0.38	&	46.20	&	5.61	&	16.07	$\pm$	0.03	&	15.09	$\pm$	0.02	&	14.72	$\pm$	0.03	&	0.84	&	0.98	&	Irr	&--- \\
170	&	J044406.51$+$445621.7	&	71.027	&	44.939	&	159.904	&	$-$0.584	&	0.33	&	10.47	&	4.93	&	16.27	$\pm$	0.03	&	15.19	$\pm$	0.02	&	14.75	$\pm$	0.03	&	1.15	&	1.02	&	$m$S	&--- \\
171	&	J044541.16$+$454528.0	&	71.421	&	45.758	&	159.464	&	0.160	&	0.19	&	$-$38.51	&	3.12	&	16.40	$\pm$	0.03	&	15.37	$\pm$	0.02	&	14.75	$\pm$	0.03	&	1.26	&	1.11	&	E/S0	&--- \\
172	&	J044645.56$+$450429.6	&	71.690	&	45.075	&	160.106	&	$-$0.140	&	0.36	&	84.91	&	5.14	&	16.07	$\pm$	0.03	&	15.12	$\pm$	0.02	&	14.76	$\pm$	0.03	&	0.88	&	0.94	&	$e$S	&--- \\
173	&	J045255.84$+$454914.2	&	73.233	&	45.821	&	160.227	&	1.169	&	0.22	&	$-$59.55	&	3.79	&	16.14	$\pm$	0.03	&	15.17	$\pm$	0.02	&	14.76	$\pm$	0.03	&	0.96	&	0.97	&	$e$S	&--- \\
174	&	J045251.64$+$441648.4	&	73.215	&	44.280	&	161.411	&	0.183	&	0.25	&	$-$14.71	&	3.83	&	15.99	$\pm$	0.02	&	15.11	$\pm$	0.02	&	14.77	$\pm$	0.03	&	0.66	&	0.94	&	E/S0	&--- \\
175	&	J044708.49$+$450527.2	&	71.785	&	45.091	&	160.137	&	$-$0.078	&	0.09	&	75.04	&	3.10	&	16.07	$\pm$	0.02	&	15.13	$\pm$	0.02	&	14.79	$\pm$	0.03	&	0.88	&	0.90	&	E/S0	&--- \\
176	&	J045433.21$+$445344.2	&	73.638	&	44.896	&	161.124	&	0.806	&	0.22	&	$-$27.96	&	3.42	&	16.30	$\pm$	0.03	&	15.34	$\pm$	0.02	&	14.80	$\pm$	0.03	&	0.89	&	1.12	&	E/S0	&--- \\
177	&	J045441.39$+$451904.9	&	73.672	&	45.318	&	160.811	&	1.090	&	0.35	&	$-$37.70	&	3.06	&	16.31	$\pm$	0.02	&	15.59	$\pm$	0.02	&	14.81	$\pm$	0.03	&	0.89	&	1.11	&	$e$S	&--- \\
178	&	J044838.24$+$445640.2	&	72.159	&	44.945	&	160.420	&	0.029	&	0.09	&	$-$68.89	&	3.45	&	16.11	$\pm$	0.02	&	15.20	$\pm$	0.02	&	14.82	$\pm$	0.03	&	0.86	&	0.92	&	E/S0	&--- \\
179	&	J044614.90$+$452745.3	&	71.562	&	45.463	&	159.753	&	0.043	&	0.29	&	13.74	&	3.22	&	16.50	$\pm$	0.03	&	15.47	$\pm$	0.02	&	14.85	$\pm$	0.03	&	1.23	&	1.11	&	$m$S	&--- \\
180	&	J045013.36$+$440839.0	&	72.556	&	44.144	&	161.210	&	$-$0.270	&	0.44	&	79.01	&	4.57	&	15.91	$\pm$	0.02	&	15.21	$\pm$	0.02	&	14.87	$\pm$	0.03	&	0.63	&	0.78	&	Irr	&	5547\\
181	&	J044938.32$+$435622.7	&	72.410	&	43.940	&	161.305	&	$-$0.479	&	0.07	&	$-$73.58	&	3.12	&	16.23	$\pm$	0.03	&	15.36	$\pm$	0.03	&	14.89	$\pm$	0.03	&	0.62	&	1.07	&	E/S0	&--- \\	
182	&	J044953.62$+$450543.4	&	72.473	&	45.095	&	160.447	&	0.296	&	0.12	&	82.40	&	3.15	&	16.38	$\pm$	0.02	&	15.38	$\pm$	0.02	&	14.90	$\pm$	0.03	&	0.99	&	1.05	&	$m$S	&--- \\
183	&	J045123.45$+$460416.4	&	72.848	&	46.071	&	159.863	&	1.121	&	0.47	&	$-$66.91	&	4.71	&	16.38	$\pm$	0.03	&	15.37	$\pm$	0.02	&	14.90	$\pm$	0.03	&	1.02	&	1.04	&	$\ell$S	&--- \\	
184	&	J045440.78$+$445807.9	&	73.670	&	44.969	&	161.082	&	0.869	&	0.23	&	80.85	&	2.66	&	16.37	$\pm$	0.02	&	15.44	$\pm$	0.02	&	14.90	$\pm$	0.03	&	0.91	&	1.07	&	E/S0	&--- \\
185	&	J044921.23$+$460126.8	&	72.338	&	46.024	&	159.674	&	0.820	&	0.33	&	6.21	&	3.62	&	16.49	$\pm$	0.03	&	15.39	$\pm$	0.03	&	14.92	$\pm$	0.03	&	1.17	&	1.07	&	Irr	&--- \\
186	&	J044738.78$+$450532.7	&	71.912	&	45.092	&	160.194	&	$-$0.009&	0.21	&	$-$64.50	&	3.95	&	16.35	$\pm$	0.03	&	15.35	$\pm$	0.02	&	14.94	$\pm$	0.03	&	0.87	&	1.03	&	$\ell$S	&--- \\
187	&	J044855.57$+$443933.7	&	72.232	&	44.659	&	160.671	&	$-$0.115&	0.72	&	4.90	&	8.79	&	16.39	$\pm$	0.03	&	15.49	$\pm$	0.03	&	14.94	$\pm$	0.04	&	0.80	&	1.10	&	$m$S	&--- \\
188	&	J044543.46$+$450440.7	&	71.431	&	45.078	&	159.985	&	$-$0.277&	0.36	&	$-$80.84	&	4.44	&	16.34	$\pm$	0.03	&	15.36	$\pm$	0.02	&	14.95	$\pm$	0.03	&	1.03	&	0.95	&	$\ell$S	&--- \\
189	&	J044952.70$+$451451.0	&	72.470	&	45.247	&	160.328	&	0.392	&	0.24	&	65.67	&	4.62	&	16.50	$\pm$	0.03	&	15.57	$\pm$	0.03	&	14.98	$\pm$	0.03	&	1.00	&	1.09	&	Irr	&--- \\
190	&	J044957.62$+$440316.6	&	72.490	&	44.055	&	161.254	&	$-$0.361&	0.33	&	11.46	&	4.17	&	16.18	$\pm$	0.03	&	15.36	$\pm$	0.02	&	14.98	$\pm$	0.03	&	0.64	&	0.93	&	$\ell$S	&--- \\
191	&	J044810.67$+$444742.0	&	72.044	&	44.795	&	160.482	&	$-$0.129&	0.38	&	32.80	&	3.89	&	16.40	$\pm$	0.03	&	15.46	$\pm$	0.03	&	14.99	$\pm$	0.04	&	0.82	&	1.05	&	$m$S	&--- \\
192	&	J044856.82$+$444953.0	&	72.237	&	44.831	&	160.542	&	$-$0.002&	0.09	&	75.80	&	3.55	&	16.35	$\pm$	0.03	&	15.37	$\pm$	0.03	&	15.03	$\pm$	0.03	&	0.83	&	0.96	&	$\ell$S	&--- \\
193	&	J044507.40$+$450214.4	&	71.281	&	45.037	&	159.947	&	$-$0.384&	0.34	&	$-$30.50	&	4.18	&	16.51	$\pm$	0.03	&	15.52	$\pm$	0.03	&	15.05	$\pm$	0.03	&	1.12	&	0.98	&	$\ell$S	&--- \\
194	&	J045420.33$+$451848.0	&	73.585	&	45.313	&	160.776	&	1.039	&	0.61	&	53.91	&	5.80	&	16.43	$\pm$	0.03	&	15.50	$\pm$	0.02	&	15.05	$\pm$	0.03	&	0.92	&	0.98	&	$e$S	&--- \\
195	&	J045443.36$+$444708.1	&	73.681	&	44.786	&	161.229	&	0.760	&	0.47	&	$-$58.34	&	4.83	&	16.47	$\pm$	0.03	&	15.55	$\pm$	0.03	&	15.05	$\pm$	0.04	&	0.88	&	1.03	&	$m$S	&--- \\
196	&	J044955.20$+$451139.7	&	72.480	&	45.194	&	160.374	&	0.363	&	0.03	&	$-$6.91	&	3.08	&	16.45	$\pm$	0.03	&	15.47	$\pm$	0.02	&	15.07	$\pm$	0.03	&	0.99	&	0.96	&	$\ell$S	&--- \\
197	&	J044737.90$+$451002.9	&	71.908	&	45.167	&	160.135	&	0.037	&	0.23	&	34.62	&	2.81	&	16.59	$\pm$	0.03	&	15.85	$\pm$	0.03	&	15.09	$\pm$	0.03	&	0.94	&	1.09	&	$\ell$S	&--- \\
198	&	J044852.61$+$445740.8	&	72.219	&	44.961	&	160.434	&	0.073	&	0.23	&	$-$14.44	&	3.24	&	16.47	$\pm$	0.03	&	15.51	$\pm$	0.02	&	15.10	$\pm$	0.03	&	0.89	&	0.98	&	$\ell$S	&--- \\
199	&	J044547.35$+$454302.2	&	71.447	&	45.717	&	159.507	&	0.148	&	0.61	&	43.30	&	5.75	&	16.72	$\pm$	0.03	&	15.77	$\pm$	0.03	&	15.12	$\pm$	0.04	&	1.27	&	1.05	&	$e$S	&--- \\
200	&	J044622.17$+$453307.3	&	71.592	&	45.552	&	159.698	&	0.117	&	0.46	&	$-$73.56	&	4.31	&	16.77	$\pm$	0.03	&	15.76	$\pm$	0.03	&	15.16	$\pm$	0.04	&	1.28	&	1.06	&	$\ell$S	&--- \\
201	&	J044545.92$+$444901.9	&	71.441	&	44.817	&	160.188	&	$-$0.441&	0.11	&	67.60	&	4.00	&	16.62	$\pm$	0.03	&	15.75	$\pm$	0.03	&	15.21	$\pm$	0.04	&	1.05	&	0.95	&	$\ell$S	&	5134\\
202	&	J045012.70$+$453746.5	&	72.553	&	45.630	&	160.072	&	0.681	&	0.21	&	$-$58.70	&	3.14	&	16.78	$\pm$	0.03	&	15.76	$\pm$	0.03	&	15.21	$\pm$	0.03	&	1.14	&	1.08	&	$\ell$S	&--- \\
203	&	J045004.97$+$450154.3	&	72.521	&	45.032	&	160.517	&	0.281	&	0.34	&	$-$22.89	&	3.77	&	16.67	$\pm$	0.03	&	15.69	$\pm$	0.02	&	15.24	$\pm$	0.03	&	0.97	&	1.01	&	$\ell$S	&--- \\
204	&	J044617.48$+$452600.3	&	71.573	&	45.433	&	159.780	&	0.030	&	0.23	&	$-$8.24	&	4.19	&	16.87	$\pm$	0.04	&	15.83	$\pm$	0.03	&	15.28	$\pm$	0.04	&	1.20	&	1.07	&	Irr	&--- \\
205	&	J045311.77$+$453648.0	&	73.299	&	45.613	&	160.417	&	1.074	&	0.18	&	22.04	&	3.19	&	16.73	$\pm$	0.03	&	15.73	$\pm$	0.02	&	15.31	$\pm$	0.03	&	1.10	&	0.94	&	Irr	&--- \\
206	&	J045329.73$+$443729.2	&	73.374	&	44.625	&	161.216	&	0.489	&	0.61	&	73.06	&	4.40	&	16.66	$\pm$	0.03	&	15.71	$\pm$	0.02	&	15.32	$\pm$	0.04	&	0.85	&	0.97	&	$m$S	&--- \\
207	&	J044445.84$+$444702.1	&	71.191	&	44.784	&	160.097	&	$-$0.597&	0.20	&	10.23	&	3.97	&	16.90	$\pm$	0.04	&	15.94	$\pm$	0.03	&	15.33	$\pm$	0.05	&	1.11	&	1.09	&	Irr	&--- \\
208	&	J045127.65$+$450922.6	&	72.865	&	45.156	&	160.577	&	0.548	&	0.56	&	56.41	&	4.66	&	16.82	$\pm$	0.03	&	15.88	$\pm$	0.03	&	15.35	$\pm$	0.03	&	0.97	&	1.06	&	$m$S	&--- \\
209	&	J044955.57$+$454444.8	&	72.482	&	45.746	&	159.951	&	0.717	&	0.47	&	17.14	&	3.51	&	17.00	$\pm$	0.03	&	16.01	$\pm$	0.03	&	15.42	$\pm$	0.03	&	1.16	&	1.09	&	$\ell$S	&--- \\
210	&	J044437.62$+$443619.0	&	71.157	&	44.605	&	160.217	&	$-$0.732&	0.28	&	25.70	&	3.64	&	16.93	$\pm$	0.03	&	15.90	$\pm$	0.03	&	15.44	$\pm$	0.04	&	1.04	&	1.04	&	Irr	&--- \\
211	&	J045330.45$+$444657.0	&	73.377	&	44.783	&	161.095	&	0.590	&	0.53	&	0.35	&	4.16	&	16.82	$\pm$	0.03	&	15.84	$\pm$	0.03	&	15.44	$\pm$	0.04	&	0.97	&	0.96	&	$e$S	&--- \\
212	&	J044756.88$+$442015.1	&	71.987	&	44.338	&	160.805	&	$-$0.456&	0.20	&	66.40	&	3.21	&	16.75	$\pm$	0.03	&	15.89	$\pm$	0.03	&	15.45	$\pm$	0.04	&	0.69	&	1.00	&	Irr	&--- \\
213	&	J044958.64$+$450152.9	&	72.494	&	45.031	&	160.505	&	0.267	&	0.31	&	$-$62.00	&	2.79	&	16.90	$\pm$	0.03	&	15.87	$\pm$	0.02	&	15.45	$\pm$	0.03	&	0.96	&	1.03	&	E/S0	&--- \\
214	&	J045225.84$+$451456.9	&	73.108	&	45.249	&	160.614	&	0.739	&	0.00	&	$-$7.64	&	3.17	&	16.85	$\pm$	0.03	&	15.87	$\pm$	0.03	&	15.48	$\pm$	0.04	&	0.90	&	0.99	&	Irr	&--- \\
215	&	J044531.48$+$443550.4	&	71.381	&	44.597	&	160.327	&	$-$0.617&	0.12	&	$-$17.07	&	2.91	&	16.96	$\pm$	0.03	&	15.98	$\pm$	0.03	&	15.49	$\pm$	0.04	&	1.04	&	1.02	&	$\ell$S	&--- \\
216	&	J044638.07$+$444609.2	&	71.659	&	44.769	&	160.324	&	$-$0.355&	0.53	&	$-$89.04	&	4.01	&	17.01	$\pm$	0.03	&	16.05	$\pm$	0.03	&	15.53	$\pm$	0.04	&	0.91	&	1.09	&	$e$S	&--- \\
217	&	J044744.40$+$445626.1	&	71.935	&	44.941	&	160.321	&	$-$0.095&	0.18	&	$-$5.80	&	2.80	&	16.89	$\pm$	0.03	&	15.99	$\pm$	0.03	&	15.53	$\pm$	0.04	&	0.85	&	1.00	&	Irr	&--- \\
218	&	J044922.06$+$453019.4	&	72.342	&	45.505	&	160.073	&	0.488	&	0.01	&	42.10	&	2.83	&	16.98	$\pm$	0.03	&	16.04	$\pm$	0.03	&	15.54	$\pm$	0.04	&	0.99	&	1.01	&	Irr	&--- \\
219	&	J045057.35$+$452532.4	&	72.739	&	45.426	&	160.312	&	0.651	&	0.14	&	9.90	&	3.04	&	17.04	$\pm$	0.03	&	15.98	$\pm$	0.03	&	15.54	$\pm$	0.04	&	1.08	&	1.03	&	$\ell$S	&--- \\
220	&	J044957.27$+$460315.1	&	72.489	&	46.054	&	159.720	&	0.920	&	0.44	&	19.55	&	4.28	&	16.93	$\pm$	0.03	&	16.07	$\pm$	0.03	&	15.56	$\pm$	0.04	&	1.12	&	0.89	&	$\ell$S	&	4942\\
221	&	J045241.41$+$452024.6	&	73.173	&	45.340	&	160.572	&	0.832	&	0.05	&	$-$62.10	&	2.92	&	16.91	$\pm$	0.03	&	16.01	$\pm$	0.03	&	15.56	$\pm$	0.04	&	0.93	&	0.95	&	$\ell$S	&--- \\
222	&	J044854.97$+$453243.7	&	72.229	&	45.545	&	159.991	&	0.454	&	0.06	&	$-$44.99	&	3.28	&	16.88	$\pm$	0.03	&	16.01	$\pm$	0.03	&	15.57	$\pm$	0.04	&	0.97	&	0.89	&	Irr	&--- \\
223	&	J045243.81$+$443819.6	&	73.183	&	44.639	&	161.119	&	0.393	&	0.24	&	$-$26.00	&	3.20	&	16.89	$\pm$	0.03	&	16.05	$\pm$	0.03	&	15.57	$\pm$	0.04	&	0.81	&	0.97	&	$\ell$S	&--- \\
224	&	J045338.36$+$452459.0	&	73.410	&	45.416	&	160.619	&	1.009	&	0.12	&	$-$50.48	&	2.87	&	17.07	$\pm$	0.03	&	16.22	$\pm$	0.03	&	15.57	$\pm$	0.03	&	0.92	&	1.10	&	Irr	&--- \\
225	&	J044959.00$+$442427.5	&	72.496	&	44.408	&	160.985	&	$-$0.132&	0.34	&	39.37	&	2.94	&	16.85	$\pm$	0.03	&	16.17	$\pm$	0.03	&	15.62	$\pm$	0.05	&	0.69	&	0.93	&	Irr	&--- \\
226	&	J045143.42$+$445733.8	&	72.931	&	44.959	&	160.758	&	0.458	&	0.23	&	74.31	&	3.34	&	16.98	$\pm$	0.03	&	16.10	$\pm$	0.03	&	15.62	$\pm$	0.04	&	0.94	&	0.95	&	Irr	&--- \\
227	&	J044812.83$+$445407.6	&	72.053	&	44.902	&	160.404	&	$-$0.055&	0.14	&	32.10	&	3.16	&	17.03	$\pm$	0.04	&	16.17	$\pm$	0.03	&	15.64	$\pm$	0.04	&	0.84	&	1.03	&	$\ell$S	&--- \\
228	&	J045038.65$+$445618.6	&	72.661	&	44.938	&	160.652	&	0.298	&	0.15	&	$-$86.90	&	2.79	&	17.02	$\pm$	0.03	&	16.02	$\pm$	0.03	&	15.65	$\pm$	0.04	&	0.92	&	0.97	&	$\ell$S	&--- \\
229	&	J045453.21$+$444522.8	&	73.722	&	44.756	&	161.270	&	0.764	&	0.22	&	48.70	&	3.10	&	17.01	$\pm$	0.04	&	16.09	$\pm$	0.03	&	15.70	$\pm$	0.05	&	0.89	&	0.93	&	Irr	&--- \\
230	&	J044822.79$+$445528.2	&	72.095	&	44.925	&	160.406	&	$-$0.018&	0.17	&	46.87	&	3.19	&	17.10	$\pm$	0.04	&	16.22	$\pm$	0.03	&	15.72	$\pm$	0.05	&	0.84	&	1.02	&	Irr	&--- \\	
231	&	J045144.68$+$455508.4	&	72.936	&	45.919	&	160.020	&	1.072	&	0.63	&	0.58	&	4.87	&	17.34	$\pm$	0.04	&	16.23	$\pm$	0.03	&	15.74	$\pm$	0.05	&	1.16	&	1.09	&	$\ell$S	&--- \\
232	&	J044706.85$+$453448.3	&	71.779	&	45.580	&	159.761	&	0.235	&	0.39	&	36.30	&	3.36	&	17.28	$\pm$	0.04	&	16.31	$\pm$	0.03	&	15.75	$\pm$	0.04	&	1.25	&	0.99	&	$\ell$S	&	4994\\
233	&	J044926.38$+$445450.8	&	72.360	&	44.914	&	160.534	&	0.119	&	0.57	&	59.34	&	4.18	&	17.10	$\pm$	0.03	&	16.21	$\pm$	0.03	&	15.75	$\pm$	0.04	&	0.92	&	0.96	&	$e$S	&--- \\
234	&	J044930.28$+$443329.0	&	72.376	&	44.558	&	160.815	&	$-$0.101&	0.51	&	$-$64.84	&	3.81	&	16.99	$\pm$	0.03	&	16.12	$\pm$	0.03	&	15.79	$\pm$	0.05	&	0.76	&	0.87	&	$\ell$S	&--- \\	
235	&	J045109.62$+$450323.9	&	72.790	&	45.057	&	160.620	&	0.444	&	0.47	&	$-$35.74	&	3.64	&	17.15	$\pm$	0.03	&	16.20	$\pm$	0.03	&	15.79	$\pm$	0.04	&	0.93	&	0.96	&	$\ell$S	&--- \\
236	&	J045517.43$+$450155.5	&	73.823	&	45.032	&	161.100	&	0.993	&	0.28	&	47.83	&	3.35	&	17.13	$\pm$	0.03	&	16.17	$\pm$	0.03	&	15.79	$\pm$	0.04	&	0.90	&	0.96	&	Irr	&--- \\
237	&	J045401.45$+$451107.1	&	73.506	&	45.185	&	160.841	&	0.916	&	0.03	&	$-$22.56	&	2.69	&	17.14	$\pm$	0.03	&	16.20	$\pm$	0.03	&	15.80	$\pm$	0.04	&	0.96	&	0.93	&	Irr	&--- \\	
238	&	J044452.40$+$443535.1	&	71.218	&	44.593	&	160.255	&	$-$0.707&	0.13	&	$-$44.30	&	2.75	&	17.23	$\pm$	0.04	&	16.27	$\pm$	0.03	&	15.83	$\pm$	0.05	&	1.01	&	0.97	&	Irr	&--- \\
239	&	J044936.24$+$450910.5	&	72.401	&	45.153	&	160.370	&	0.294	&	0.16	&	$-$26.71	&	2.74	&	17.24	$\pm$	0.04	&	16.27	$\pm$	0.03	&	15.84	$\pm$	0.04	&	0.97	&	0.97	&	Irr	&--- \\
240	&	J044727.80$+$444723.1	&	71.866	&	44.790	&	160.400	&	$-$0.230&	0.68	&	$-$44.49	&	6.25	&	17.08	$\pm$	0.04	&	16.32	$\pm$	0.04	&	15.87	$\pm$	0.07	&	0.88	&	0.83	&	Irr	&	7288\\
241	&	J044428.93$+$443630.5	&	71.121	&	44.608	&	160.197	&	$-$0.750&	0.07	&	41.10	&	2.86	&	17.40	$\pm$	0.04	&	16.46	$\pm$	0.04	&	15.97	$\pm$	0.06	&	1.06	&	0.96	&	Irr	&--- \\
242	&	J044746.91$+$453420.6	&	71.945	&	45.572	&	159.843	&	0.319	&	0.30	&	$-$41.46	&	3.09	&	17.48	$\pm$	0.04	&	16.45	$\pm$	0.03	&	15.99	$\pm$	0.05	&	1.23	&	0.96	&	Irr	&--- \\
243	&	J045143.79$+$444623.9	&	72.932	&	44.773	&	160.902	&	0.341	&	0.45	&	43.24	&	3.42	&	17.32	$\pm$	0.04	&	16.37	$\pm$	0.03	&	16.02	$\pm$	0.06	&	0.88	&	0.91	&	Irr	&--- \\
244	&	J044846.14$+$455955.0	&	72.192	&	45.999	&	159.628	&	0.725	&	0.34	&	29.77	&	3.04	&	17.61	$\pm$	0.05	&	16.60	$\pm$	0.04	&	16.10	$\pm$	0.05	&	1.18	&	1.01	&	$\ell$S	&--- \\	
245	&	J044942.24$+$452830.7	&	72.426	&	45.475	&	160.134	&	0.514	&	0.07	&	$-$32.30	&	2.56	&	17.45	$\pm$	0.04	&	16.49	$\pm$	0.03	&	16.10	$\pm$	0.05	&	1.04	&	0.90	&	Irr	&--- \\
246	&	J045533.75$+$445954.9	&	73.891	&	44.999	&	161.157	&	1.009	&	0.20	&	$-$60.40	&	2.72	&	17.44	$\pm$	0.04	&	16.50	$\pm$	0.03	&	16.10	$\pm$	0.05	&	0.91	&	0.94	&	Irr	&--- \\
247	&	J045024.44$+$453101.5	&	72.602	&	45.517	&	160.181	&	0.636	&	0.11	&	$-$61.08	&	2.70	&	17.78	$\pm$	0.05	&	16.77	$\pm$	0.04	&	16.28	$\pm$	0.07	&	1.16	&	1.00	&	Irr	&--- \\
248	&	J044758.33$+$440708.0	&	71.993	&	44.119	&	160.975	&	$-$0.593&	0.53	&	$-$81.40	&	3.07	&	17.66	$\pm$	0.04	&	16.84	$\pm$	0.04	&	16.38	$\pm$	0.06	&	0.64	&	1.00	&	$\ell$S	&--- \\
249	&	J044644.16$+$442005.1	&	71.684	&	44.335	&	160.667	&	$-$0.623&	0.56	&	$-$74.98	&	3.12	&	17.66	$\pm$	0.04	&	16.81	$\pm$	0.04	&	16.47	$\pm$	0.07	&	0.73	&	0.88	&	$m$S	&	5654\\
250	&	J044602.99$+$445836.4	&	71.512	&	44.977	&	160.099	&	$-$0.299&	0.33	&	$-$63.40	&	3.07	&	17.87	$\pm$	0.05	&	17.24	$\pm$	0.05	&	16.47	$\pm$	0.07	&	0.96	&	0.99	&	Irr	&--- \\
251	&	J045222.42$+$450342.7	&	73.093	&	45.062	&	160.752	&	0.612	&	0.33	&	$-$74.00	&	2.64	&	18.03	$\pm$	0.05	&	17.02	$\pm$	0.04	&	16.57	$\pm$	0.07	&	0.91	&	1.07	&	Irr	&--- \\
252	&	J045430.84$+$444644.9	&	73.628	&	44.779	&	161.211	&	0.727	&	0.68	&	5.66	&	2.85	&	18.45	$\pm$	0.06	&	17.50	$\pm$	0.05	&	17.08	$\pm$	0.09	&	0.88	&	0.99	&	$\ell$S	&--- \\
253	&	J045302.36$+$454133.4	&	73.260	&	45.693	&	160.338	&	1.102	&	0.74	&	$-$73.61	&	3.78	&	18.56	$\pm$	0.06	&	17.68	$\pm$	0.05	&	17.09	$\pm$	0.08	&	1.11	&	1.00	&	Irr	&--- \\
254	&	J044817.36$+$445352.2	&	72.072	&	44.898	&	160.416	&	$-$0.048&	0.95	&	$-$33.50	&	11.71	&	18.42	$\pm$	0.09	&	17.67	$\pm$	0.09	&	17.17	$\pm$	0.15	&	0.84	&	0.89	&	Irr	&--- \\	
255	&	J045034.26$+$435930.6	&	72.643	&	43.992	&	161.370	&	$-$0.320&	0.60	&	$-$48.07	&	2.10	&	19.24	$\pm$	0.10	&	18.93	$\pm$	0.13	&	17.81	$\pm$	0.13	&	0.64	&	1.16	&	Irr	&	5724\\
256	&	J044602.10$+$443424.5	&	71.509	&	44.573	&	160.400	&	$-$0.560&	0.05	&	68.79	&	1.51	&	19.00	$\pm$	0.11	&	18.07	$\pm$	0.09	&	17.85	$\pm$	0.19	&	0.94	&	0.74	&	Irr	&	6281\\
257	&	J045426.93$+$444941.5	&	73.612	&	44.828	&	161.170	&	0.750	&	0.24	&	69.67	&	1.38	&	19.29	$\pm$	0.09	&	18.44	$\pm$	0.08	&	18.05	$\pm$	0.15	&	0.89	&	0.85	&	Irr	&	5072\\
258	&	J045217.47$+$450839.9	&	73.073	&	45.144	&	160.680	&	0.650	&	0.56	&	43.93	&	1.78	&	19.72	$\pm$	0.11	&	18.84	$\pm$	0.09	&	18.27	$\pm$	0.15	&	0.87	&	1.07	&	Irr	&	6680\\
259	&	J044644.76$+$444734.7	&	71.687	&	44.793	&	160.320	&	$-$0.320&	0.19	&	75.89	&	0.94	&	19.83	$\pm$	0.12	&	18.84	$\pm$	0.09	&	18.80	$\pm$	0.24	&	0.89	&	0.65	&	Irr	&	5709\\
260	&	J044838.16$+$454506.4	&	72.159	&	45.752	&	159.800	&	0.550	&	0.35	&	$-$77.10	&	1.07	&	21.83	$\pm$	0.74	&	19.89	$\pm$	0.27	&	18.96	$\pm$	0.24	&	1.14	&	2.37	&	Irr	&	5745\\
261	&	J045026.76$+$451130.7	&	72.612	&	45.192	&	160.440	&	0.430	&	0.63	&	$-$53.75	&	0.73	&	21.77	$\pm$	0.35	&	20.25	$\pm$	0.17	&	20.07	$\pm$	0.39	&	1.02	&	1.26	&	Irr	&	4639\\
\hline
\end{longtable}
\end{landscape}
\end{centering}

%\documentclass{article}
%\usepackage{graphicx}
%\usepackage{epsfig}
%\usepackage{color}
%\usepackage{amssymb}
%\usepackage{txfonts}
%\usepackage{lscape}
%\usepackage{float}
%\usepackage{amsmath}
%\usepackage{natbib}
%\usepackage{longtable}
%\usepackage{journal_names}
%\usepackage{rotating}
%\usepackage{enumitem}
%\usepackage{subcaption}
%\usepackage{xcolor}
%\usepackage[export]{adjustbox}
%\usepackage{caption}
%\DeclareCaptionLabelFormat{continued}{Figure \ref{postage}}
%\captionsetup[ContinuedFloat]{labelformat=continued}

%\begin{document}

\newpage

\begin{figure*}
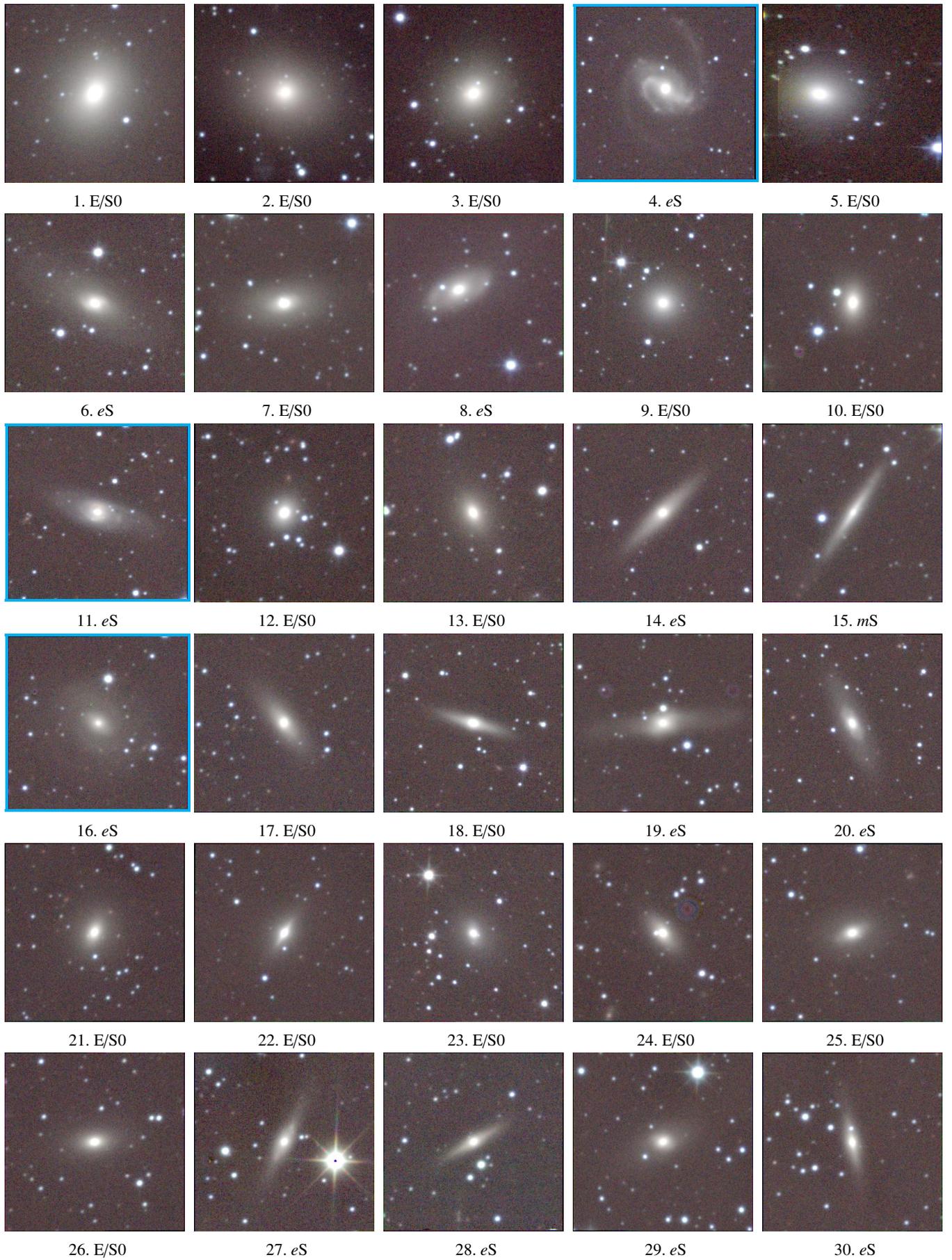

\centering

\begin{subfigure}[b]{0.19\textwidth}\includegraphics[width=3.4cm, height=3.4cm]{s36.jpeg}\caption*{1. E/S0}\end{subfigure}
\begin{subfigure}[b]{0.19\textwidth}\includegraphics[width=3.4cm, height=3.4cm]{s76.jpeg}\caption*{2. E/S0}\end{subfigure}
\begin{subfigure}[b]{0.19\textwidth}\includegraphics[width=3.4cm, height=3.4cm]{s51.jpeg}\caption*{3. E/S0}\end{subfigure}
\begin{subfigure}[b]{0.19\textwidth}\includegraphics[width=3.4cm,height=3.3cm,cfbox=cyan 1.5pt 0.0pt]{s3.jpeg}\caption*{4. $e$S}\end{subfigure}
\begin{subfigure}[b]{0.19\textwidth}\includegraphics[width=3.4cm, height=3.4cm]{s18.jpeg}\caption*{5. E/S0}\end{subfigure}

\begin{subfigure}[b]{0.19\textwidth}\includegraphics[width=3.4cm, height=3.4cm]{s29.jpeg}\caption*{6. $e$S}\end{subfigure}
\begin{subfigure}[b]{0.19\textwidth}\includegraphics[width=3.4cm, height=3.4cm]{s10.jpeg}\caption*{7. E/S0}\end{subfigure}
\begin{subfigure}[b]{0.19\textwidth}\includegraphics[width=3.4cm, height=3.4cm]{s24.jpeg}\caption*{8. $e$S}\end{subfigure}
\begin{subfigure}[b]{0.19\textwidth}\includegraphics[width=3.4cm, height=3.4cm]{s21.jpeg}\caption*{9. E/S0}\end{subfigure}
\begin{subfigure}[b]{0.19\textwidth}\includegraphics[width=3.4cm, height=3.4cm]{s9.jpeg}\caption*{10. E/S0}\end{subfigure}

\begin{subfigure}[b]{0.19\textwidth}\includegraphics[width=3.4cm,height=3.3cm,cfbox=cyan 1.5pt 0.0pt]{s26.jpeg}\caption*{11. $e$S}\end{subfigure}
\begin{subfigure}[b]{0.19\textwidth}\includegraphics[width=3.4cm, height=3.4cm]{s5.jpeg}\caption*{12. E/S0}\end{subfigure}
\begin{subfigure}[b]{0.19\textwidth}\includegraphics[width=3.4cm, height=3.4cm]{s15.jpeg}\caption*{13. E/S0}\end{subfigure}
\begin{subfigure}[b]{0.19\textwidth}\includegraphics[width=3.4cm, height=3.4cm]{s13.jpeg}\caption*{14. $e$S}\end{subfigure}
\begin{subfigure}[b]{0.19\textwidth}\includegraphics[width=3.4cm, height=3.4cm]{s108.jpeg}\caption*{15. $m$S}\end{subfigure}

\begin{subfigure}[b]{0.19\textwidth}\includegraphics[width=3.4cm,height=3.3cm,cfbox=cyan 1.5pt 0.0pt]{s123.jpeg}\caption*{16. $e$S}\end{subfigure}
\begin{subfigure}[b]{0.19\textwidth}\includegraphics[width=3.4cm, height=3.4cm]{s39.jpeg}\caption*{17. E/S0}\end{subfigure}
\begin{subfigure}[b]{0.19\textwidth}\includegraphics[width=3.4cm, height=3.4cm]{s6.jpeg}\caption*{18. E/S0}\end{subfigure}
\begin{subfigure}[b]{0.19\textwidth}\includegraphics[width=3.4cm, height=3.4cm]{s72.jpeg}\caption*{19. $e$S}\end{subfigure}
\begin{subfigure}[b]{0.19\textwidth}\includegraphics[width=3.4cm, height=3.4cm]{s35.jpeg}\caption*{20. $e$S}\end{subfigure}

\begin{subfigure}[b]{0.19\textwidth}\includegraphics[width=3.4cm, height=3.4cm]{s20.jpeg}\caption*{21. E/S0}\end{subfigure}
\begin{subfigure}[b]{0.19\textwidth}\includegraphics[width=3.4cm, height=3.4cm]{s7.jpeg}\caption*{22. E/S0}\end{subfigure}
\begin{subfigure}[b]{0.19\textwidth}\includegraphics[width=3.4cm, height=3.4cm]{s45.jpeg}\caption*{23. E/S0}\end{subfigure}
\begin{subfigure}[b]{0.19\textwidth}\includegraphics[width=3.4cm, height=3.4cm]{s11.jpeg}\caption*{24. E/S0}\end{subfigure}
\begin{subfigure}[b]{0.19\textwidth}\includegraphics[width=3.4cm, height=3.4cm]{s49.jpeg}\caption*{25. E/S0}\end{subfigure}

\begin{subfigure}[b]{0.19\textwidth}\includegraphics[width=3.4cm, height=3.4cm]{s19.jpeg}\caption*{26. E/S0}\end{subfigure}
\begin{subfigure}[b]{0.19\textwidth}\includegraphics[width=3.4cm, height=3.4cm]{s38.jpeg}\caption*{27. $e$S}\end{subfigure}
\begin{subfigure}[b]{0.19\textwidth}\includegraphics[width=3.4cm, height=3.4cm]{s47.jpeg}\caption*{28. $e$S}\end{subfigure}
\begin{subfigure}[b]{0.19\textwidth}\includegraphics[width=3.4cm, height=3.4cm]{s60.jpeg}\caption*{29. $e$S}\end{subfigure}
\begin{subfigure}[b]{0.19\textwidth}\includegraphics[width=3.4cm, height=3.4cm]{s40.jpeg}\caption*{30. $e$S}\end{subfigure}
\caption*{\textbf{Figure B.} The false-colour $-$ $J$ (blue), $H$ (green) and $K$ (red) representation (1.2\arcmin\ $\times$ 1.2\arcmin) of galaxy candidates of the 3C\,129 cluster. The cyan frames indicate H\textsc{i} detection in the WSRT H\textsc{i}-survey.}
\end{figure*}

\begin{figure*}
\ContinuedFloat
\centering
\begin{subfigure}[b]{0.19\textwidth}\includegraphics[width=3.4cm, height=3.4cm]{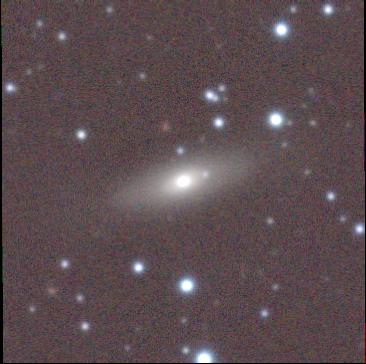}\caption*{31. E/S0}\end{subfigure}
\begin{subfigure}[b]{0.19\textwidth}\includegraphics[width=3.4cm, height=3.4cm]{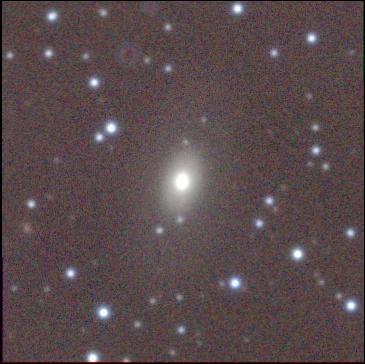}\caption*{32. E/S0}\end{subfigure}
\begin{subfigure}[b]{0.19\textwidth}\includegraphics[width=3.4cm, height=3.4cm]{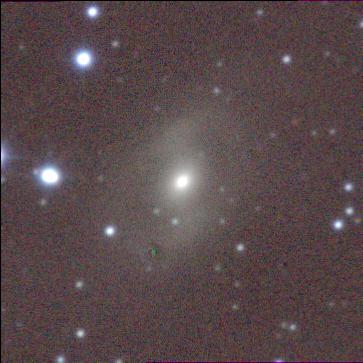}\caption*{33. $m$S}\end{subfigure}
\begin{subfigure}[b]{0.19\textwidth}\includegraphics[width=3.4cm, height=3.4cm]{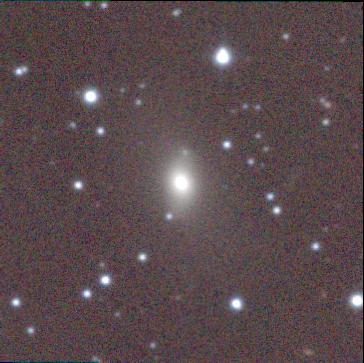}\caption*{34. E/S0}\end{subfigure}
\begin{subfigure}[b]{0.19\textwidth}\includegraphics[width=3.4cm, height=3.4cm]{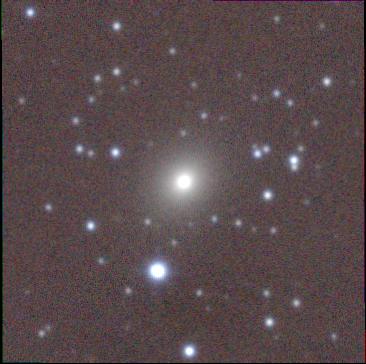}\caption*{35. E/S0}\end{subfigure}

\begin{subfigure}[b]{0.19\textwidth}\includegraphics[width=3.4cm, height=3.4cm]{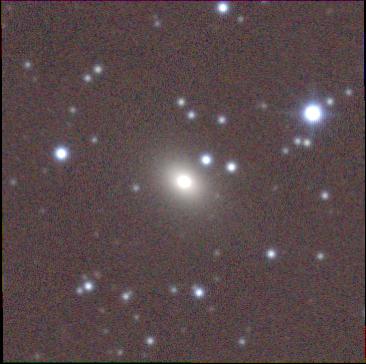}\caption*{36. E/S0}\end{subfigure}
\begin{subfigure}[b]{0.19\textwidth}\includegraphics[width=3.4cm,height=3.3cm,cfbox=cyan 1.5pt 0.0pt]{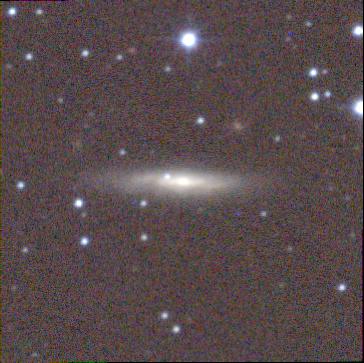}\caption*{37. $e$S}\end{subfigure}
\begin{subfigure}[b]{0.19\textwidth}\includegraphics[width=3.4cm, height=3.4cm]{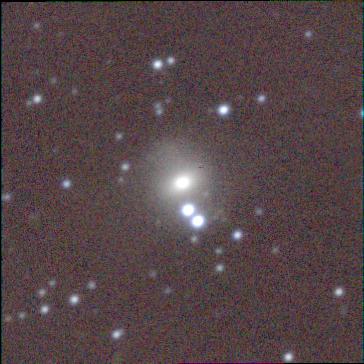}\caption*{38. $e$S}\end{subfigure}
\begin{subfigure}[b]{0.19\textwidth}\includegraphics[width=3.4cm, height=3.4cm]{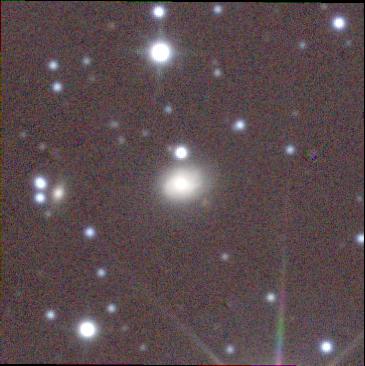}\caption*{39. $e$S}\end{subfigure}
\begin{subfigure}[b]{0.19\textwidth}\includegraphics[width=3.4cm, height=3.4cm]{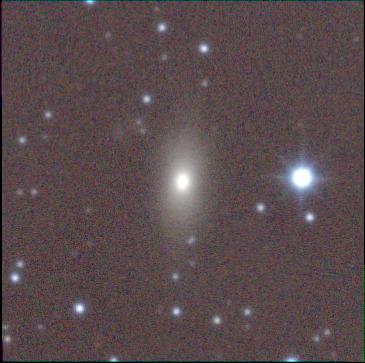}\caption*{40. $e$S}\end{subfigure}

\begin{subfigure}[b]{0.19\textwidth}\includegraphics[width=3.4cm, height=3.4cm]{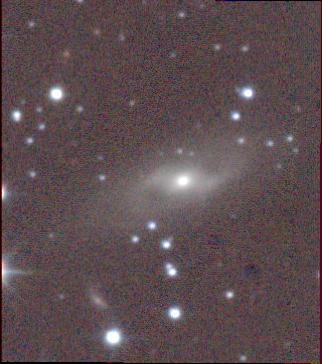}\caption*{41. $e$S}\end{subfigure}
\begin{subfigure}[b]{0.19\textwidth}\includegraphics[width=3.4cm, height=3.4cm]{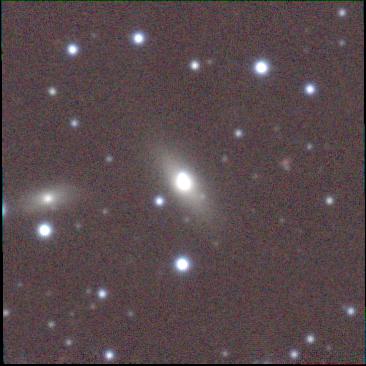}\caption*{42. E/S0}\end{subfigure}
\begin{subfigure}[b]{0.19\textwidth}\includegraphics[width=3.4cm, height=3.4cm]{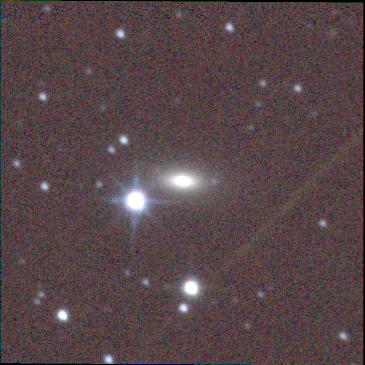}\caption*{43. E/S0}\end{subfigure}
\begin{subfigure}[b]{0.19\textwidth}\includegraphics[width=3.4cm, height=3.4cm]{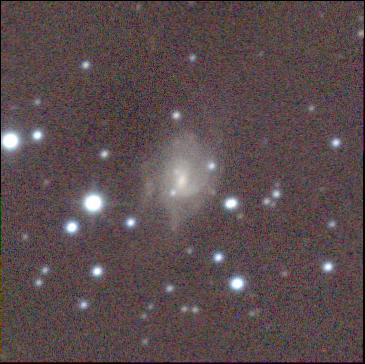}\caption*{44. $m$S}\end{subfigure}
\begin{subfigure}[b]{0.19\textwidth}\includegraphics[width=3.4cm, height=3.4cm]{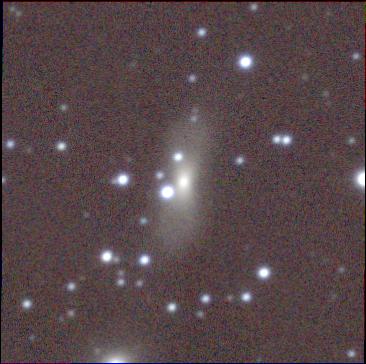}\caption*{45. $e$S}\end{subfigure}

\begin{subfigure}[b]{0.19\textwidth}\includegraphics[width=3.4cm, height=3.4cm]{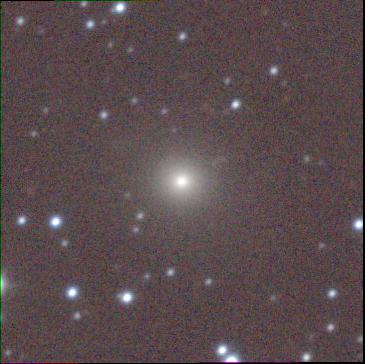}\caption*{46. E/S0}\end{subfigure}
\begin{subfigure}[b]{0.19\textwidth}\includegraphics[width=3.4cm, height=3.4cm]{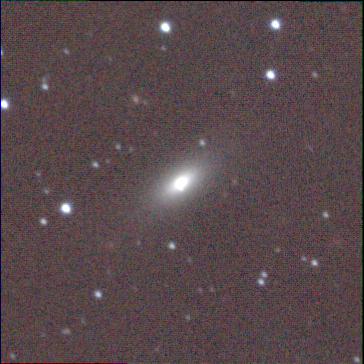}\caption*{47. E/S0}\end{subfigure}
\begin{subfigure}[b]{0.19\textwidth}\includegraphics[width=3.4cm, height=3.4cm]{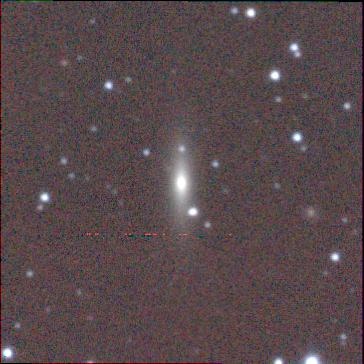}\caption*{48. $e$S}\end{subfigure}
\begin{subfigure}[b]{0.19\textwidth}\includegraphics[width=3.4cm, height=3.4cm]{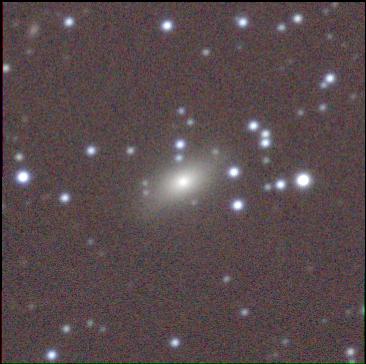}\caption*{49. $e$S}\end{subfigure}
\begin{subfigure}[b]{0.19\textwidth}\includegraphics[width=3.4cm, height=3.4cm]{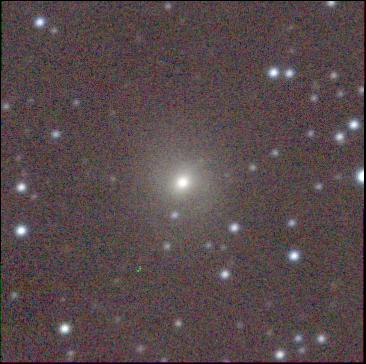}\caption*{50. E/S0}\end{subfigure}

\begin{subfigure}[b]{0.19\textwidth}\includegraphics[width=3.4cm, height=3.4cm]{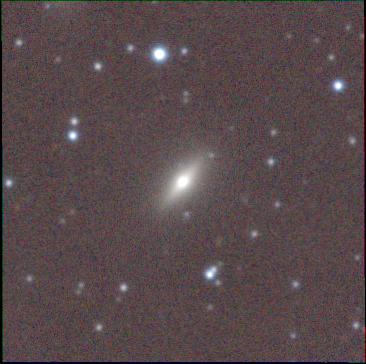}\caption*{51. $e$S}\end{subfigure}
\begin{subfigure}[b]{0.19\textwidth}\includegraphics[width=3.4cm, height=3.4cm]{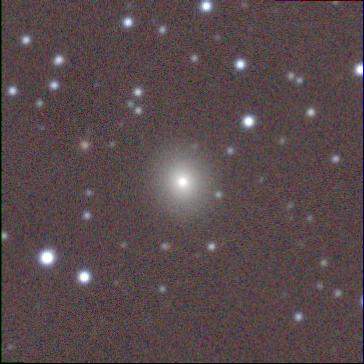}\caption*{52. E/S0}\end{subfigure}
\begin{subfigure}[b]{0.19\textwidth}\includegraphics[width=3.4cm, height=3.4cm]{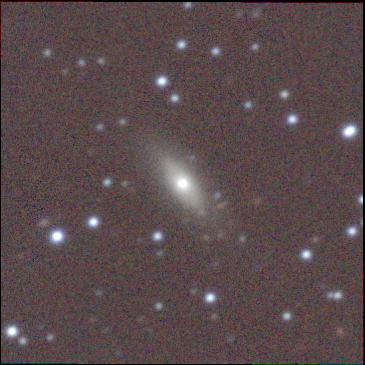}\caption*{53. E/S0}\end{subfigure}
\begin{subfigure}[b]{0.19\textwidth}\includegraphics[width=3.4cm, height=3.4cm]{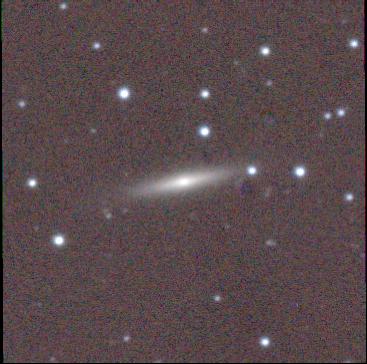}\caption*{54. $e$S}\end{subfigure}
\begin{subfigure}[b]{0.19\textwidth}\includegraphics[width=3.4cm, height=3.4cm]{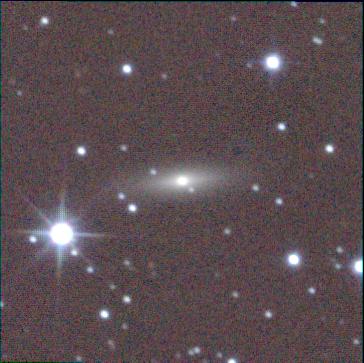}\caption*{55. $e$S}\end{subfigure}

\begin{subfigure}[b]{0.19\textwidth}\includegraphics[width=3.4cm, height=3.4cm]{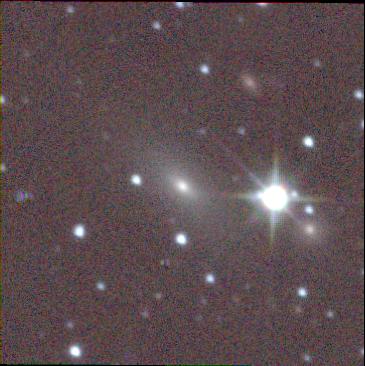}\caption*{56. $m$S}\end{subfigure}
\begin{subfigure}[b]{0.19\textwidth}\includegraphics[width=3.4cm, height=3.4cm]{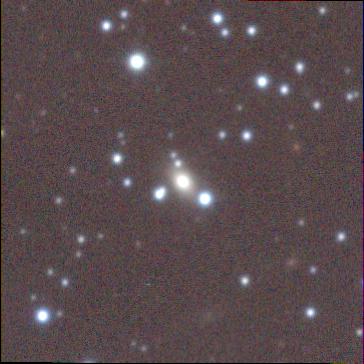}\caption*{57. E/S0}\end{subfigure}
\begin{subfigure}[b]{0.19\textwidth}\includegraphics[width=3.4cm, height=3.4cm]{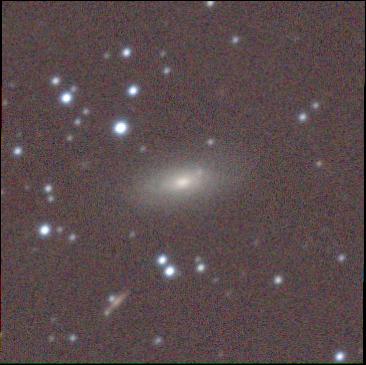}\caption*{58. $m$S}\end{subfigure}
\begin{subfigure}[b]{0.19\textwidth}\includegraphics[width=3.4cm, height=3.4cm]{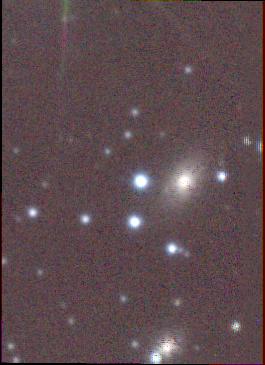}\caption*{59. E/S0}\end{subfigure}
\begin{subfigure}[b]{0.19\textwidth}\includegraphics[width=3.4cm, height=3.4cm]{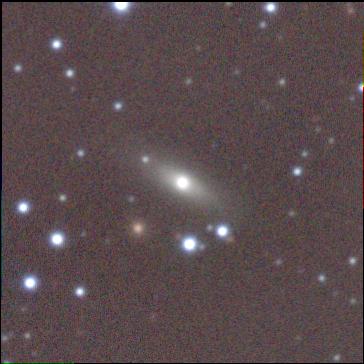}\caption*{60. $e$S}\end{subfigure}
\caption*{\textbf{Figure B} -- Continued.}
\end{figure*}

\begin{figure*}
\ContinuedFloat
\centering

\begin{subfigure}[b]{0.19\textwidth}\includegraphics[width=3.4cm, height=3.4cm]{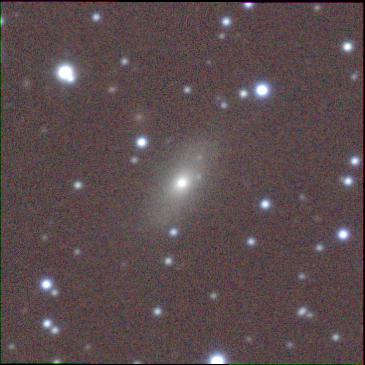}\caption*{61. $e$S}\end{subfigure}
\begin{subfigure}[b]{0.19\textwidth}\includegraphics[width=3.4cm, height=3.4cm]{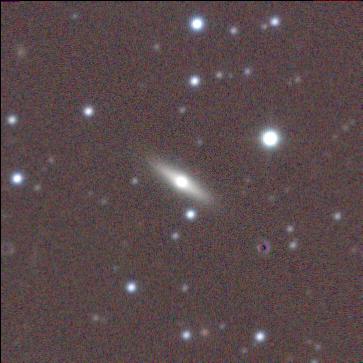}\caption*{62. $e$S}\end{subfigure}
\begin{subfigure}[b]{0.19\textwidth}\includegraphics[width=3.4cm, height=3.4cm]{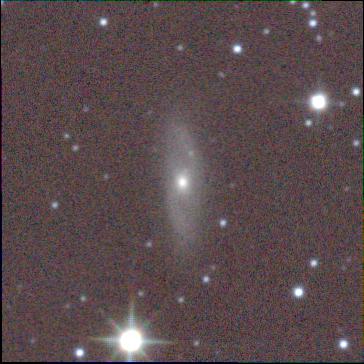}\caption*{63. $m$S}\end{subfigure}
\begin{subfigure}[b]{0.19\textwidth}\includegraphics[width=3.4cm, height=3.4cm]{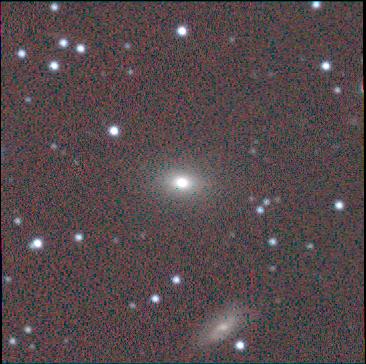}\caption*{64. E/S0}\end{subfigure}
\begin{subfigure}[b]{0.19\textwidth}\includegraphics[width=3.4cm, height=3.4cm]{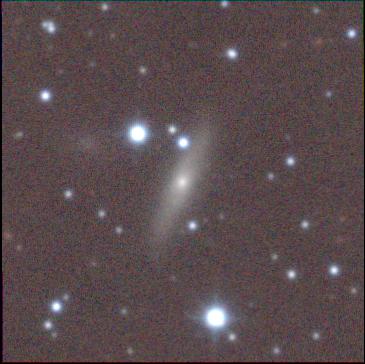}\caption*{65. $m$S}\end{subfigure}

\begin{subfigure}[b]{0.19\textwidth}\includegraphics[width=3.4cm, height=3.4cm]{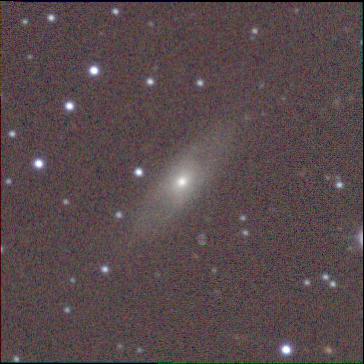}\caption*{66. $m$S}\end{subfigure}
\begin{subfigure}[b]{0.19\textwidth}\includegraphics[width=3.4cm, height=3.4cm]{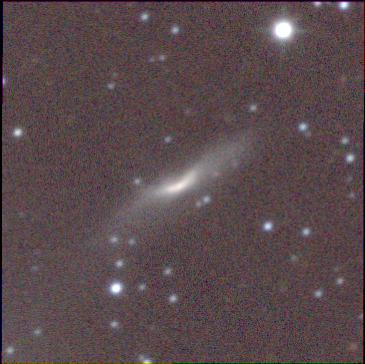}\caption*{67. $e$S}\end{subfigure}
\begin{subfigure}[b]{0.19\textwidth}\includegraphics[width=3.4cm, height=3.4cm]{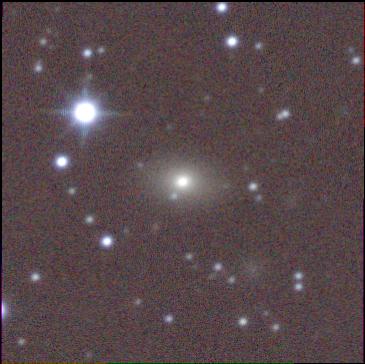}\caption*{68. E/S0}\end{subfigure}
\begin{subfigure}[b]{0.19\textwidth}\includegraphics[width=3.4cm, height=3.4cm]{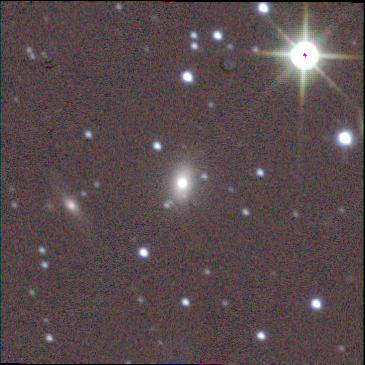}\caption*{69. $e$S}\end{subfigure}
\begin{subfigure}[b]{0.19\textwidth}\includegraphics[width=3.4cm,height=3.3cm,cfbox=cyan 1.5pt 0.0pt]{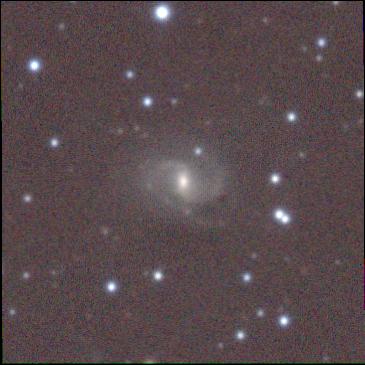}\caption*{70. $m$S}\end{subfigure}

\begin{subfigure}[b]{0.19\textwidth}\includegraphics[width=3.4cm, height=3.4cm]{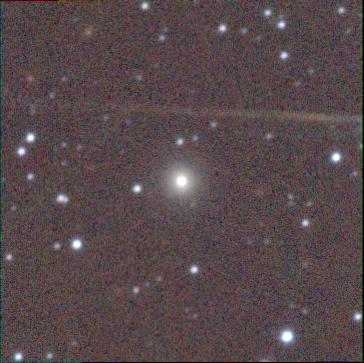}\caption*{71. $e$S}\end{subfigure}
\begin{subfigure}[b]{0.19\textwidth}\includegraphics[width=3.4cm, height=3.4cm]{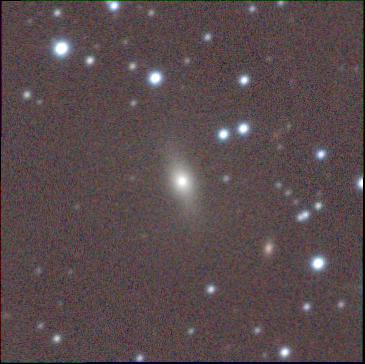}\caption*{72. $e$S}\end{subfigure}
\begin{subfigure}[b]{0.19\textwidth}\includegraphics[width=3.4cm, height=3.4cm]{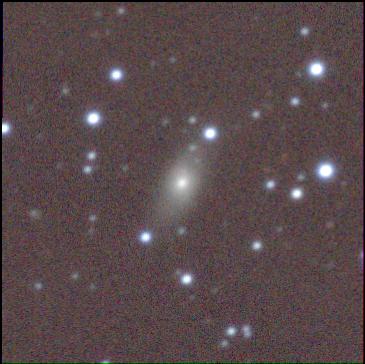}\caption*{73. $e$S}\end{subfigure}
\begin{subfigure}[b]{0.19\textwidth}\includegraphics[width=3.4cm, height=3.4cm]{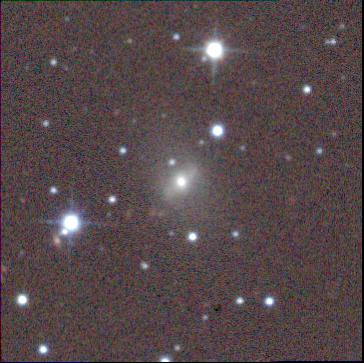}\caption*{74. $e$S}\end{subfigure}
\begin{subfigure}[b]{0.19\textwidth}\includegraphics[width=3.4cm, height=3.4cm]{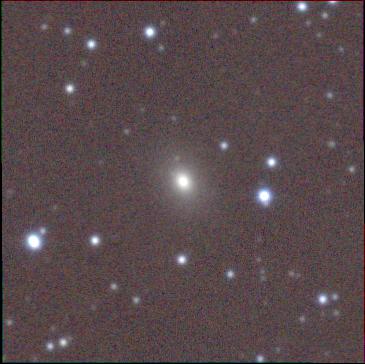}\caption*{75. E/S0}\end{subfigure}

\begin{subfigure}[b]{0.19\textwidth}\includegraphics[width=3.4cm, height=3.4cm]{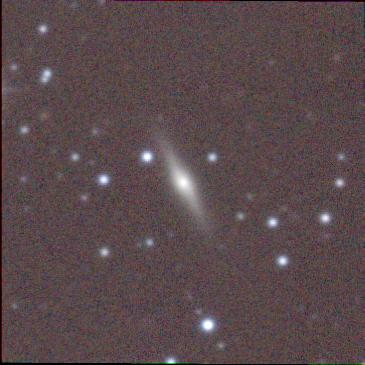}\caption*{76. $e$S}\end{subfigure}
\begin{subfigure}[b]{0.19\textwidth}\includegraphics[width=3.4cm, height=3.4cm]{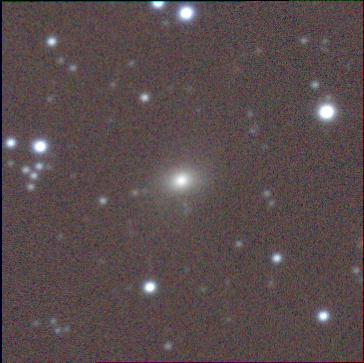}\caption*{77. E/S0}\end{subfigure}
\begin{subfigure}[b]{0.19\textwidth}\includegraphics[width=3.4cm, height=3.4cm]{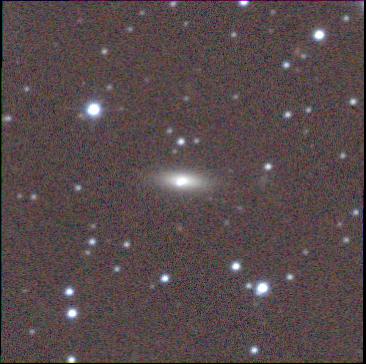}\caption*{78. $e$S}\end{subfigure}
\begin{subfigure}[b]{0.19\textwidth}\includegraphics[width=3.4cm, height=3.4cm]{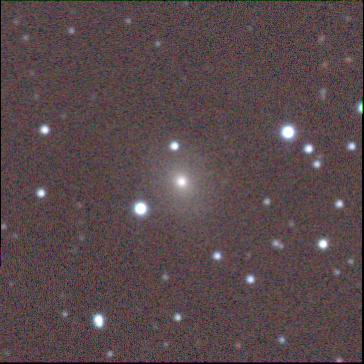}\caption*{79. E/S0}\end{subfigure}
\begin{subfigure}[b]{0.19\textwidth}\includegraphics[width=3.4cm, height=3.4cm]{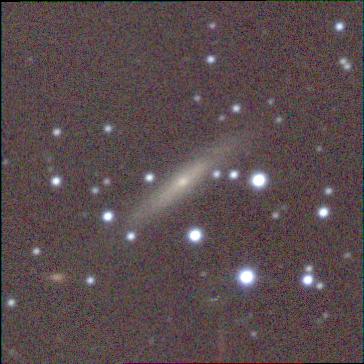}\caption*{80. $m$S}\end{subfigure}

\begin{subfigure}[b]{0.19\textwidth}\includegraphics[width=3.4cm, height=3.4cm]{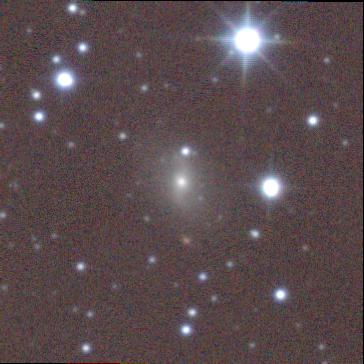}\caption*{81. $m$S}\end{subfigure}
\begin{subfigure}[b]{0.19\textwidth}\includegraphics[width=3.4cm, height=3.4cm]{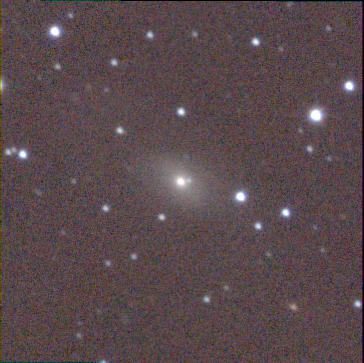}\caption*{82. $e$S}\end{subfigure}
\begin{subfigure}[b]{0.19\textwidth}\includegraphics[width=3.4cm, height=3.4cm]{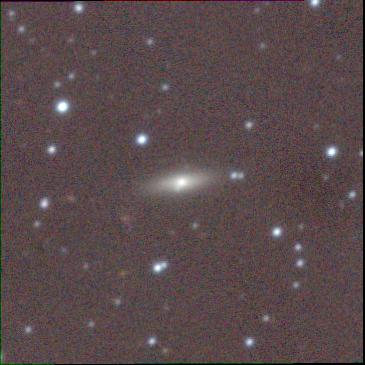}\caption*{83. $e$S}\end{subfigure}
\begin{subfigure}[b]{0.19\textwidth}\includegraphics[width=3.4cm,height=3.3cm,cfbox=cyan 1.5pt 0.0pt]{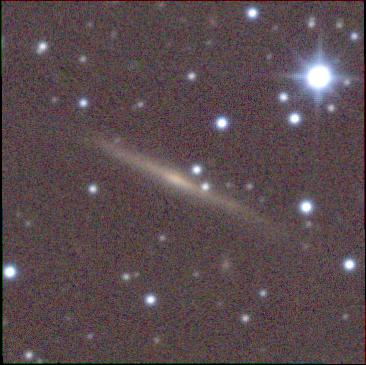}\caption*{84. $m$S}\end{subfigure}
\begin{subfigure}[b]{0.19\textwidth}\includegraphics[width=3.4cm, height=3.4cm]{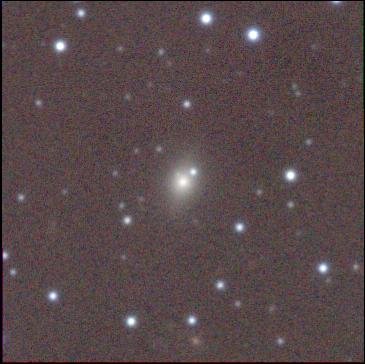}\caption*{85. E/S0}\end{subfigure}

\begin{subfigure}[b]{0.19\textwidth}\includegraphics[width=3.4cm, height=3.4cm]{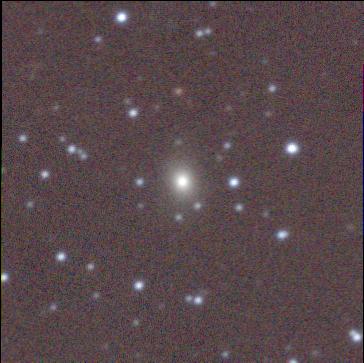}\caption*{86. E/S0}\end{subfigure}
\begin{subfigure}[b]{0.19\textwidth}\includegraphics[width=3.4cm, height=3.4cm]{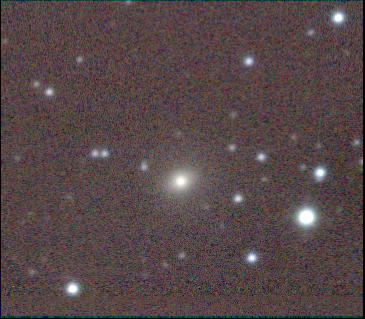}\caption*{87. E/S0}\end{subfigure}
\begin{subfigure}[b]{0.19\textwidth}\includegraphics[width=3.4cm, height=3.4cm]{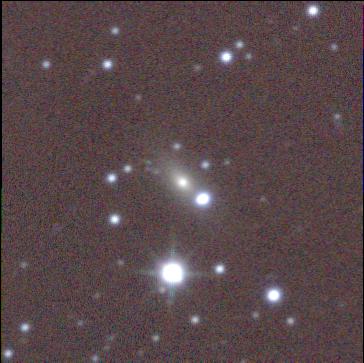}\caption*{88. $m$S}\end{subfigure}
\begin{subfigure}[b]{0.19\textwidth}\includegraphics[width=3.4cm, height=3.4cm]{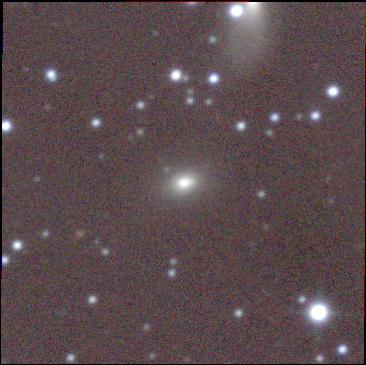}\caption*{89. E/S0}\end{subfigure}
\begin{subfigure}[b]{0.19\textwidth}\includegraphics[width=3.4cm, height=3.4cm]{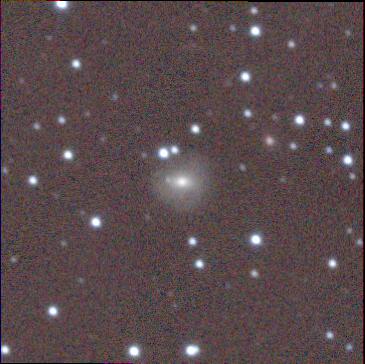}\caption*{90. $e$S}\end{subfigure}
\caption*{\textbf{Figure B} -- Continued.}
\end{figure*}

\begin{figure*}
\ContinuedFloat
\centering

\begin{subfigure}[b]{0.19\textwidth}\includegraphics[width=3.4cm, height=3.4cm]{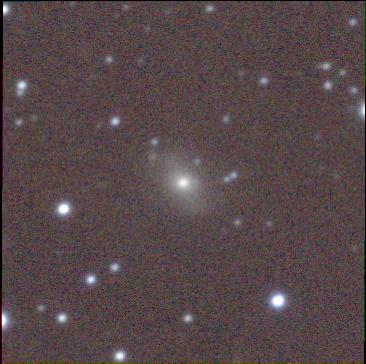}\caption*{91. $e$S}\end{subfigure}
\begin{subfigure}[b]{0.19\textwidth}\includegraphics[width=3.4cm, height=3.4cm]{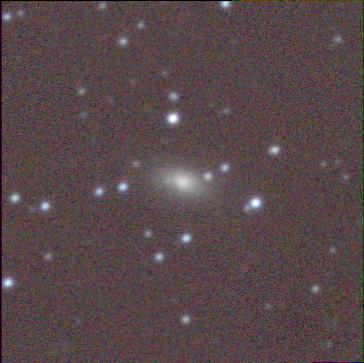}\caption*{92. $e$S}\end{subfigure}
\begin{subfigure}[b]{0.19\textwidth}\includegraphics[width=3.4cm, height=3.4cm]{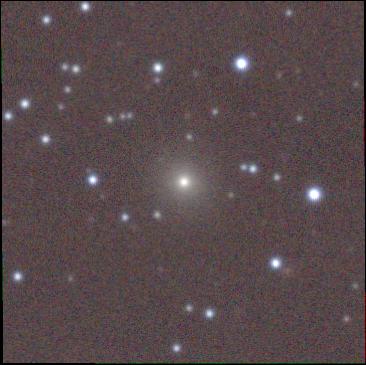}\caption*{93. E/S0}\end{subfigure}
\begin{subfigure}[b]{0.19\textwidth}\includegraphics[width=3.4cm, height=3.4cm]{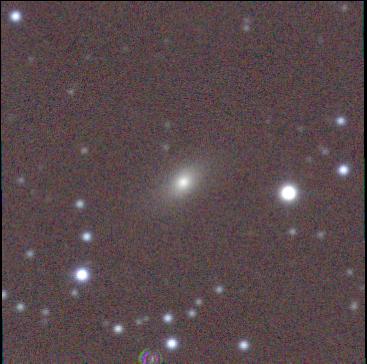}\caption*{94. $e$S}\end{subfigure}
\begin{subfigure}[b]{0.19\textwidth}\includegraphics[width=3.4cm, height=3.4cm]{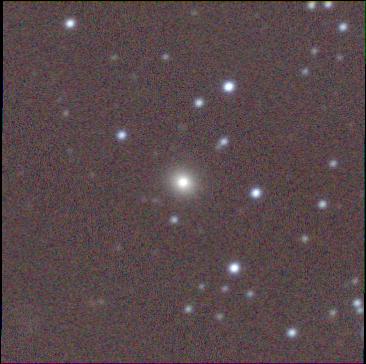}\caption*{95. E/S0}\end{subfigure}

\begin{subfigure}[b]{0.19\textwidth}\includegraphics[width=3.4cm, height=3.4cm]{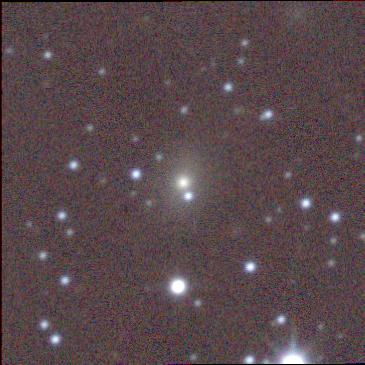}\caption*{96. E/S0}\end{subfigure}
\begin{subfigure}[b]{0.19\textwidth}\includegraphics[width=3.4cm, height=3.4cm]{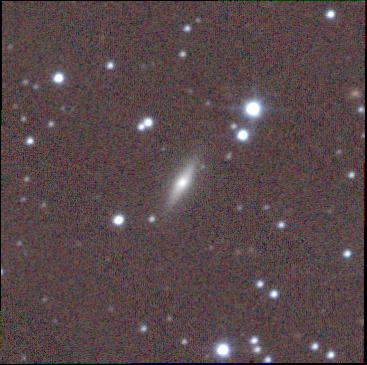}\caption*{97. $e$S}\end{subfigure}
\begin{subfigure}[b]{0.19\textwidth}\includegraphics[width=3.4cm, height=3.4cm]{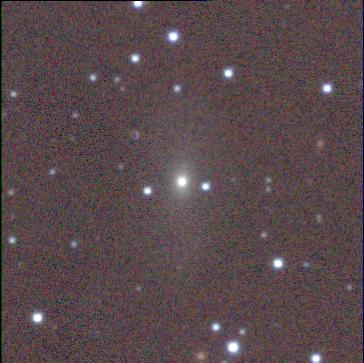}\caption*{98. E/S0}\end{subfigure}
\begin{subfigure}[b]{0.19\textwidth}\includegraphics[width=3.4cm, height=3.4cm]{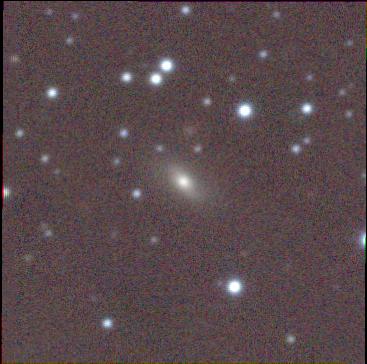}\caption*{99. $e$S}\end{subfigure}
\begin{subfigure}[b]{0.19\textwidth}\includegraphics[width=3.4cm, height=3.4cm]{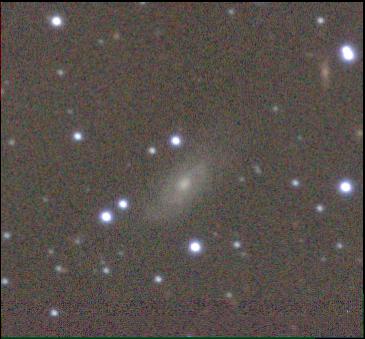}\caption*{100. $m$S}\end{subfigure}

\begin{subfigure}[b]{0.19\textwidth}\includegraphics[width=3.4cm, height=3.4cm]{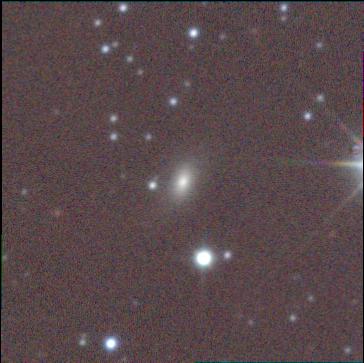}\caption*{101. E/S0}\end{subfigure}
\begin{subfigure}[b]{0.19\textwidth}\includegraphics[width=3.4cm, height=3.4cm]{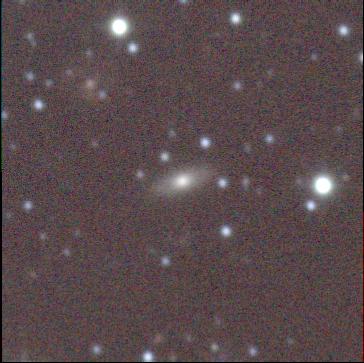}\caption*{102. $m$S}\end{subfigure}
\begin{subfigure}[b]{0.19\textwidth}\includegraphics[width=3.4cm, height=3.4cm]{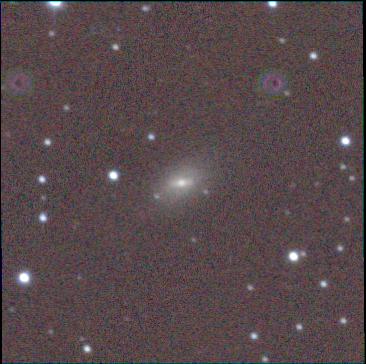}\caption*{103. $m$S}\end{subfigure}
\begin{subfigure}[b]{0.19\textwidth}\includegraphics[width=3.4cm, height=3.4cm]{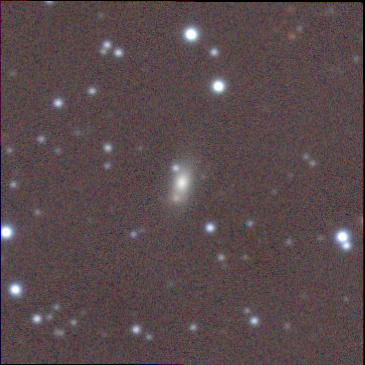}\caption*{104. E/S0}\end{subfigure}
\begin{subfigure}[b]{0.19\textwidth}\includegraphics[width=3.4cm, height=3.4cm]{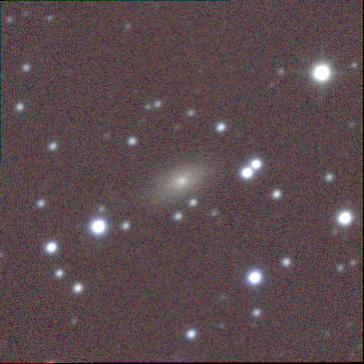}\caption*{105. $e$S}\end{subfigure}

\begin{subfigure}[b]{0.19\textwidth}\includegraphics[width=3.4cm, height=3.4cm]{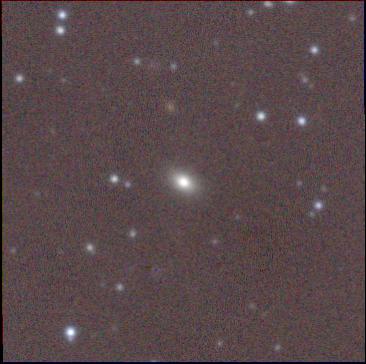}\caption*{106. E/S0}\end{subfigure}
\begin{subfigure}[b]{0.19\textwidth}\includegraphics[width=3.4cm, height=3.4cm]{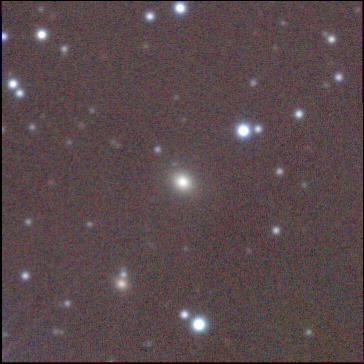}\caption*{107. E/S0}\end{subfigure}
\begin{subfigure}[b]{0.19\textwidth}\includegraphics[width=3.4cm, height=3.4cm]{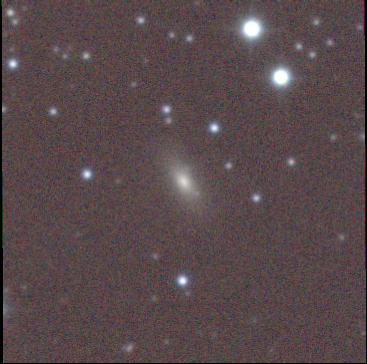}\caption*{108. $m$S}\end{subfigure}
\begin{subfigure}[b]{0.19\textwidth}\includegraphics[width=3.4cm, height=3.4cm]{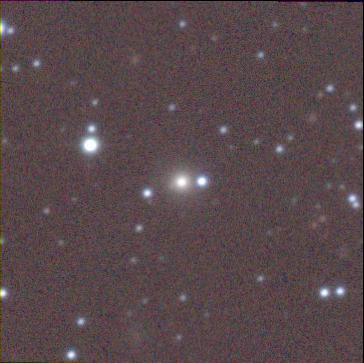}\caption*{109. E/S0}\end{subfigure}
\begin{subfigure}[b]{0.19\textwidth}\includegraphics[width=3.4cm,height=3.3cm,cfbox=cyan 1.5pt 0.0pt]{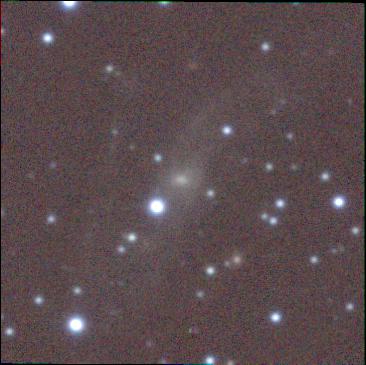}\caption*{110. $m$S}\end{subfigure}

\begin{subfigure}[b]{0.19\textwidth}\includegraphics[width=3.4cm, height=3.4cm]{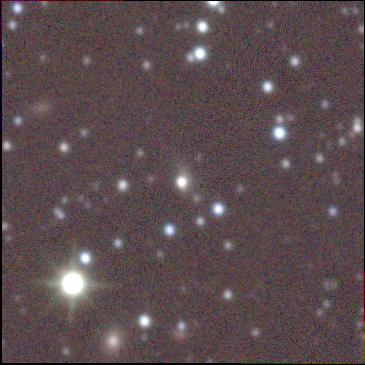}\caption*{111. E/S0}\end{subfigure}
\begin{subfigure}[b]{0.19\textwidth}\includegraphics[width=3.4cm,height=3.3cm,cfbox=cyan 1.5pt 0.0pt]{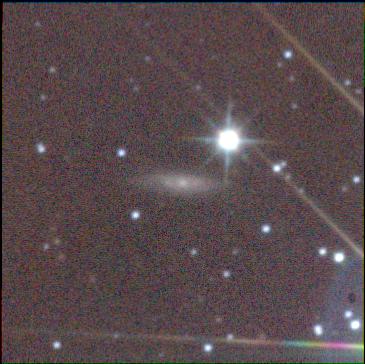}\caption*{112. $m$S}\end{subfigure}
\begin{subfigure}[b]{0.19\textwidth}\includegraphics[width=3.4cm, height=3.4cm]{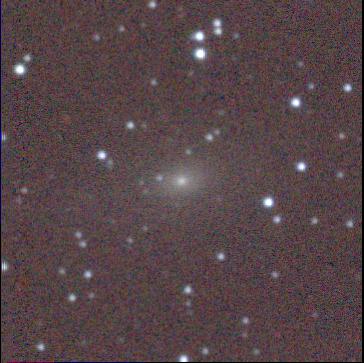}\caption*{113. $m$S}\end{subfigure}
\begin{subfigure}[b]{0.19\textwidth}\includegraphics[width=3.4cm, height=3.4cm]{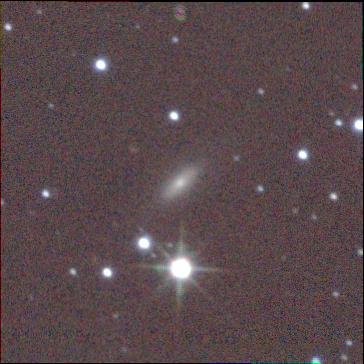}\caption*{114. $m$S}\end{subfigure}
\begin{subfigure}[b]{0.19\textwidth}\includegraphics[width=3.4cm, height=3.4cm]{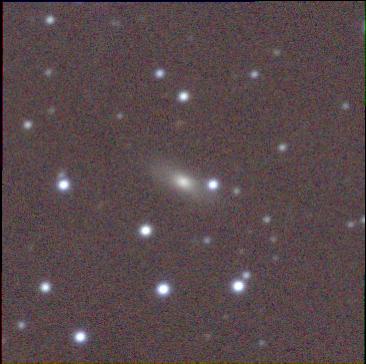}\caption*{115. $m$S}\end{subfigure}

\begin{subfigure}[b]{0.19\textwidth}\includegraphics[width=3.4cm,height=3.3cm,cfbox=cyan 1.5pt 0.0pt]{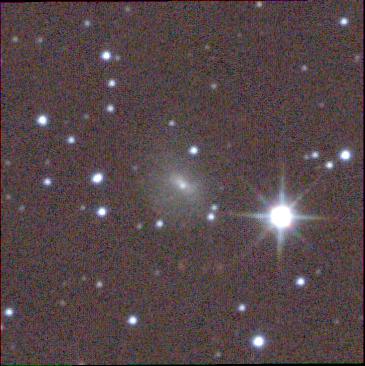}\caption*{116. $m$S}\end{subfigure}
\begin{subfigure}[b]{0.19\textwidth}\includegraphics[width=3.4cm, height=3.4cm]{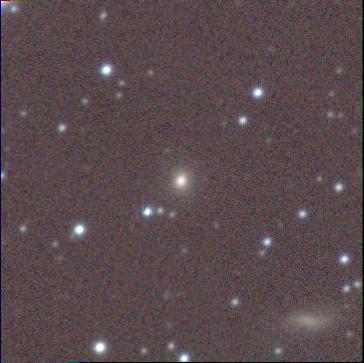}\caption*{117. E/S0}\end{subfigure}
\begin{subfigure}[b]{0.19\textwidth}\includegraphics[width=3.4cm, height=3.4cm]{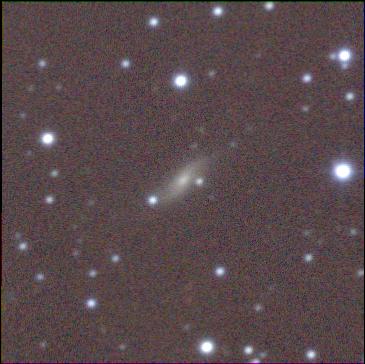}\caption*{118. $m$S}\end{subfigure}
\begin{subfigure}[b]{0.19\textwidth}\includegraphics[width=3.4cm, height=3.4cm]{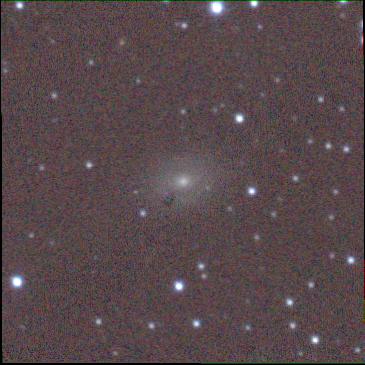}\caption*{119. $m$S}\end{subfigure}
\begin{subfigure}[b]{0.19\textwidth}\includegraphics[width=3.4cm, height=3.4cm]{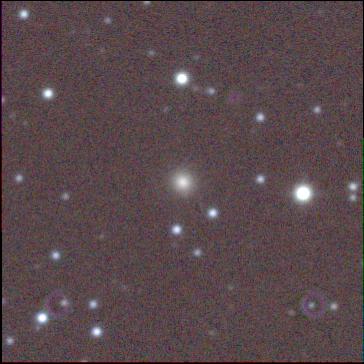}\caption*{120. E/S0}\end{subfigure}
\caption*{\textbf{Figure B} -- Continued.}
\end{figure*}

\begin{figure*}
\ContinuedFloat
\centering

\begin{subfigure}[b]{0.19\textwidth}\includegraphics[width=3.4cm, height=3.4cm]{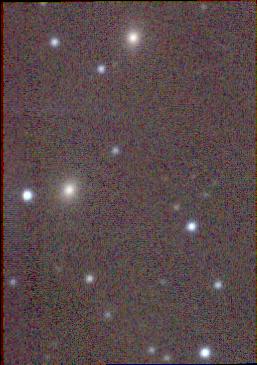}\caption*{121. E/S0}\end{subfigure}
\begin{subfigure}[b]{0.19\textwidth}\includegraphics[width=3.4cm, height=3.4cm]{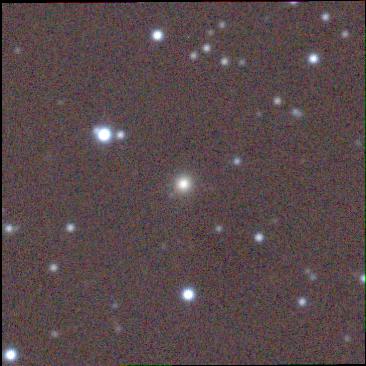}\caption*{122. E/S0}\end{subfigure}
\begin{subfigure}[b]{0.19\textwidth}\includegraphics[width=3.4cm, height=3.4cm]{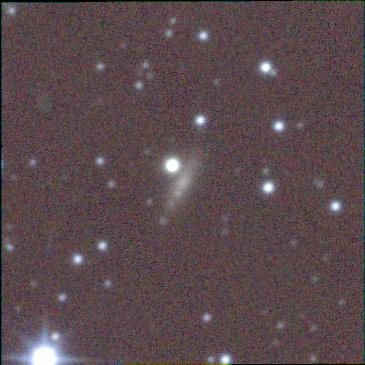}\caption*{123. $m$S}\end{subfigure}
\begin{subfigure}[b]{0.19\textwidth}\includegraphics[width=3.4cm, height=3.4cm]{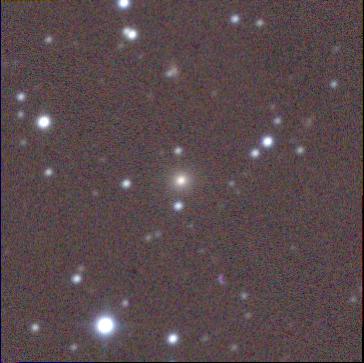}\caption*{124. E/S0}\end{subfigure}
\begin{subfigure}[b]{0.19\textwidth}\includegraphics[width=3.4cm, height=3.4cm]{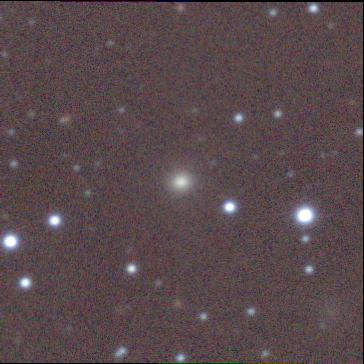}\caption*{125. E/S0}\end{subfigure}

\begin{subfigure}[b]{0.19\textwidth}\includegraphics[width=3.4cm, height=3.4cm]{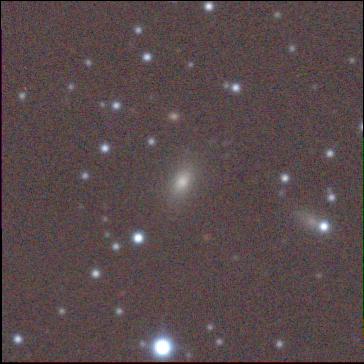}\caption*{126. $m$S}\end{subfigure}
\begin{subfigure}[b]{0.19\textwidth}\includegraphics[width=3.4cm, height=3.4cm]{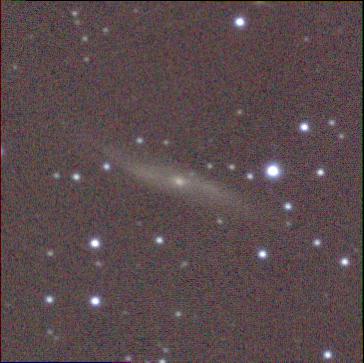}\caption*{127. $m$S}\end{subfigure}
\begin{subfigure}[b]{0.19\textwidth}\includegraphics[width=3.4cm, height=3.4cm]{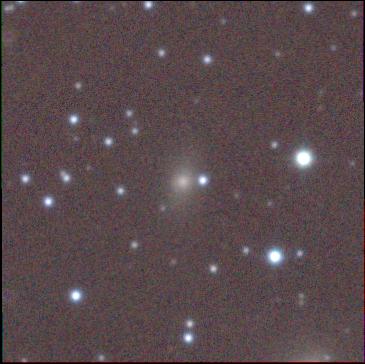}\caption*{128. $m$S}\end{subfigure}
\begin{subfigure}[b]{0.19\textwidth}\includegraphics[width=3.4cm, height=3.4cm]{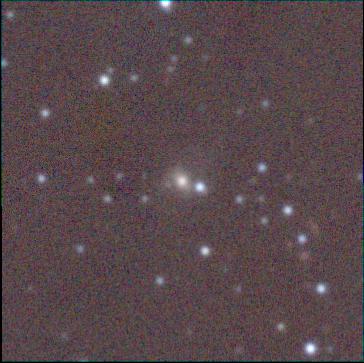}\caption*{129. E/S0}\end{subfigure}
\begin{subfigure}[b]{0.19\textwidth}\includegraphics[width=3.4cm, height=3.4cm]{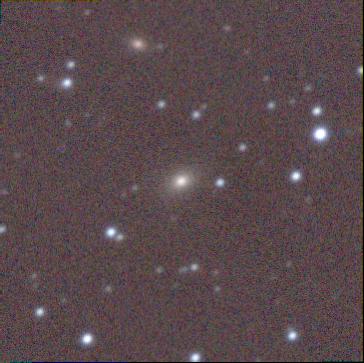}\caption*{130. E/S0}\end{subfigure}

\begin{subfigure}[b]{0.19\textwidth}\includegraphics[width=3.4cm, height=3.4cm]{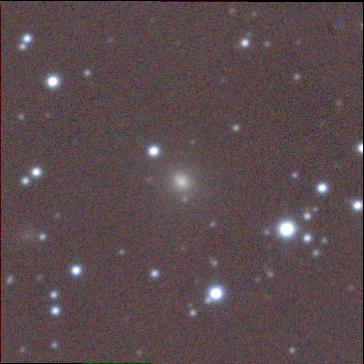}\caption*{131. $m$S}\end{subfigure}
\begin{subfigure}[b]{0.19\textwidth}\includegraphics[width=3.4cm, height=3.4cm]{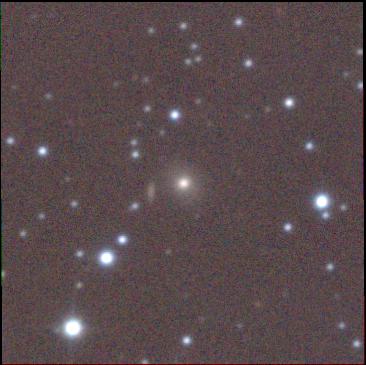}\caption*{132. $e$S}\end{subfigure}
\begin{subfigure}[b]{0.19\textwidth}\includegraphics[width=3.4cm, height=3.4cm]{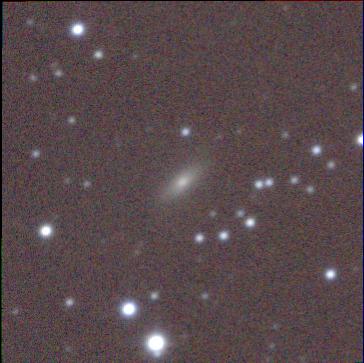}\caption*{133. $m$S}\end{subfigure}
\begin{subfigure}[b]{0.19\textwidth}\includegraphics[width=3.4cm, height=3.4cm]{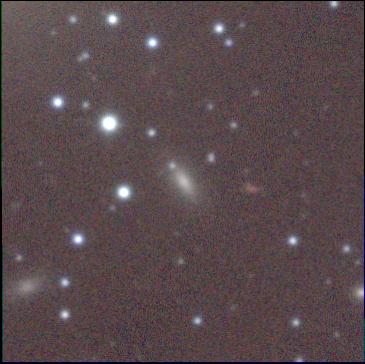}\caption*{134. $e$S}\end{subfigure}
\begin{subfigure}[b]{0.19\textwidth}\includegraphics[width=3.4cm, height=3.4cm]{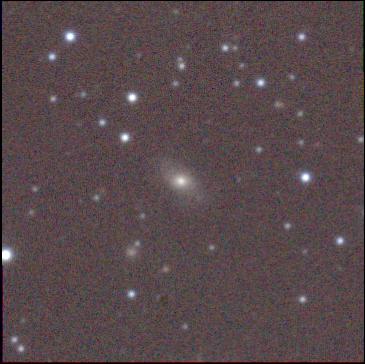}\caption*{135. $e$S}\end{subfigure}

\begin{subfigure}[b]{0.19\textwidth}\includegraphics[width=3.4cm, height=3.4cm]{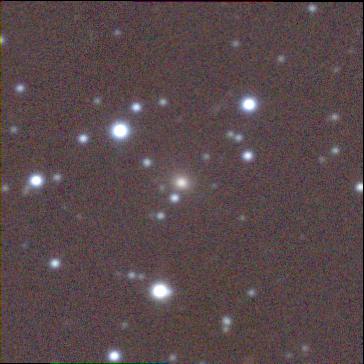}\caption*{136. E/S0}\end{subfigure}
\begin{subfigure}[b]{0.19\textwidth}\includegraphics[width=3.4cm,height=3.3cm,cfbox=cyan 1.5pt 0.0pt]{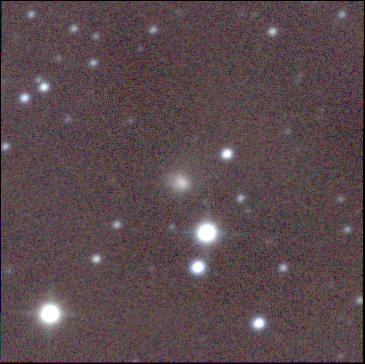}\caption*{137. $m$S}\end{subfigure}
\begin{subfigure}[b]{0.19\textwidth}\includegraphics[width=3.4cm, height=3.4cm]{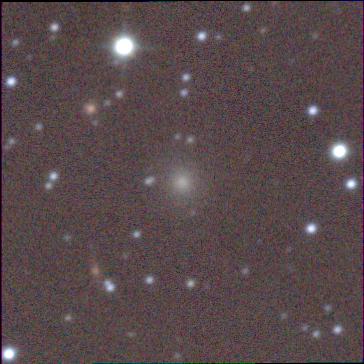}\caption*{138. $m$S}\end{subfigure}
\begin{subfigure}[b]{0.19\textwidth}\includegraphics[width=3.4cm, height=3.4cm]{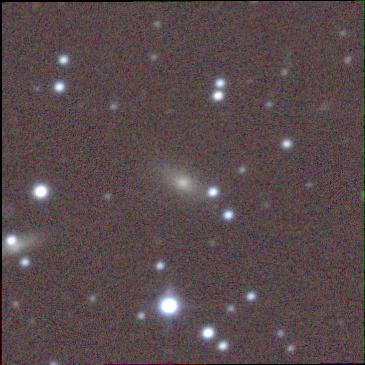}\caption*{139. $m$S}\end{subfigure}
\begin{subfigure}[b]{0.19\textwidth}\includegraphics[width=3.4cm, height=3.4cm]{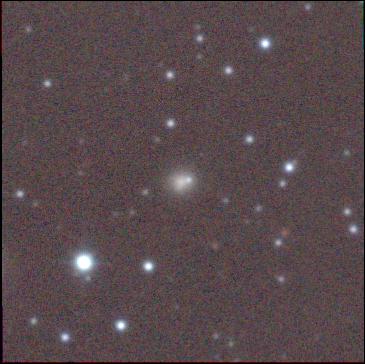}\caption*{140. Irr}\end{subfigure}

\begin{subfigure}[b]{0.19\textwidth}\includegraphics[width=3.4cm, height=3.4cm]{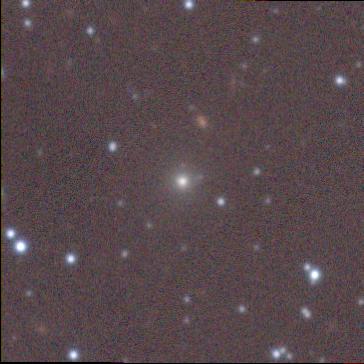}\caption*{141. E/S0}\end{subfigure}
\begin{subfigure}[b]{0.19\textwidth}\includegraphics[width=3.4cm, height=3.4cm]{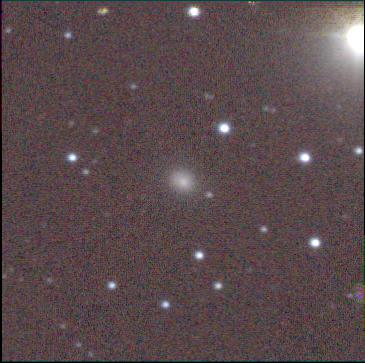}\caption*{142. $m$S}\end{subfigure}
\begin{subfigure}[b]{0.19\textwidth}\includegraphics[width=3.4cm, height=3.4cm]{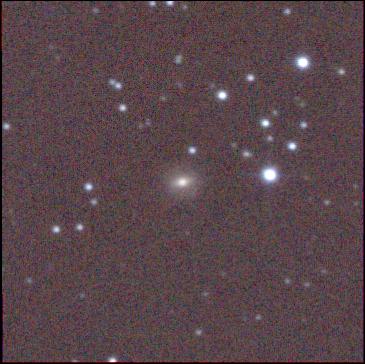}\caption*{143. $e$S}\end{subfigure}
\begin{subfigure}[b]{0.19\textwidth}\includegraphics[width=3.4cm, height=3.4cm]{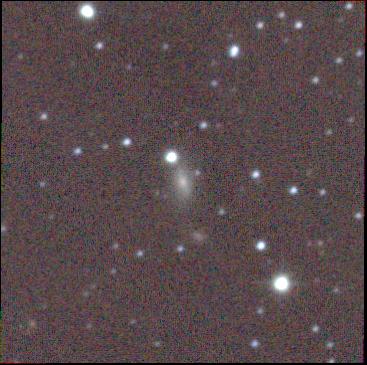}\caption*{144. $m$S}\end{subfigure}
\begin{subfigure}[b]{0.19\textwidth}\includegraphics[width=3.4cm, height=3.4cm]{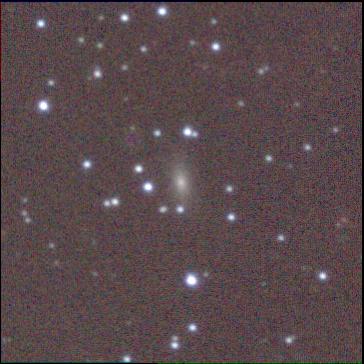}\caption*{145. $m$S}\end{subfigure}

\begin{subfigure}[b]{0.19\textwidth}\includegraphics[width=3.4cm, height=3.4cm]{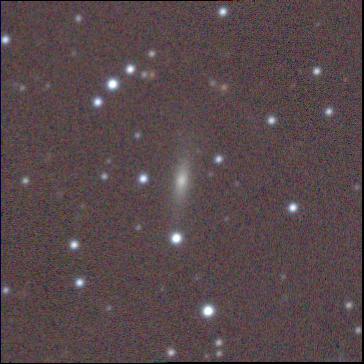}\caption*{146. $e$S}\end{subfigure}
\begin{subfigure}[b]{0.19\textwidth}\includegraphics[width=3.4cm, height=3.4cm]{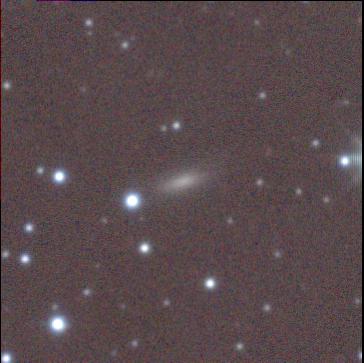}\caption*{147. $m$S}\end{subfigure}
\begin{subfigure}[b]{0.19\textwidth}\includegraphics[width=3.4cm, height=3.4cm]{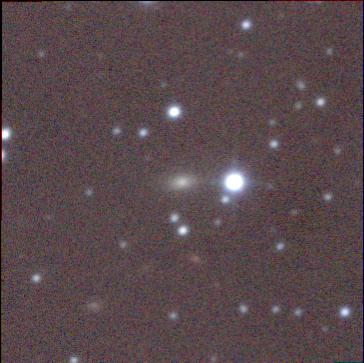}\caption*{148. $\ell$S}\end{subfigure}
\begin{subfigure}[b]{0.19\textwidth}\includegraphics[width=3.4cm, height=3.4cm]{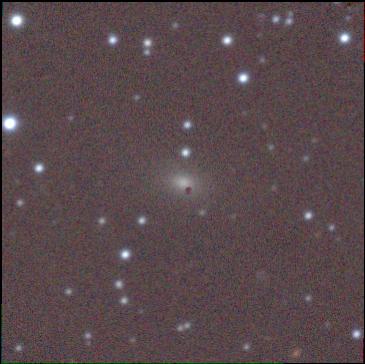}\caption*{149. E/S0}\end{subfigure}
\begin{subfigure}[b]{0.19\textwidth}\includegraphics[width=3.4cm, height=3.4cm]{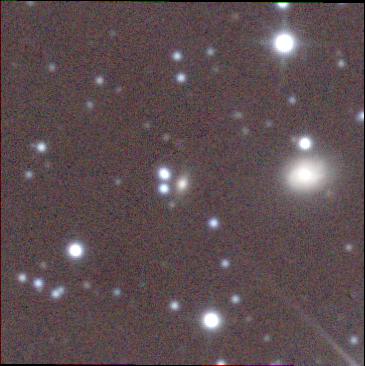}\caption*{150. E/S0}\end{subfigure}
\caption*{\textbf{Figure B} -- Continued.}
\end{figure*}

\begin{figure*}
\ContinuedFloat
\centering

\begin{subfigure}[b]{0.19\textwidth}\includegraphics[width=3.4cm, height=3.4cm]{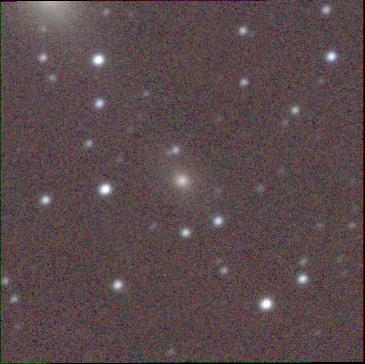}\caption*{151. E/S0}\end{subfigure}
\begin{subfigure}[b]{0.19\textwidth}\includegraphics[width=3.4cm, height=3.4cm]{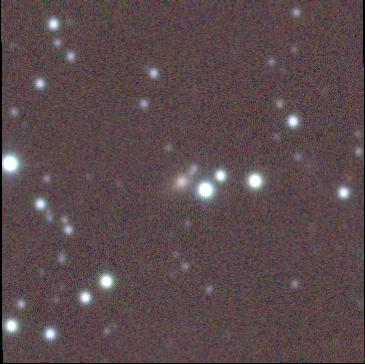}\caption*{152. E/S0}\end{subfigure}
\begin{subfigure}[b]{0.19\textwidth}\includegraphics[width=3.4cm, height=3.4cm]{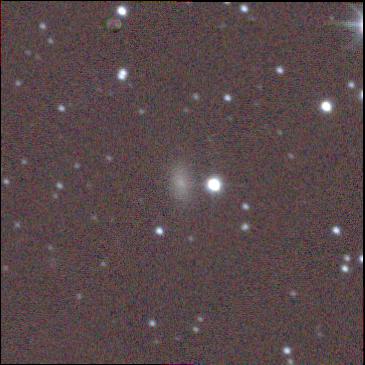}\caption*{153. $\ell$S}\end{subfigure}
\begin{subfigure}[b]{0.19\textwidth}\includegraphics[width=3.4cm,height=3.3cm,cfbox=cyan 1.5pt 0.0pt]{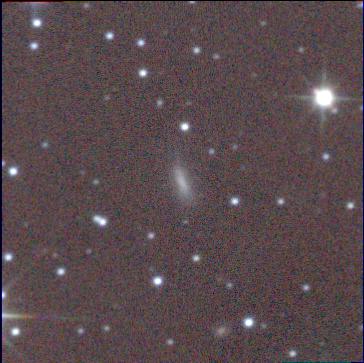}\caption*{154. $\ell$S}\end{subfigure}
\begin{subfigure}[b]{0.19\textwidth}\includegraphics[width=3.4cm, height=3.4cm]{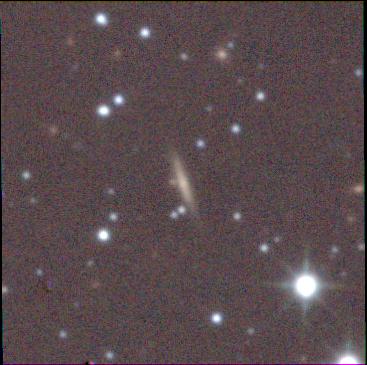}\caption*{155. $m$S}\end{subfigure}

\begin{subfigure}[b]{0.19\textwidth}\includegraphics[width=3.4cm, height=3.4cm]{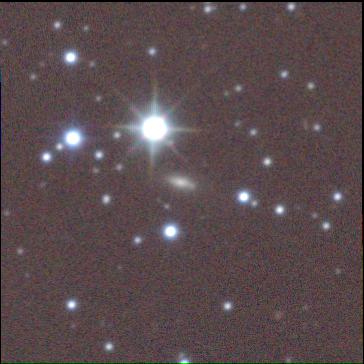}\caption*{156. $e$S}\end{subfigure}
\begin{subfigure}[b]{0.19\textwidth}\includegraphics[width=3.4cm, height=3.4cm]{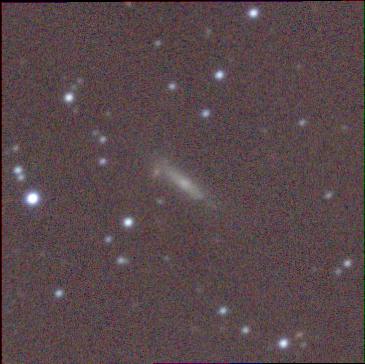}\caption*{157. $m$S}\end{subfigure}
\begin{subfigure}[b]{0.19\textwidth}\includegraphics[width=3.4cm, height=3.4cm]{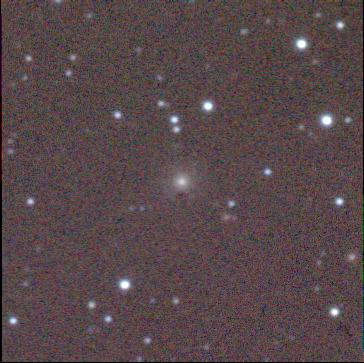}\caption*{158. E/S0}\end{subfigure}
\begin{subfigure}[b]{0.19\textwidth}\includegraphics[width=3.4cm, height=3.4cm]{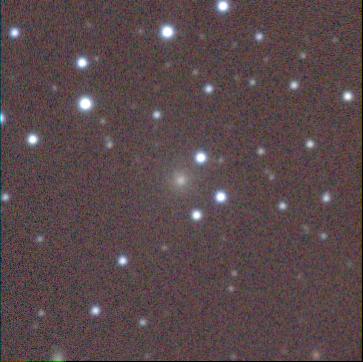}\caption*{159. $m$S}\end{subfigure}
\begin{subfigure}[b]{0.19\textwidth}\includegraphics[width=3.4cm, height=3.4cm]{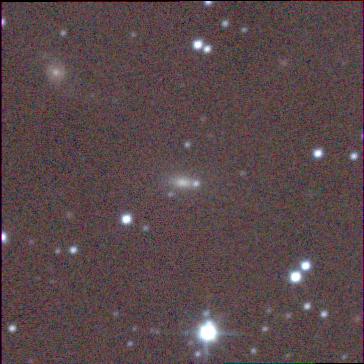}\caption*{160. $\ell$S}\end{subfigure}

\begin{subfigure}[b]{0.19\textwidth}\includegraphics[width=3.4cm, height=3.4cm]{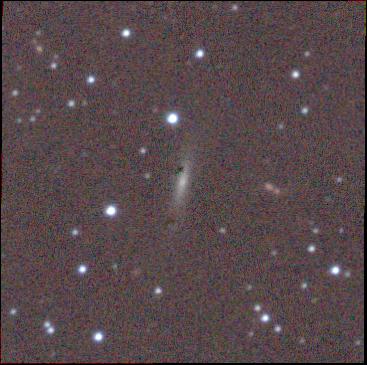}\caption*{161. $m$S}\end{subfigure}
\begin{subfigure}[b]{0.19\textwidth}\includegraphics[width=3.4cm, height=3.4cm]{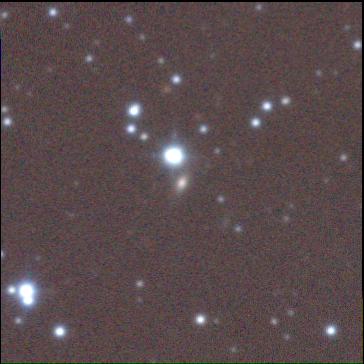}\caption*{162. E/S0}\end{subfigure}
\begin{subfigure}[b]{0.19\textwidth}\includegraphics[width=3.4cm, height=3.4cm]{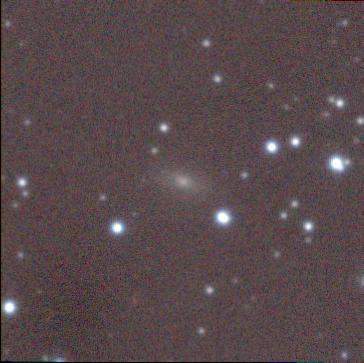}\caption*{163. $e$S}\end{subfigure}
\begin{subfigure}[b]{0.19\textwidth}\includegraphics[width=3.4cm, height=3.4cm]{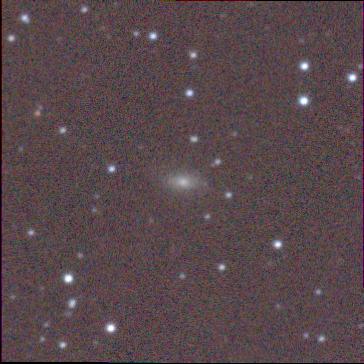}\caption*{164. $m$S}\end{subfigure}
\begin{subfigure}[b]{0.19\textwidth}\includegraphics[width=3.4cm, height=3.4cm]{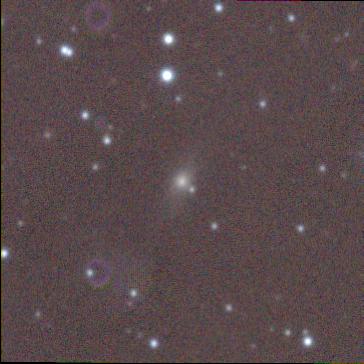}\caption*{165. $m$S}\end{subfigure}

\begin{subfigure}[b]{0.19\textwidth}\includegraphics[width=3.4cm, height=3.4cm]{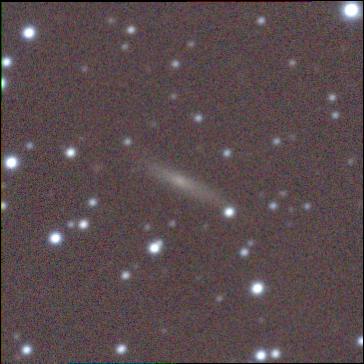}\caption*{166. $m$S}\end{subfigure}
\begin{subfigure}[b]{0.19\textwidth}\includegraphics[width=3.4cm, height=3.4cm]{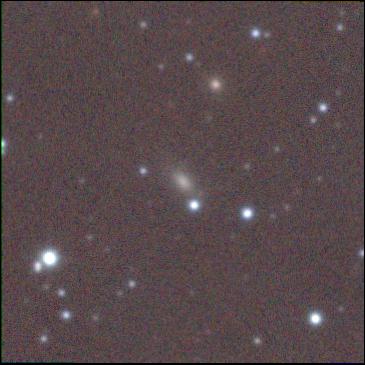}\caption*{167. $m$S}\end{subfigure}
\begin{subfigure}[b]{0.19\textwidth}\includegraphics[width=3.4cm, height=3.4cm]{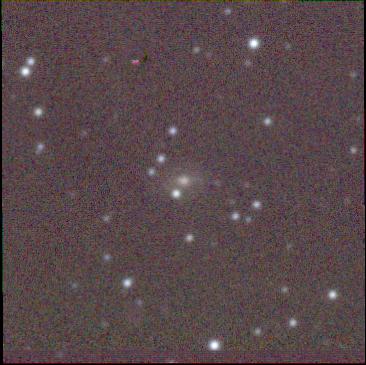}\caption*{168. $e$S}\end{subfigure}
\begin{subfigure}[b]{0.19\textwidth}\includegraphics[width=3.4cm, height=3.4cm]{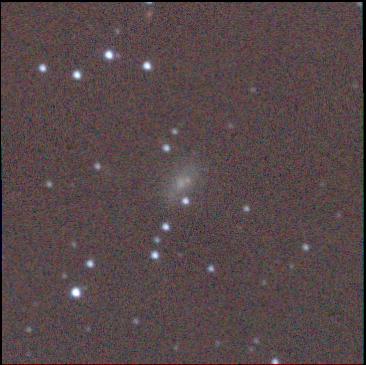}\caption*{169. Irr}\end{subfigure}
\begin{subfigure}[b]{0.19\textwidth}\includegraphics[width=3.4cm, height=3.4cm]{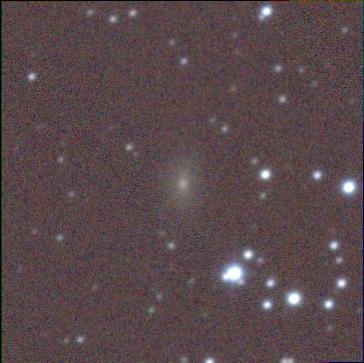}\caption*{170. $m$S}\end{subfigure}

\begin{subfigure}[b]{0.19\textwidth}\includegraphics[width=3.4cm, height=3.4cm]{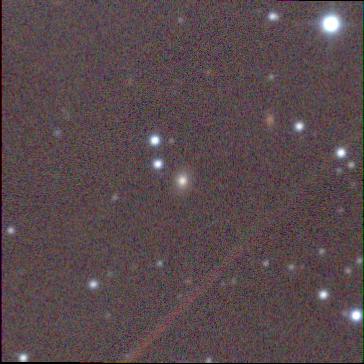}\caption*{171. E/S0}\end{subfigure}
\begin{subfigure}[b]{0.19\textwidth}\includegraphics[width=3.4cm, height=3.4cm]{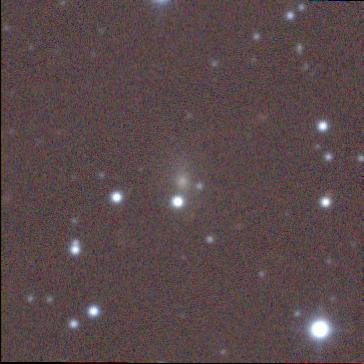}\caption*{172. $e$S}\end{subfigure}
\begin{subfigure}[b]{0.19\textwidth}\includegraphics[width=3.4cm, height=3.4cm]{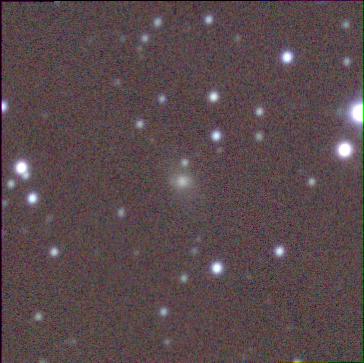}\caption*{173. $e$S}\end{subfigure}
\begin{subfigure}[b]{0.19\textwidth}\includegraphics[width=3.4cm, height=3.4cm]{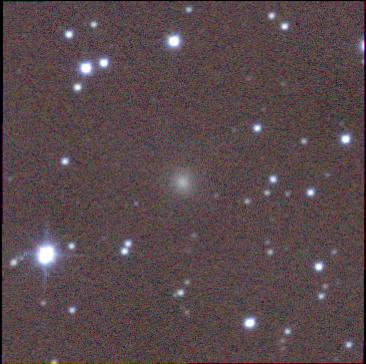}\caption*{174. E/S0}\end{subfigure}
\begin{subfigure}[b]{0.19\textwidth}\includegraphics[width=3.4cm, height=3.4cm]{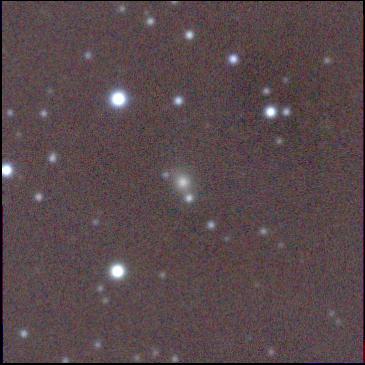}\caption*{175. E/S0}\end{subfigure}

\begin{subfigure}[b]{0.19\textwidth}\includegraphics[width=3.4cm, height=3.4cm]{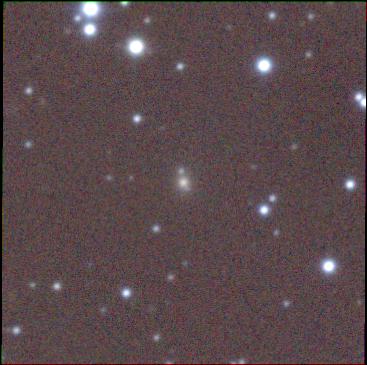}\caption*{176. E/S0}\end{subfigure}
\begin{subfigure}[b]{0.19\textwidth}\includegraphics[width=3.4cm, height=3.4cm]{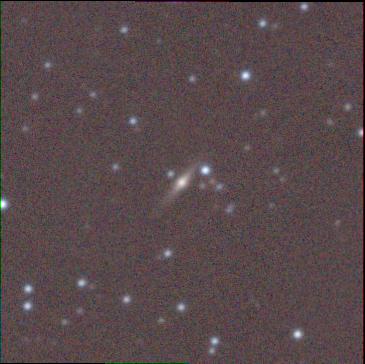}\caption*{177. $e$S}\end{subfigure}
\begin{subfigure}[b]{0.19\textwidth}\includegraphics[width=3.4cm, height=3.4cm]{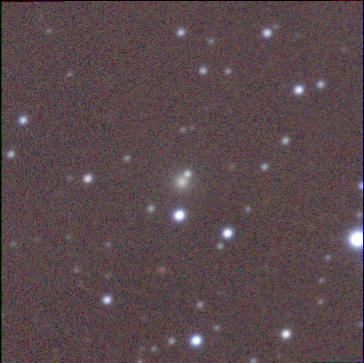}\caption*{178. E/S0}\end{subfigure}
\begin{subfigure}[b]{0.19\textwidth}\includegraphics[width=3.4cm, height=3.4cm]{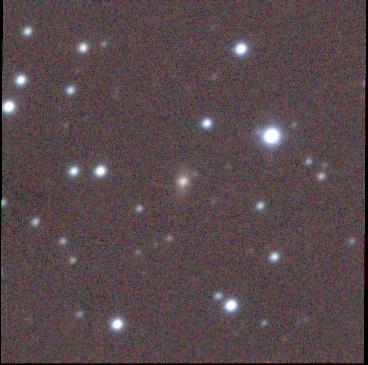}\caption*{179. $m$S}\end{subfigure}
\begin{subfigure}[b]{0.19\textwidth}\includegraphics[width=3.4cm,height=3.3cm,cfbox=cyan 1.5pt 0.0pt]{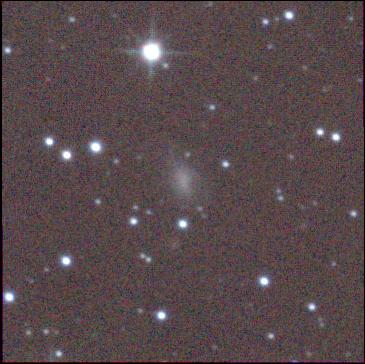}\caption*{180. Irr}\end{subfigure}
\caption*{\textbf{Figure B} -- Continued.}
\end{figure*}

\begin{figure*}
\ContinuedFloat
\centering

\begin{subfigure}[b]{0.19\textwidth}\includegraphics[width=3.4cm, height=3.4cm]{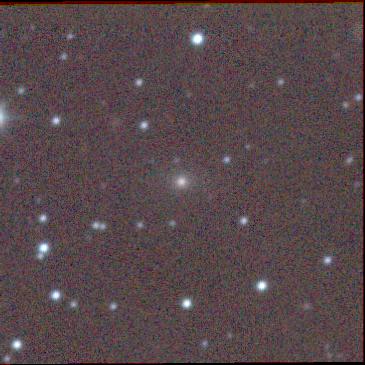}\caption*{181. E/S0}\end{subfigure}
\begin{subfigure}[b]{0.19\textwidth}\includegraphics[width=3.4cm, height=3.4cm]{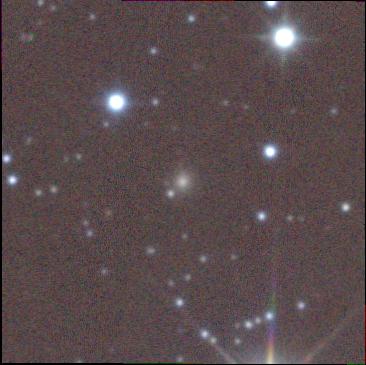}\caption*{182. $m$S}\end{subfigure}
\begin{subfigure}[b]{0.19\textwidth}\includegraphics[width=3.4cm, height=3.4cm]{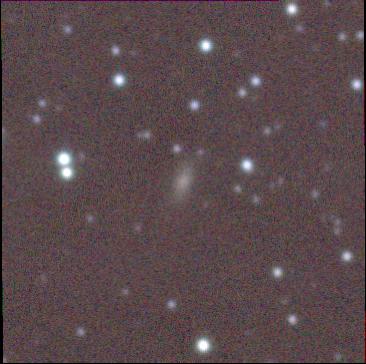}\caption*{183. $\ell$S}\end{subfigure}
\begin{subfigure}[b]{0.19\textwidth}\includegraphics[width=3.4cm, height=3.4cm]{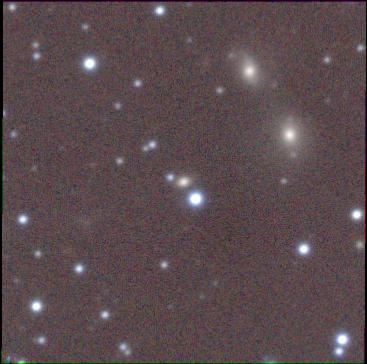}\caption*{184. E/S0}\end{subfigure}
\begin{subfigure}[b]{0.19\textwidth}\includegraphics[width=3.4cm, height=3.4cm]{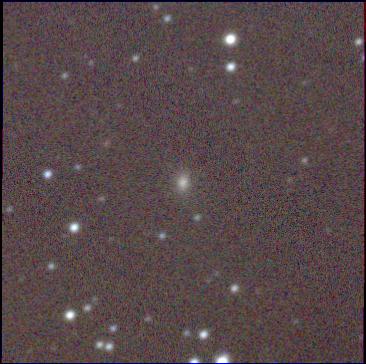}\caption*{185. Irr}\end{subfigure}

\begin{subfigure}[b]{0.19\textwidth}\includegraphics[width=3.4cm, height=3.4cm]{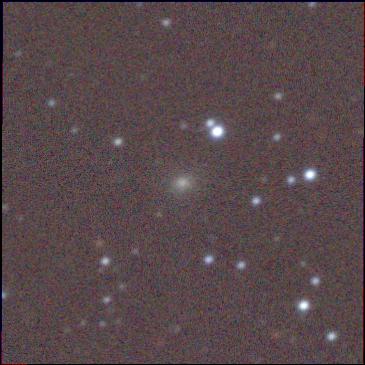}\caption*{186. $\ell$S}\end{subfigure}
\begin{subfigure}[b]{0.19\textwidth}\includegraphics[width=3.4cm, height=3.4cm]{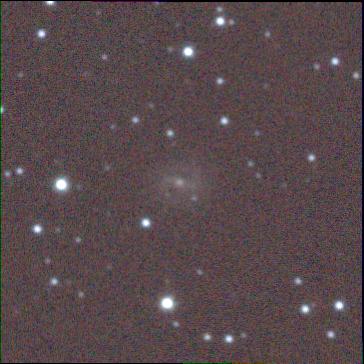}\caption*{187. $m$S}\end{subfigure}
\begin{subfigure}[b]{0.19\textwidth}\includegraphics[width=3.4cm, height=3.4cm]{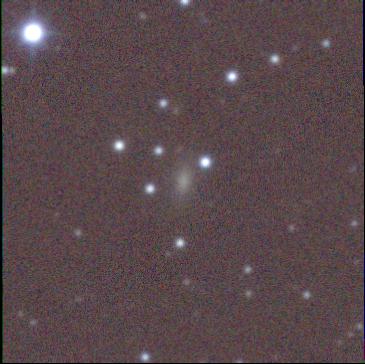}\caption*{188. $\ell$S}\end{subfigure}
\begin{subfigure}[b]{0.19\textwidth}\includegraphics[width=3.4cm, height=3.4cm]{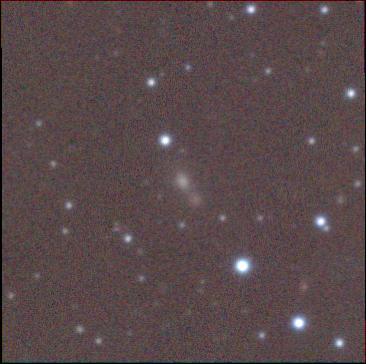}\caption*{189. Irr}\end{subfigure}
\begin{subfigure}[b]{0.19\textwidth}\includegraphics[width=3.4cm, height=3.4cm]{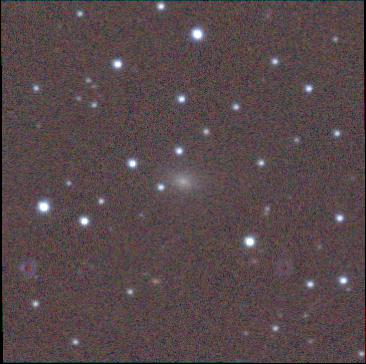}\caption*{190. $\ell$S}\end{subfigure}

\begin{subfigure}[b]{0.19\textwidth}\includegraphics[width=3.4cm, height=3.4cm]{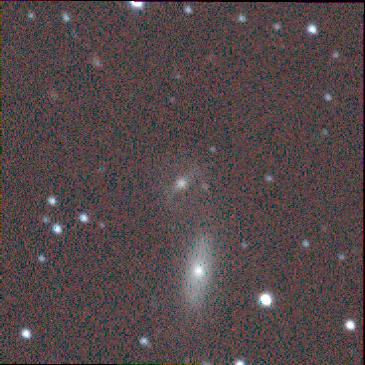}\caption*{191. $m$S}\end{subfigure}
\begin{subfigure}[b]{0.19\textwidth}\includegraphics[width=3.4cm, height=3.4cm]{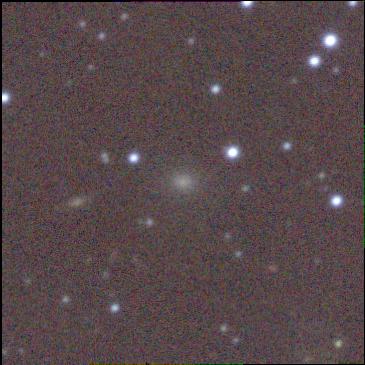}\caption*{192. $\ell$S}\end{subfigure}
\begin{subfigure}[b]{0.19\textwidth}\includegraphics[width=3.4cm, height=3.4cm]{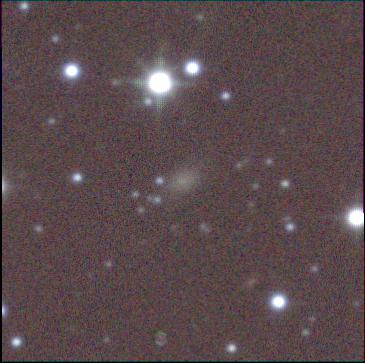}\caption*{193. $\ell$S}\end{subfigure}
\begin{subfigure}[b]{0.19\textwidth}\includegraphics[width=3.4cm, height=3.4cm]{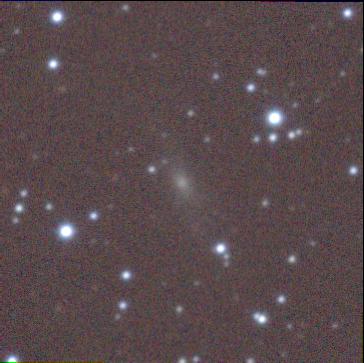}\caption*{194. $e$S}\end{subfigure}
\begin{subfigure}[b]{0.19\textwidth}\includegraphics[width=3.4cm, height=3.4cm]{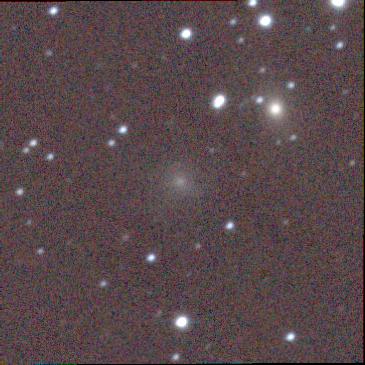}\caption*{195. $m$S}\end{subfigure}

\begin{subfigure}[b]{0.19\textwidth}\includegraphics[width=3.4cm, height=3.4cm]{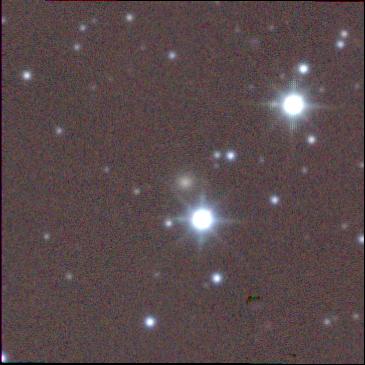}\caption*{196. $\ell$S}\end{subfigure}
\begin{subfigure}[b]{0.19\textwidth}\includegraphics[width=3.4cm, height=3.4cm]{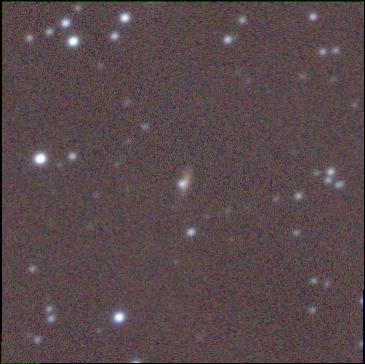}\caption*{197. $\ell$S}\end{subfigure}
\begin{subfigure}[b]{0.19\textwidth}\includegraphics[width=3.4cm, height=3.4cm]{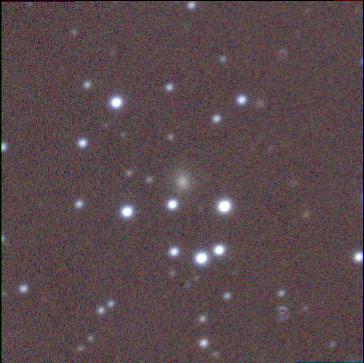}\caption*{198. $\ell$S}\end{subfigure}
\begin{subfigure}[b]{0.19\textwidth}\includegraphics[width=3.4cm, height=3.4cm]{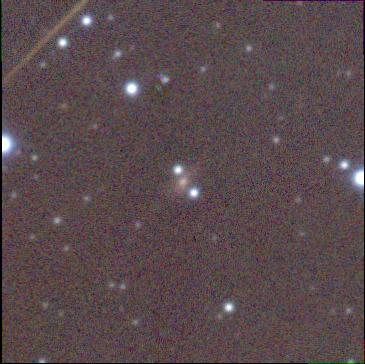}\caption*{199. $e$S}\end{subfigure}
\begin{subfigure}[b]{0.19\textwidth}\includegraphics[width=3.4cm, height=3.4cm]{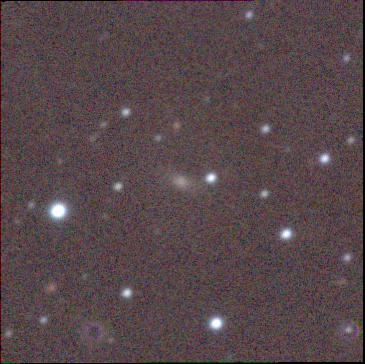}\caption*{200. $\ell$S}\end{subfigure}

\begin{subfigure}[b]{0.19\textwidth}\includegraphics[width=3.4cm,height=3.3cm,cfbox=cyan 1.5pt 0.0pt]{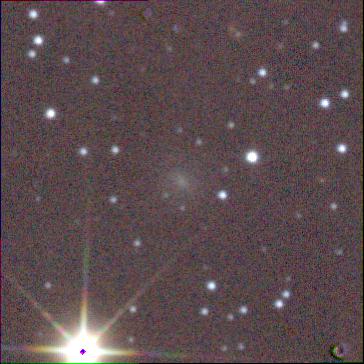}\caption*{201. $\ell$S}\end{subfigure}
\begin{subfigure}[b]{0.19\textwidth}\includegraphics[width=3.4cm, height=3.4cm]{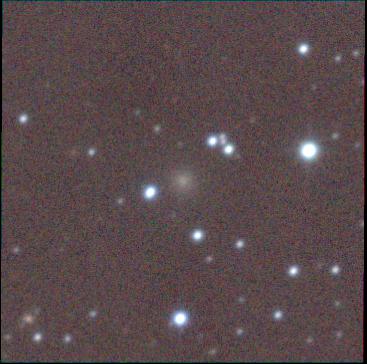}\caption*{202. $\ell$S}\end{subfigure}
\begin{subfigure}[b]{0.19\textwidth}\includegraphics[width=3.4cm, height=3.4cm]{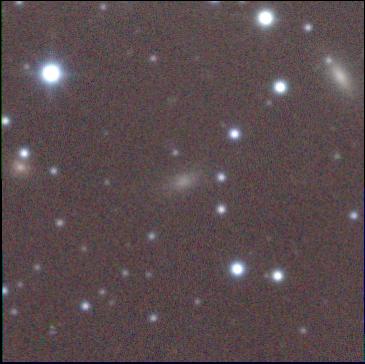}\caption*{203. $\ell$S}\end{subfigure}
\begin{subfigure}[b]{0.19\textwidth}\includegraphics[width=3.4cm, height=3.4cm]{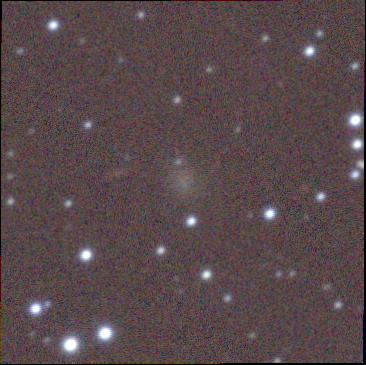}\caption*{204. Irr}\end{subfigure}
\begin{subfigure}[b]{0.19\textwidth}\includegraphics[width=3.4cm, height=3.4cm]{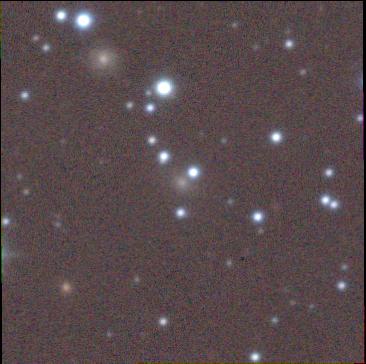}\caption*{205. Irr}\end{subfigure}

\begin{subfigure}[b]{0.19\textwidth}\includegraphics[width=3.4cm, height=3.4cm]{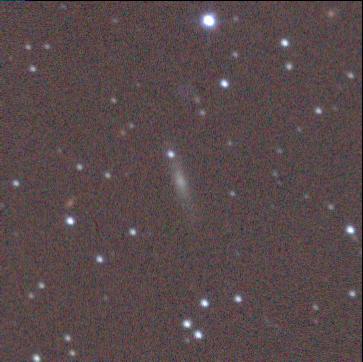}\caption*{206. $m$S}\end{subfigure}
\begin{subfigure}[b]{0.19\textwidth}\includegraphics[width=3.4cm, height=3.4cm]{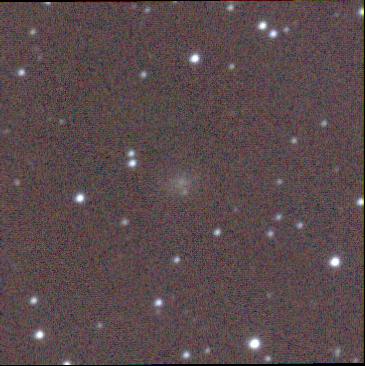}\caption*{207. Irr}\end{subfigure}
\begin{subfigure}[b]{0.19\textwidth}\includegraphics[width=3.4cm, height=3.4cm]{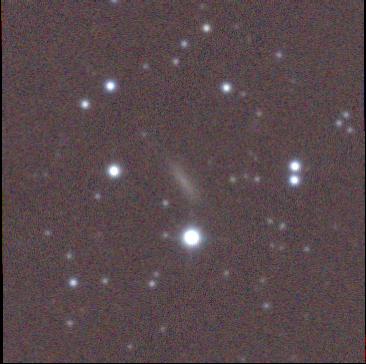}\caption*{208. $m$S}\end{subfigure}
\begin{subfigure}[b]{0.19\textwidth}\includegraphics[width=3.4cm, height=3.4cm]{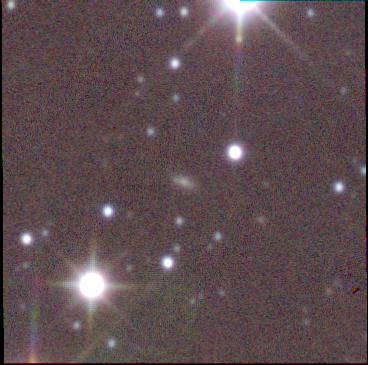}\caption*{209. $\ell$S}\end{subfigure}
\begin{subfigure}[b]{0.19\textwidth}\includegraphics[width=3.4cm, height=3.4cm]{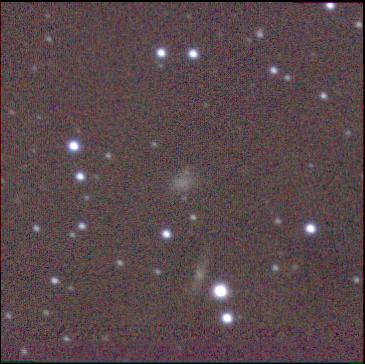}\caption*{210. Irr}\end{subfigure}
\caption*{\textbf{Figure B} -- Continued.}
\end{figure*}

\begin{figure*}
\ContinuedFloat
\centering

\begin{subfigure}[b]{0.19\textwidth}\includegraphics[width=3.4cm, height=3.4cm]{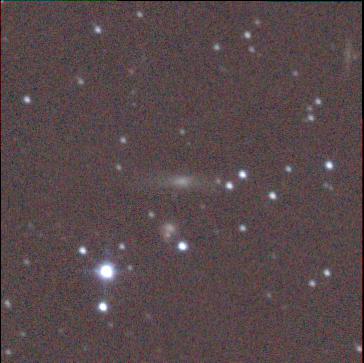}\caption*{211. $e$S}\end{subfigure}
\begin{subfigure}[b]{0.19\textwidth}\includegraphics[width=3.4cm, height=3.4cm]{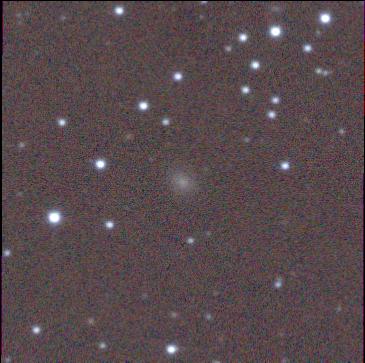}\caption*{212. Irr}\end{subfigure}
\begin{subfigure}[b]{0.19\textwidth}\includegraphics[width=3.4cm, height=3.4cm]{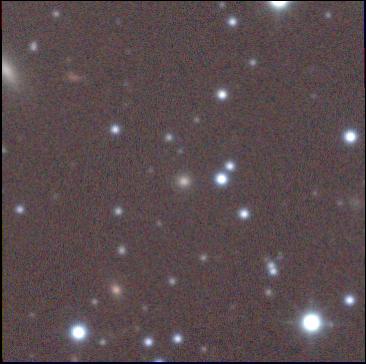}\caption*{213. E/S0}\end{subfigure}
\begin{subfigure}[b]{0.19\textwidth}\includegraphics[width=3.4cm, height=3.4cm]{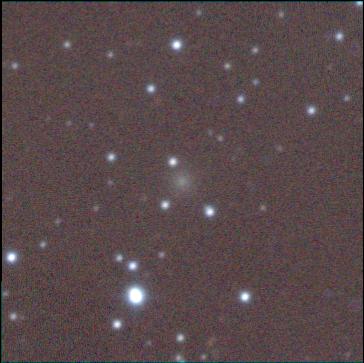}\caption*{214. Irr}\end{subfigure}
\begin{subfigure}[b]{0.19\textwidth}\includegraphics[width=3.4cm, height=3.4cm]{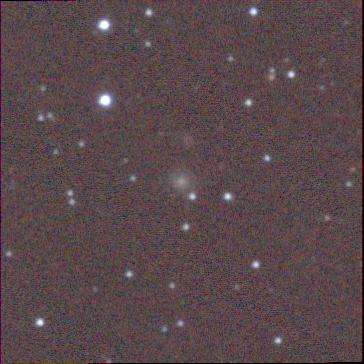}\caption*{215. $\ell$S}\end{subfigure}

\begin{subfigure}[b]{0.19\textwidth}\includegraphics[width=3.4cm, height=3.4cm]{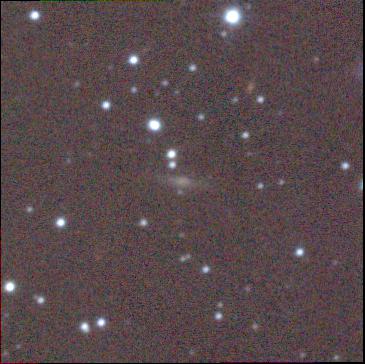}\caption*{216. $e$S}\end{subfigure}
\begin{subfigure}[b]{0.19\textwidth}\includegraphics[width=3.4cm, height=3.4cm]{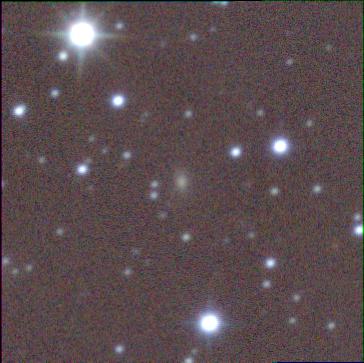}\caption*{217. Irr}\end{subfigure}
\begin{subfigure}[b]{0.19\textwidth}\includegraphics[width=3.4cm, height=3.4cm]{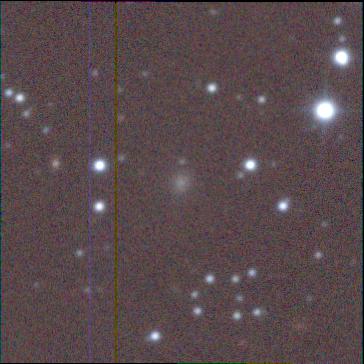}\caption*{218. Irr}\end{subfigure}
\begin{subfigure}[b]{0.19\textwidth}\includegraphics[width=3.4cm, height=3.4cm]{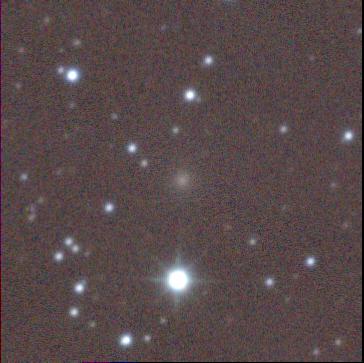}\caption*{219. $\ell$S}\end{subfigure}
\begin{subfigure}[b]{0.19\textwidth}\includegraphics[width=3.4cm,height=3.3cm,cfbox=cyan 1.5pt 0.0pt]{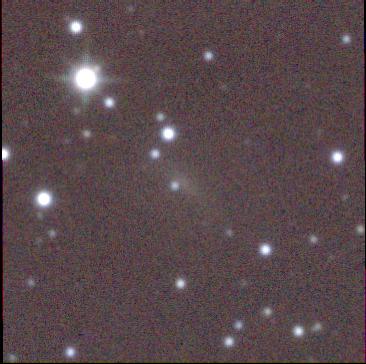}\caption*{220. $\ell$S}\end{subfigure}

\begin{subfigure}[b]{0.19\textwidth}\includegraphics[width=3.4cm, height=3.4cm]{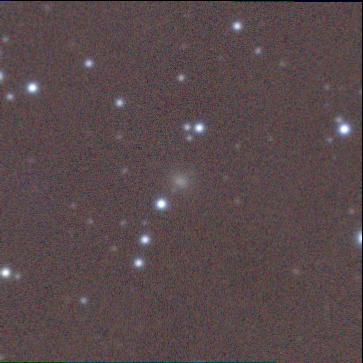}\caption*{221. $\ell$S}\end{subfigure}
\begin{subfigure}[b]{0.19\textwidth}\includegraphics[width=3.4cm, height=3.4cm]{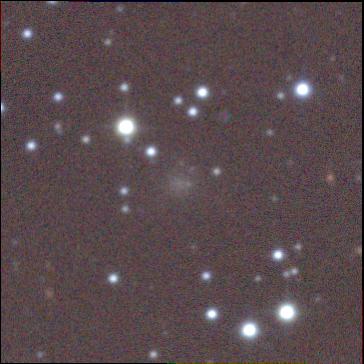}\caption*{222. Irr}\end{subfigure}
\begin{subfigure}[b]{0.19\textwidth}\includegraphics[width=3.4cm, height=3.4cm]{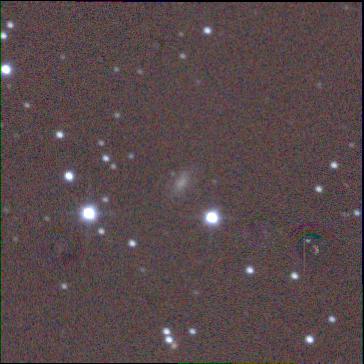}\caption*{223. $\ell$S}\end{subfigure}
\begin{subfigure}[b]{0.19\textwidth}\includegraphics[width=3.4cm, height=3.4cm]{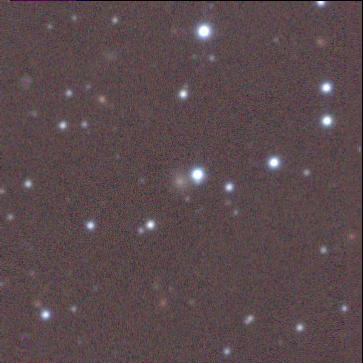}\caption*{224. Irr}\end{subfigure}
\begin{subfigure}[b]{0.19\textwidth}\includegraphics[width=3.4cm, height=3.4cm]{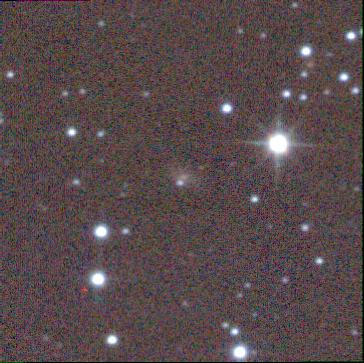}\caption*{225. Irr}\end{subfigure}

\begin{subfigure}[b]{0.19\textwidth}\includegraphics[width=3.4cm, height=3.4cm]{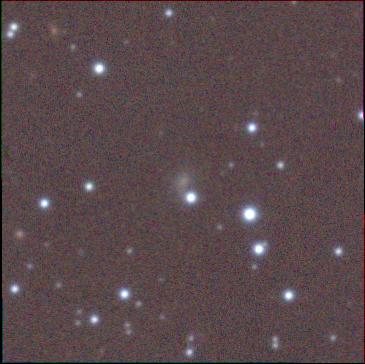}\caption*{226. Irr}\end{subfigure}
\begin{subfigure}[b]{0.19\textwidth}\includegraphics[width=3.4cm, height=3.4cm]{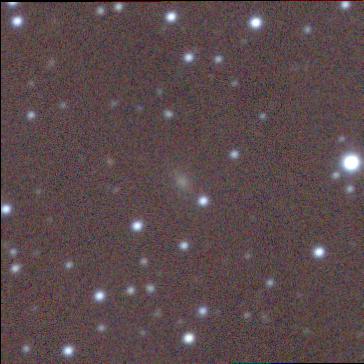}\caption*{227. $\ell$S}\end{subfigure}
\begin{subfigure}[b]{0.19\textwidth}\includegraphics[width=3.4cm, height=3.4cm]{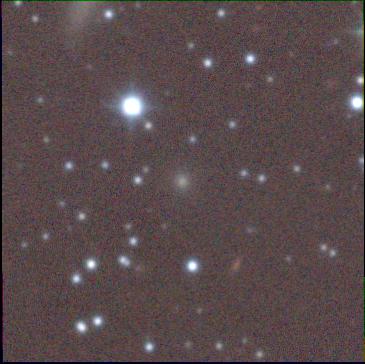}\caption*{228. $\ell$S}\end{subfigure}
\begin{subfigure}[b]{0.19\textwidth}\includegraphics[width=3.4cm, height=3.4cm]{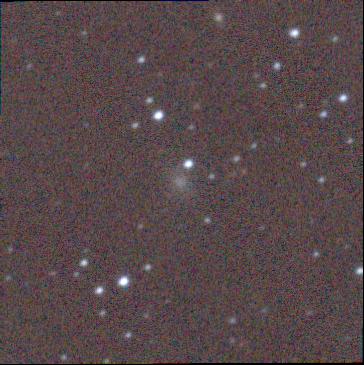}\caption*{229. Irr}\end{subfigure}
\begin{subfigure}[b]{0.19\textwidth}\includegraphics[width=3.4cm, height=3.4cm]{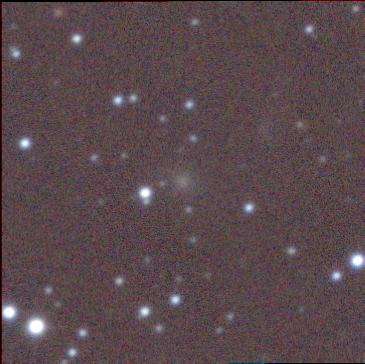}\caption*{230. Irr}\end{subfigure}

\begin{subfigure}[b]{0.19\textwidth}\includegraphics[width=3.4cm, height=3.4cm]{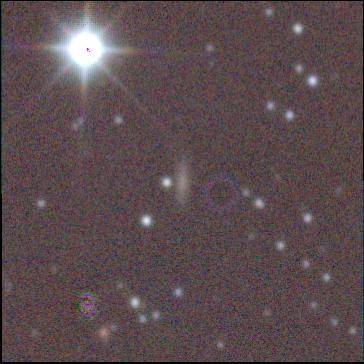}\caption*{231. $\ell$S}\end{subfigure}
\begin{subfigure}[b]{0.19\textwidth}\includegraphics[width=3.4cm,height=3.3cm,cfbox=cyan 1.5pt 0.0pt]{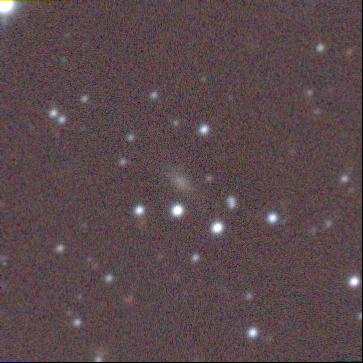}\caption*{232. $\ell$S}\end{subfigure}
\begin{subfigure}[b]{0.19\textwidth}\includegraphics[width=3.4cm, height=3.4cm]{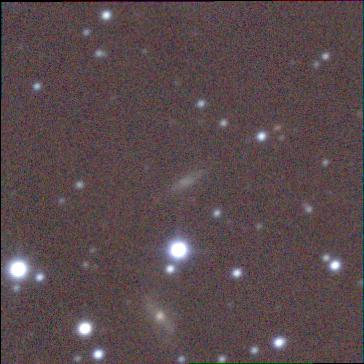}\caption*{233. $e$S}\end{subfigure}
\begin{subfigure}[b]{0.19\textwidth}\includegraphics[width=3.4cm, height=3.4cm]{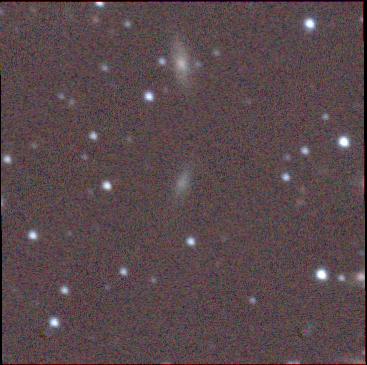}\caption*{234. $\ell$S}\end{subfigure}
\begin{subfigure}[b]{0.19\textwidth}\includegraphics[width=3.4cm, height=3.4cm]{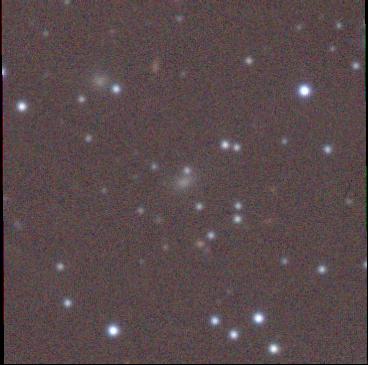}\caption*{235. $\ell$S}\end{subfigure}

\begin{subfigure}[b]{0.19\textwidth}\includegraphics[width=3.4cm, height=3.4cm]{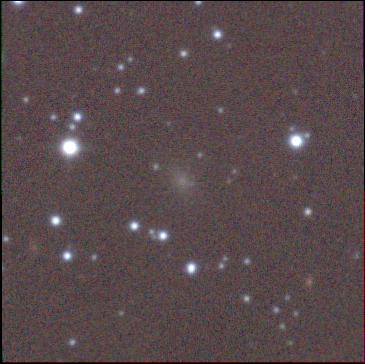}\caption*{236. Irr}\end{subfigure}
\begin{subfigure}[b]{0.19\textwidth}\includegraphics[width=3.4cm, height=3.4cm]{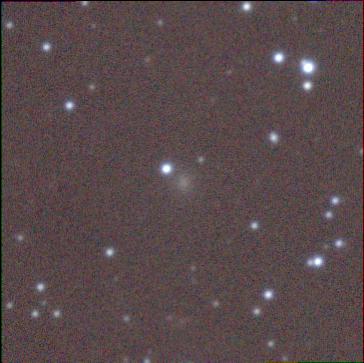}\caption*{237. Irr}\end{subfigure}
\begin{subfigure}[b]{0.19\textwidth}\includegraphics[width=3.4cm, height=3.4cm]{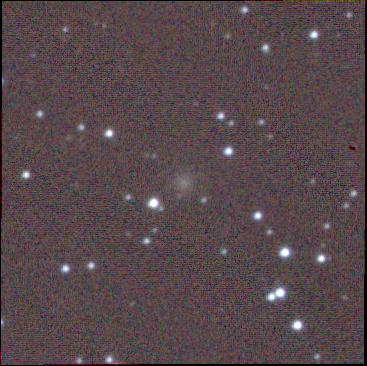}\caption*{238. Irr}\end{subfigure}
\begin{subfigure}[b]{0.19\textwidth}\includegraphics[width=3.4cm, height=3.4cm]{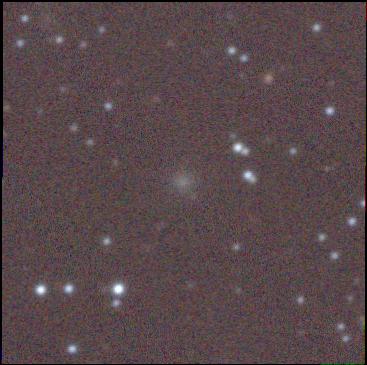}\caption*{239. Irr}\end{subfigure}
\begin{subfigure}[b]{0.19\textwidth}\includegraphics[width=3.4cm,height=3.3cm,cfbox=cyan 1.5pt 0.0pt]{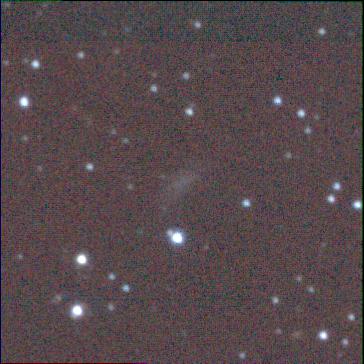}\caption*{240. Irr}\end{subfigure}
\caption*{\textbf{Figure B} -- Continued.}
\end{figure*}

\begin{figure*}
\ContinuedFloat
\centering
\begin{subfigure}[b]{0.19\textwidth}\includegraphics[width=3.4cm, height=3.4cm]{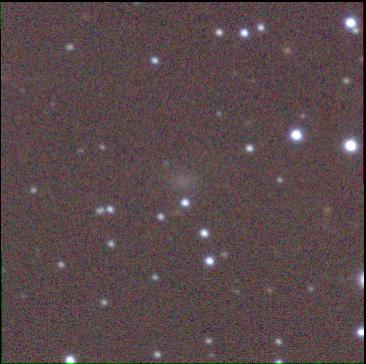}\caption*{241. Irr}\end{subfigure}
\begin{subfigure}[b]{0.19\textwidth}\includegraphics[width=3.4cm, height=3.4cm]{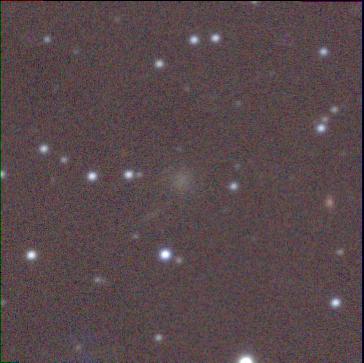}\caption*{242. Irr}\end{subfigure}
\begin{subfigure}[b]{0.19\textwidth}\includegraphics[width=3.4cm, height=3.4cm]{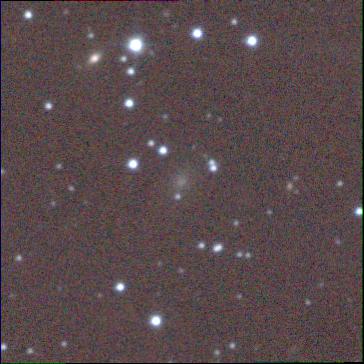}\caption*{243. Irr}\end{subfigure}
\begin{subfigure}[b]{0.19\textwidth}\includegraphics[width=3.4cm, height=3.4cm]{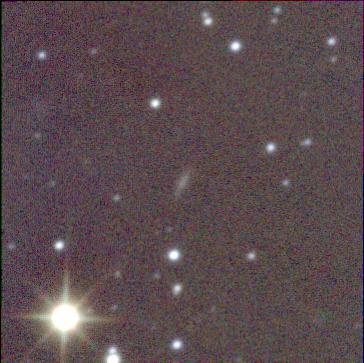}\caption*{244. $\ell$S}\end{subfigure}
\begin{subfigure}[b]{0.19\textwidth}\includegraphics[width=3.4cm, height=3.4cm]{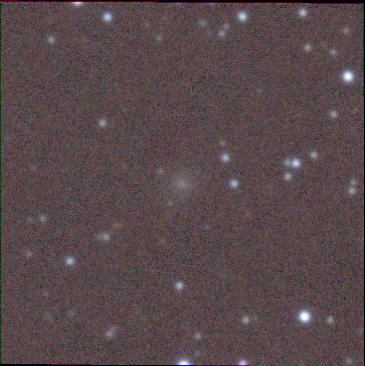}\caption*{245. Irr}\end{subfigure}

\begin{subfigure}[b]{0.19\textwidth}\includegraphics[width=3.4cm, height=3.4cm]{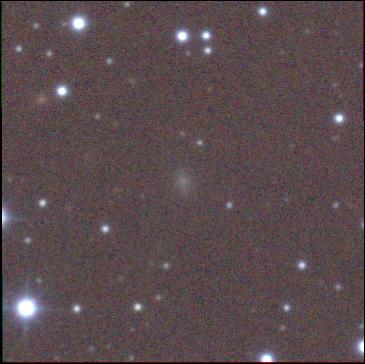}\caption*{246. Irr}\end{subfigure}
\begin{subfigure}[b]{0.19\textwidth}\includegraphics[width=3.4cm, height=3.4cm]{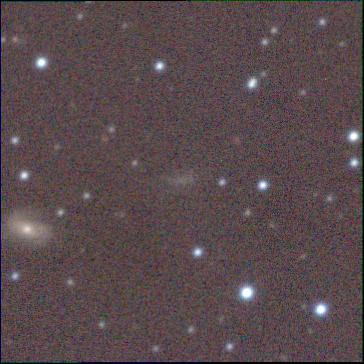}\caption*{247. Irr}\end{subfigure}
\begin{subfigure}[b]{0.19\textwidth}\includegraphics[width=3.4cm, height=3.4cm]{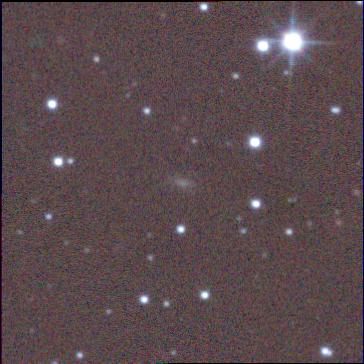}\caption*{248. $\ell$S}\end{subfigure}
\begin{subfigure}[b]{0.19\textwidth}\includegraphics[width=3.4cm,height=3.3cm,cfbox=cyan 1.5pt 0.0pt]{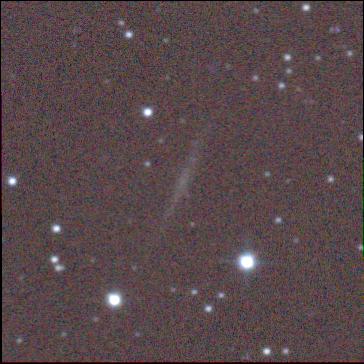}\caption*{249. $m$S}\end{subfigure}
\begin{subfigure}[b]{0.19\textwidth}\includegraphics[width=3.4cm, height=3.4cm]{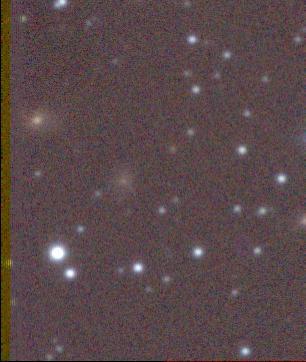}\caption*{250. Irr}\end{subfigure}

\begin{subfigure}[b]{0.19\textwidth}\includegraphics[width=3.4cm, height=3.4cm]{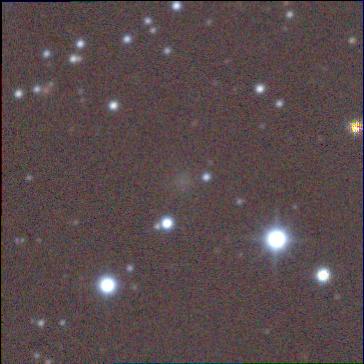}\caption*{251. Irr}\end{subfigure}
\begin{subfigure}[b]{0.19\textwidth}\includegraphics[width=3.4cm, height=3.4cm]{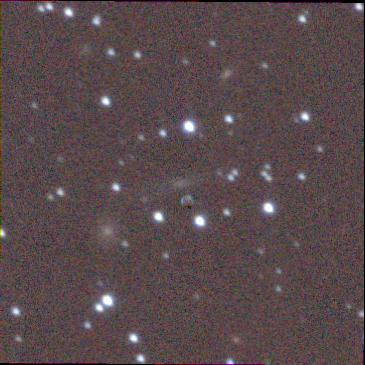}\caption*{252. $\ell$S}\end{subfigure}
\begin{subfigure}[b]{0.19\textwidth}\includegraphics[width=3.4cm, height=3.4cm]{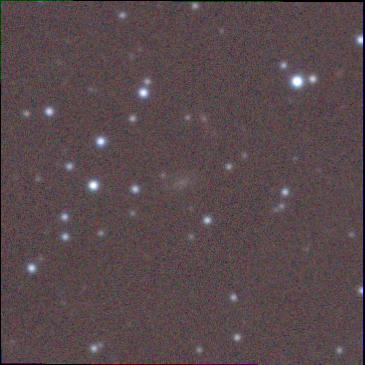}\caption*{253. Irr}\end{subfigure}
\begin{subfigure}[b]{0.19\textwidth}\includegraphics[width=3.4cm, height=3.4cm]{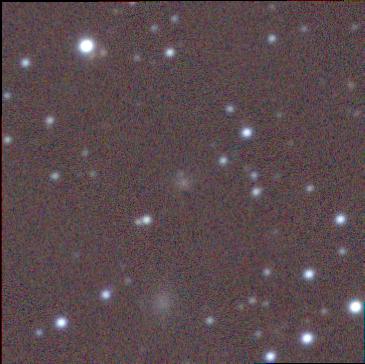}\caption*{254. Irr}\end{subfigure}
\begin{subfigure}[b]{0.19\textwidth}\includegraphics[width=3.4cm,height=3.3cm,cfbox=cyan 1.5pt 0.0pt]{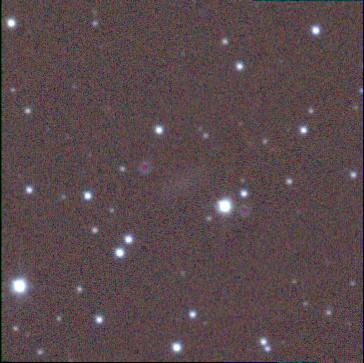}\caption*{255. Irr}\end{subfigure}

\begin{subfigure}[b]{0.19\textwidth}\includegraphics[width=3.4cm,height=3.3cm,cfbox=cyan 1.5pt 0.0pt]{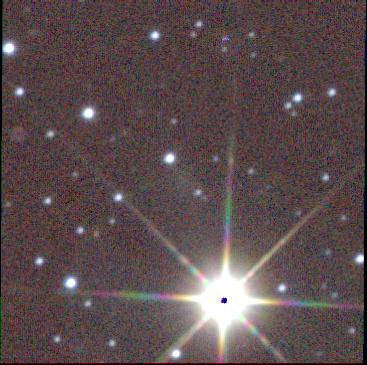}\caption*{256. Irr}\end{subfigure}
\begin{subfigure}[b]{0.19\textwidth}\includegraphics[width=3.4cm,height=3.3cm,cfbox=cyan 1.5pt 0.0pt]{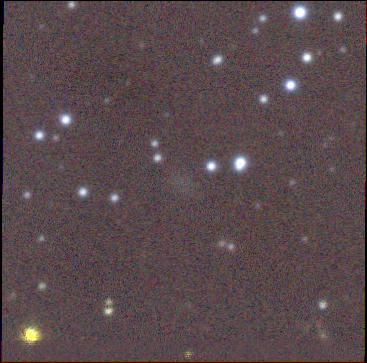}\caption*{257. Irr}\end{subfigure}
\begin{subfigure}[b]{0.19\textwidth}\includegraphics[width=3.4cm,height=3.3cm,cfbox=cyan 1.5pt 0.0pt]{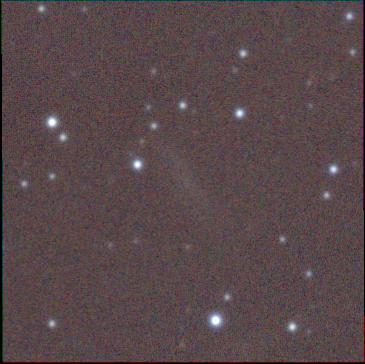}\caption*{258. Irr}\end{subfigure}
\begin{subfigure}[b]{0.19\textwidth}\includegraphics[width=3.4cm,height=3.3cm,cfbox=cyan 1.5pt 0.0pt]{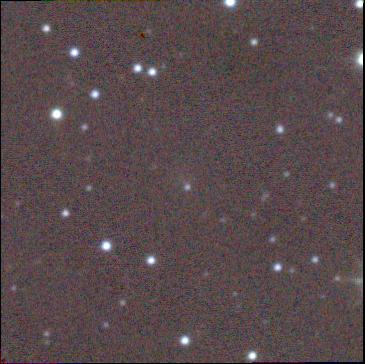}\caption*{259. Irr}\end{subfigure}
\begin{subfigure}[b]{0.19\textwidth}\includegraphics[width=3.4cm,height=3.3cm,cfbox=cyan 1.5pt 0.0pt]{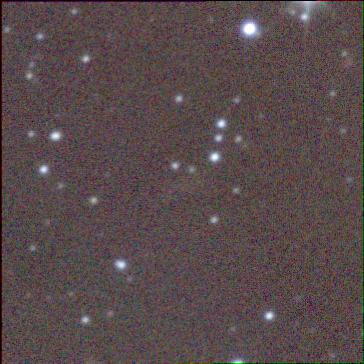}\caption*{260. Irr}\end{subfigure}

\begin{subfigure}[b]{0.19\textwidth}\includegraphics[width=3.4cm,height=3.3cm,cfbox=cyan 1.5pt 0.0pt]{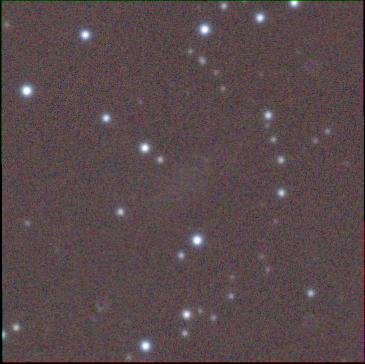}\caption*{261. Irr}\end{subfigure}
\caption*{\textbf{Figure B} -- Continued.}
\end{figure*}

\end{document}